\documentclass[times,final]{elsarticle}

\usepackage{jcomp}
\usepackage{framed,multirow}

\usepackage{amssymb}
\usepackage{latexsym}

\usepackage{lipsum}
\usepackage{amsmath,accents}
\usepackage[labelfont={small}, font={normal}]{subfig}
\usepackage{epstopdf}
\usepackage{graphicx}

\usepackage{lineno}
\usepackage{enumitem}
\usepackage[utf8]{inputenc}

\usepackage[dvipsnames]{xcolor}
\usepackage{url}
\usepackage{xcolor}
\definecolor{newcolor}{rgb}{.8,.349,.1}
\journal{Journal of Computational Physics}

\begin{document}

\verso{Sharma and Koch }

\begin{frontmatter}

\title{Finite difference method in prolate spheroidal coordinates for freely suspended spheroidal particles in linear flows of viscous and viscoelastic fluids}
\author[a]{Arjun Sharma}
\address[a]{Sibley School of Mechanical and Aerospace Engineering, Cornell University, Ithaca, NY, 14853, USA}
\author[b]{Donald L. Koch}
\address[b]{Robert Frederick Smith School of Chemical and Biomolecular Engineering, Cornell University, Ithaca, NY 14853, USA}

\begin{abstract}
	A finite difference scheme is used to develop a numerical method to solve the flow of an unbounded viscoelastic fluid with zero to moderate inertia around a prolate spheroidal particle. The equations are written in prolate spheroidal coordinates, and the shape of the particle is exactly resolved as one of the coordinate surfaces representing the inner boundary of the computational domain. As the prolate spheroidal grid is naturally clustered near the particle surface, good resolution is obtained in the regions where the gradients of relevant flow variables are most significant. This coordinate system also allows large domain sizes with a reasonable number of mesh points to simulate unbounded fluid around a particle. Changing the aspect ratio of the inner computational boundary enables simulations of different particle shapes ranging from a sphere to a slender fiber. Numerical studies of the latter particle shape allow testing of slender body theories. The mass and momentum equations are solved with a Schur complement approach allowing us to solve the zero inertia case necessary to isolate the viscoelastic effects. The singularities associated with the coordinate system are overcome using L'Hopital's rule. A straightforward imposition of conditions representing a time-varying combination of linear flows on the outer boundary allows us to study various flows with the same computational domain geometry. {For the special but important case of zero fluid and particle inertia we obtain a novel formulation that satisfies the force- and torque-free constraint in an iteration-free manner.}  The numerical method is demonstrated for various flows of Newtonian and viscoelastic fluids around spheres and spheroids (including those with large aspect ratio).  Good agreement is demonstrated with existing theoretical and numerical results.
\end{abstract}
\end{frontmatter}

\section{Introduction}\label{sec:Introduction}
The flow of viscoelastic or polymeric fluids around solid particles of various shapes is important in many industrial processes. A particle shape is chosen to achieve the manufactured product's specific purpose or property. For example, fibers allow the desired anisotropy in the roll-to-roll manufacturing of high aspect ratio, low resistance films for flexible and transparent electronics \cite{mutiso2013integrating,yin2010inkjet}. In hydraulic fracturing \cite{barbati2016complex}, spheres may be used as proppants to keep the pores of fractured rocks from closing. In extrusion molding and fiber spinning  \cite{breitenbach2002melt,huang2003review,nakajima1994advanced,chae2008making} spheres or fibers may be added to the fluid to impart strength to the finished product. A spheroid is a convenient shape to span a range of aspect ratios that can be synthesized by dispersing {polystyrene} spheres in a {solution of polyvinyl alcohol}, followed by drying the solution into thin sheets. The particles obtained after heating, stretching, and cooling these sheets \cite{ho1993preparation} may be used in experiments \cite{subramony2017employing} or industrial applications (such as the ones discussed above) involving the flow of viscoelastic fluids around particles. 
Viscoelasticity arises in these fluids from the underlying polymer molecules. Viscoelastic fluids exhibit several properties such as shear thinning and first normal stress difference (rod-climbing) due to the response of the polymers to the imposed flow field. Therefore, numerical computations using specific polymer constitutive models are very useful to isolate the origins of novel flow physics from the interaction of the polymers with the suspended particles. Furthermore, due to several parameters defining the properties of a viscoelastic fluid, such as the polymer mobility, its maximum extensibility, and relaxation time \cite{bird2016polymer}, computations complement laboratory experiments by exploring a wider range of these parameters. Studies concerning moderate inertial effects such that the flow is not turbulent but leads to mechanisms (absent in Stokes flow) such as a significant Saffman lift force \cite{saffman1965lift} on a particle are relevant to several engineering applications and natural phenomena that involve particulate flows. These include air and water pollution, pneumatic and slurry transport, fluidized bed combustion, mineral separation, hemodynamics, and sedimentation in rivers \cite{bagchi2002effect,shi2019lift}.

The numerical challenges in studying the flow of particle suspensions in the applications mentioned above are two-fold: fluid-particle interaction and particle-particle interaction. To incorporate both these effects, numerical computations involving more than one particle resort to the immersed boundary method (IBM) \cite{mittal2005immersed,griffith2020immersed}. These are useful in studying dense particle suspensions. However, the no-slip condition on the particle surface is not directly imposed in IBM. Instead, the particle region is modeled using fictitious forces required to enforce the necessary no-slip condition. The diffuse particle-solid interface does not fully resolve the large polymer stress gradients near the surface. In several of the scenarios above, the particle concentration is dilute enough, so the particle-particle interaction is rare. {Thus, dilute particle suspensions can be modeled as an ensemble of several isolated particles in an unbounded fluid.} The fluid-particle interaction is often analytical in Stokes flow (i.e. flow of inertia-less Newtonian fluid). However, it is complex for a Newtonian fluid with moderate inertia or a viscoelastic fluid because the particle-induced disturbance affects the velocity field in a way that alters the interplay of viscous and inertial or viscous and elastic forces. Therefore, valuable and more accurate physical insight is obtained by studying the flow of a fluid around an isolated particle where the no-slip on the particle surface is imposed. Such numerical studies can also {qualitatively} complement IBM simulations {(where the particle-particle interaction is incorporated)} for dense suspensions, such as in \cite{jain2021transient}. {Furthermore, accurate force and torque coefficients for spheroids of different aspect ratios obtained from single particle simulations can be useful in Lagrangian models.} In this paper, we describe a numerical method based on the finite-difference approximations to model the flow of viscoelastic fluid around a prolate spheroidal particle. The equations are solved in a particle-fixed reference frame which rotates and translates with the spheroid.

Computational fluid mechanics of viscoelastic fluids, which first started in the 1970s, is now a well-established research field \cite{shaqfeh2019rheology,alves2021numerical}. Viscoelasticity of linear polymers is modeled through continuum equations governing second moment of the polymer end-to-end distance averaged over the polymer configuration \cite{bird2016polymer}.
Interesting and novel physical phenomena arise in the industrially relevant parameter regimes when the polymer stretch is significant, leading to significant polymer stress and its gradients. It is in these parameter regimes where unique numerical challenges also arise that have required ingenious solutions in the past, such as the log-conformation formulation by Fattal and Kupferman (2004) \cite{fattal2004constitutive} and algebraic numerical treatment of the polymer stretch by Richter et al. (2010) \cite{richter2010simulations} to ensure finite polymer length and hence finite polymer stress. 
Accurately resolving the polymeric flow around particles with large aspect ratios provides yet another numerical challenge as it requires high spatial resolution to accurately model the large polymer stress gradients along with the thin particle surface. Therefore, this paper uses a prolate spheroidal coordinate system to discretize the governing equations spatially. This exactly models the particle surface as one of the coordinate surfaces and is well suited to study the flow around a prolate spheroid, just as spherical \cite{santelli2021finite} or cylindrical \cite{verzicco1996finite,morinishi2004fully,desjardins2008high} coordinate systems are beneficial in studying the flow around a sphere and in a cylindrical pipe respectively. Furthermore, even a uniform grid in our chosen coordinate system is naturally more clustered (than a Cartesian grid) near the particle surface in the Euclidean sense, allowing enhanced spatial resolution in the regions that require it the most.

The flow of viscoelastic fluids around particles in the aforementioned industrial applications undergoes a series of local linear flows with time in a Lagrangian reference frame. For example, in fiber spinning, the material is first sheared within the spinneret and then pulled by the drawing mechanism leading to a strong uni-axial extensional flow before solidifying to form a fiber. The computational domain consists of the prolate spheroidal surface of the solid particle as the inner boundary and a nearly spherical outer surface where the imposed flow boundary conditions are applied. On the outer boundary of our computational domain, we can apply any choice of imposed stationary, time-varying, or alternating linear flow fields that mimic industrial scenarios. This allows us greater flexibility in the choice of imposed flow conditions as compared to previous numerical techniques where the computational domain is problem specific such as using parallel, oppositely moving walls to obtain simple shear flow \cite{yang2016numerical,d2014bistability} or a cylindrical outer surface for a uniform far-field flow \cite{binagia2020swimming}. In our method, boundary conditions can be changed with time within a simulation.

Theoretical studies and numerical simulations are complementary. Theoretical studies are however often done either for spherical \cite{koch2006stress,einarsson2017spherical}, or for slender particles \cite{harlen1993simple,leal1975slow}. For the latter, a matched asymptotic expansion in particle aspect ratio, also known as slender body theory, \cite{cox1970motion,cox1971motion,batchelor1970slender} is used to obtain useful physical insight by considering the particle aspect ratio to be very large. Furthermore, flow around a particle is solved in an unbounded fluid in such theoretical developments. The choice of a prolate spheroidal coordinate system allows us to simulate the flow around a large aspect ratio particle in a very large computational domain with good spatial resolution in the regions near the particle surface where the velocity and polymer stress gradients are expected to be large while maintaining a reasonable number of mesh points. Therefore, our numerical method is a suitable testing ground for slender body theories. As we demonstrate later, we can simulate the flow around a sphere by allowing the particle aspect ratio to be 1+$\epsilon$, where $\epsilon$ is a very small positive number.

We use a finite difference method to discretize the spatial gradients. Polymer viscosity is large in polymer melts and concentrated polymer solutions. An increase in the solvent viscosity in dilute polymer solutions leads to a large polymer relaxation time, leading to interesting mechanisms that are numerically challenging to resolve. Therefore, viscoelastic fluids are generally highly viscous, and numerical studies with negligible fluid and particle inertia are appropriate to study the effects of viscoelasticity \cite{yang2016numerical,d2010numerical,d2014bistability}. Furthermore, the solutions of uniform and linear flows of an unbounded inertia-less Newtonian fluid around spheres and spheroids \cite{chwang1975hydromechanics} are analytically known. Thus, numerical studies around such particles in the absence of inertia are more relevant in the presence of viscoelasticity.

Some previous investigations of the flow of viscoelastic fluids, such as \cite{yang2016numerical} and \cite{pimenta2017stabilization}, intending to ignore inertia, have used small but finite values of Reynolds number, $Re$ (ratio of inertial to viscous forces). In such numerical solvers, the momentum conservation and incompressibility (mass conservation) equations are solved via a splitting method where momentum equations are advanced in time. The incompressibility is imposed via a pressure Poisson equation. Therefore, additional boundary conditions are required for the pressure field, and Neumann boundary conditions are usually used at the solid surface \cite{gresho1987pressure,sani2006pressure}. The splitting method is more appropriate for large $Re$  flows where the splitting errors due to the introduction of artificial boundary conditions are not dominant. However, when $Re$ is small, splitting errors increase with the dominance of the viscous forces \cite{cai2014efficient}.
To avoid this issue, we solve the coupled system of momentum and incompressibility equation iteratively using a Schur complement method with GMRES \cite{furuichi2011development}. In addition, unlike the splitting method, we can solve a flow with $Re=0$ where the momentum equation is quasi-steady, and the method used does not require us to use time marching.

For a finite $Re$, we incorporate the inertial terms in the momentum equation within the Schur complement method similar to \cite{cai2014efficient}. Therefore, in addition to solving the flow of inertia-less viscoelastic fluid, our method allows us to access the effects of small to moderate inertia (with or without viscoelasticity).
Accurate simulations for small (but nonzero) inertia made possible by the large computational domain enable us to test perturbation theories for the limit $Re \ll 1$. Such simulations of a Newtonian fluid with finite inertia for larger aspect ratio prolate spheroids can allow us to test the slender body theories, such as in \cite{subramanian2005inertial}, describing the effect of inertia on fibers.

In the next section, we present governing equations for the mass and the momentum conservation in the fluid along with the polymer constitutive equations in their original and log-conformal \cite{fattal2004constitutive} formulation. We also present the equations governing particle dynamics and treatment of the boundary conditions. Section \ref{sec:TemporalDiscretization} deals with the temporal discretization of the governing equations. This entails the description of the Schur complement method \cite{furuichi2011development} for solving mass and momentum equations, the technique of \cite{richter2010simulations} to treat finite extensibility of a polymer, quaternion formulation to account for the particle orientation, and separate methodologies for zero and finite particle inertia to account for the particle motion. In section \ref{sec:SpatialDiscretization} we illustrate the spatial discretization of the equations using the finite difference method. In this section, we also present the method to treat the coordinate system dependent axis singularities that arise when the governing equations are expressed in the prolate spheroidal coordinate system. This is motivated by a similar method \cite{morinishi2004fully,desjardins2008high} developed for the cylindrical coordinate system. In section \ref{sec:Examples} we demonstrate the robustness and versatility of our numerical solver through examples of a variety of flows past a particle in (a) Stokes flow (inertia-less flows of a Newtonian fluid), (b) flow of a Newtonian fluid with finite inertia, and (c) inertia-less flows of viscoelastic fluids with different constitutive models. Comparison with previous numerical studies or analytical results is provided for each case. Our numerical solver is parallelized using the domain decomposition method implemented in the Message Passing Interface (MPI). All the examples presented are run on more than one processing unit. Finally, we present the conclusions in section \ref{sec:Conclusions}.

\section{Governing equations}\label{sec:GoverningEquations}
\begin{figure}[h!]
\centering
\includegraphics[width=0.5\linewidth]{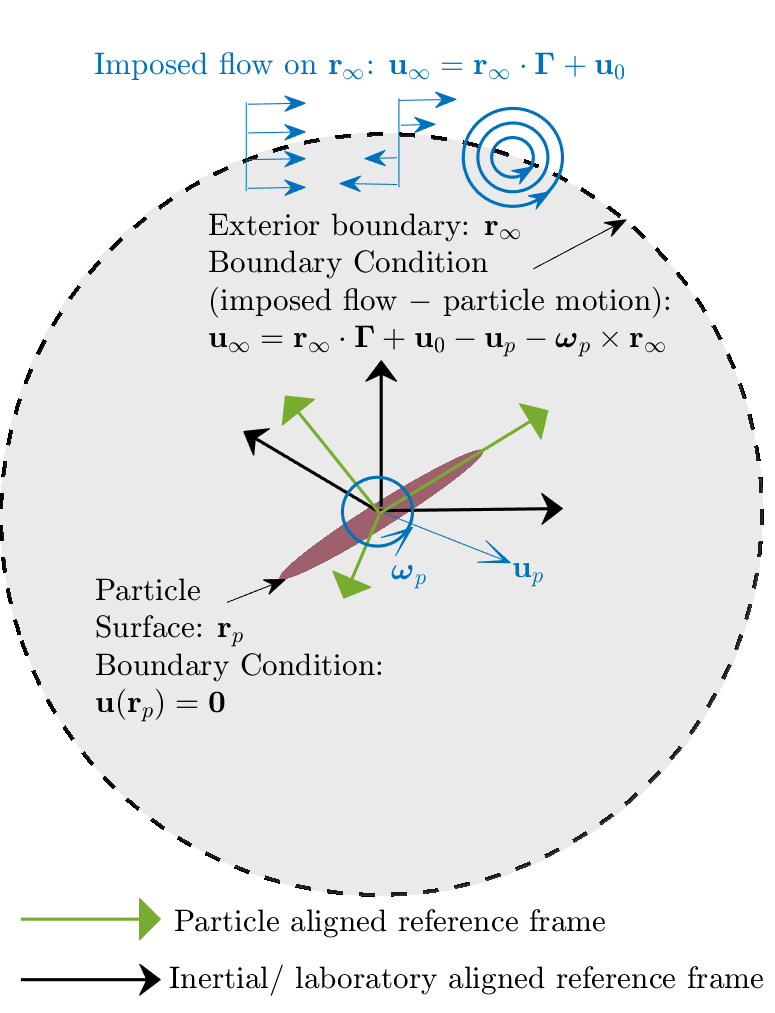}
\caption{Computational domain for the flow around a prolate spheroid: particle surface, $\mathbf{r}_p$ is the inner boundary, and the exterior boundary is $\mathbf{r}_\infty$ indicated with a dashed black curve. $\mathbf{r}_\infty$ is at a large distance from the particle's center and represents a surface in the far-field where the velocity boundary conditions are applied. A velocity field, $\mathbf{r}\cdot \boldsymbol{\Gamma}+\mathbf{u}_0$ with time varying $\boldsymbol{\Gamma}$ and $\mathbf{u}_0$ indicated by blue arrows and curves can be imposed at $\mathbf{r}_\infty$. The gray shaded region is the interior fluid region over which we solve the governing equations.}
\label{fig:computationaldomain}
\end{figure}

We consider the flow of an incompressible viscoelastic fluid around a prolate spheroidal particle in a reference frame rotating and translating with the particle (figure \ref{fig:computationaldomain}).

The equations governing the conservation of mass (incompressibility) and momentum in this rotating reference frame are
\begin{equation}
\nabla \cdot \mathbf{u}=0,\label{eq:MassConsv}
\end{equation}
\begin{equation}
\rho_f \big(\frac{\partial \mathbf{u}}{\partial t}+\textbf{ADV}(\mathbf{u},\mathbf{u}_p,\boldsymbol{\omega}_p;\mathbf{r})\big)=\nabla \cdot \boldsymbol{\sigma},\label{eq:MomentumConsv}
\end{equation}
where
\begin{equation}
\textbf{ADV}(\mathbf{u},\mathbf{u}_p,\boldsymbol{\omega}_p;\mathbf{r})=\frac{d \mathbf{u}_p}{dt}+\mathbf{u}\cdot\nabla \mathbf{u}+2\boldsymbol{\omega}_p\times\mathbf{u}+\boldsymbol{\omega}_p\times\boldsymbol{\omega}_p\times\mathbf{r}+\frac{d \boldsymbol{\omega}_p}{d t}\times\mathbf{r}.\label{eq:AdvectionTerms}
\end{equation}
$\boldsymbol{\omega}_p$ is the angular velocity of the particle, $\mathbf{r}$ and $\mathbf{u}$ are the position and the fluid velocity vector relative to the centroid of the particle, and $\boldsymbol{\sigma}$  is the stress tensor field in the fluid. The stress in a viscoelastic fluid is the sum of the Newtonian solvent stress, $\boldsymbol{\tau}$, (with viscosity $\mu$) and the polymer, $\boldsymbol{\Pi}$, stress,
\begin{equation}
\boldsymbol{\sigma}=\boldsymbol{\tau}+\boldsymbol{\Pi}=-p\boldsymbol{\delta}+2\mu\boldsymbol{e}+\boldsymbol{\Pi},\label{eq:StressDefinition}
\end{equation}
where $p$ is the reduced pressure (the difference between the hydrodynamic and the hydrostatic pressure), $\boldsymbol{e}=(\nabla\mathbf{u}+(\nabla\mathbf{u})^T)/2$ is the strain rate tensor, and $\rho_f$ is the fluid density. A rotating frame fixed with the particle avoids the need to introduce a mesh velocity but leads to the first, third, fourth, and fifth terms on the RHS of equation \eqref{eq:AdvectionTerms}. These arise due to the non-inertial (rotating) reference frame and represent the linear acceleration, Coriolis force, centrifugal force, and angular acceleration. 
As shown in figure \ref{fig:computationaldomain} the computational domain for these equations is bounded by $\mathbf{r}_p$ (representing the particle surface) on the inside, and $\mathbf{r}_\infty$ (representing the far-field) on the outside. The boundary conditions on the velocity field are the imposed flow conditions in the far-field and no-slip and no-penetration on the particle surface. In the frame of reference rotating and translating with the particle, these are,
\begin{align}\begin{split}
\mathbf{u}=\mathbf{0}, \text{ on particle surface}, \hspace{0.2in}
\mathbf{u}=\mathbf{u}_\infty(\mathbf{r})=\mathbf{r}\cdot\boldsymbol{\Gamma}+\mathbf{u}_0-\mathbf{u}_p-\boldsymbol{\omega}_p\times\mathbf{r}, \text{ as }|\mathbf{r}|\rightarrow \infty\approx\mathbf{r}_\infty ,
\end{split}\label{eq:VelBCs}\end{align}
where $\boldsymbol{\Gamma}$ and $\mathbf{u}_0$ are the imposed velocity gradient and uniform velocity field.
The angular and translational velocities of the particle, $\boldsymbol{\omega}_p$ and $\mathbf{u}_p$, may either be imposed or obtained from relevant equations governing the motion of the particle due to the moments and forces acting on the particle. In the latter case, Newton's equations govern the particle motion,
\begin{equation}
\rho_p V_p\frac{d \mathbf{u}_p}{dt}={\mathbf{f}_\text{fluid}+\mathbf{f}_\text{ext.}}
,\label{eq:ForceOnParticle}
\end{equation}
\begin{equation}
\mathbf{I}_p\cdot\frac{d \boldsymbol{\omega}_p}{dt}=\mathbf{q}_\text{fluid}+\mathbf{q}_\text{ext.},\label{eq:TorqueOnParticle}
\end{equation}
{where $V_p$, $\rho_p$ and $\mathbf{I}_p$ are the volume, density and moment of inertia tensor of the particle. $\mathbf{f}_\text{ext.}$ and $\mathbf{q}_\text{ext.}$ are the external imposed force and torque that can be prescribed. In case gravity is present, $\mathbf{f}_\text{ext.}$ includes the buoyancy force acting on the particle due to the difference in particle and fluid densities represented by $(\rho_p-\rho_f)V_p {\mathbf{g}}$. Here, ${\mathbf{g}}$ is the gravity vector in the particle fixed frame. Although it is likely fixed in the inertial reference frame, the orientations of $\mathbf{f}_\text{ext.}$, $\mathbf{q}_\text{ext.}$ and $\mathbf{g}$ vary in the particle reference frame with time for a rotating particle.} 
For a prolate spheroid with minor radius, $r_\text{minor}$ and aspect ratio, $\kappa$,
\begin{equation}
V_p=\frac{4}{3}\pi r_\text{minor}^3\kappa, \hspace{0.2in}\mathbf{I}_p=\frac{1}{5}\rho_pV_pr_\text{minor}^2\begin{bmatrix}
1+\kappa^2&0&0\\0&1+\kappa^2&0\\0&0&2
\end{bmatrix}.
\end{equation}
$\mathbf{f}_\text{fluid}=\mathbf{f}(\boldsymbol{\sigma})$ and $\mathbf{q}_\text{fluid}=\mathbf{q}(\boldsymbol{\sigma})$ are the fluid stress dependent hydrodynamic force and torque acting on the particle defined as,
\begin{equation}
\mathbf{f}(\boldsymbol{\sigma})=\int_{\mathbf{r}_p} dS\hspace{0.1in}\boldsymbol{\sigma}\cdot\mathbf{n},\hspace{0.2in}\mathbf{q}(\boldsymbol{\sigma})=\int_{\mathbf{r}_p} dS\hspace{0.1in}\mathbf{r}\times (\boldsymbol{\sigma}\cdot\mathbf{n}),\label{eq:FluidTorqueandForce}
\end{equation}
where $\mathbf{n}$ is the surface normal. In a viscoelastic fluid $\mathbf{f}_\text{fluid}=\mathbf{f}(\boldsymbol{\sigma})$ and $\mathbf{q}_\text{fluid}=\mathbf{q}(\boldsymbol{\sigma})$ are thus sums of contributions from the Newtonian solvent and polymeric stress,
\begin{equation}
\mathbf{f}_\text{fluid}=\mathbf{f}(\boldsymbol{\tau})+\mathbf{f}(\boldsymbol{\Pi}),\hspace{0.1in}\mathbf{q}_\text{fluid}=\mathbf{q}(\boldsymbol{\tau})+\mathbf{q}(\boldsymbol{\Pi}),\label{eq:TorqueDecompose}
\end{equation}

As considered in section \ref{sec:FiniteParticleInertial} later, for finite particle inertia ($\rho_p\ne0$) equations \eqref{eq:ForceOnParticle} and \eqref{eq:TorqueOnParticle} are numerically integrated in time after discretizing the time derivative on the left-hand side of these equations. However, in the limit of zero particle inertia ($\rho_p=0$), the appropriate conditions are zero net force and torque on the particle. In that case, as discussed in section \ref{sec:ZeroParticleInertial}, {either a secant iteration method (for a massless particle in a fluid with finite density, $\rho_f$) or a novel decomposition of the inertia-less non-Newtonian momentum equation combined with a resistivity formulation} is used to ensure that various components on the right-hand side of these equations balance each other.

Finite difference discretization of the governing equations in the Cartesian coordinate system is exactly satisfied by a uniform flow, but this is not the case in curvilinear coordinates because the spatial gradients of the relevant variables such as velocity, pressure, and polymer stress involve coordinate system dependent metric derivatives \cite{jiang2014free}. Free-stream preservation of the imposed linear flow field far away from the particle is particularly important for the simulations of our interest since we use a large computational domain. Various previously proposed techniques to treat this issue are in \cite{jiang2014free,visbal2002use} and references therein. However, by simply simulating the deviation of the velocity field from its far-field value, the discretization errors associated with the violation of free-stream preservation are removed. In other words, we do not need to numerically simulate the analytical value of the far-field velocity, $\mathbf{u}_\infty$, and pressure, $p_\infty$.

Since the fluid is considered to be incompressible, in the imposed velocity gradient tensor, $\boldsymbol{\Gamma}$, we require $\text{trace}({\boldsymbol{\Gamma}})=0$ and hence we have $\nabla\cdot\mathbf{u}_\infty=0$. The far-field flow (relative to particle motion), $\mathbf{u}(\mathbf{r})_\infty=\mathbf{u}_\infty$ is linear in position, the polymer stress generated by the linear imposed velocity field is a spatially constant value, $\boldsymbol{\Pi}_\infty$. Therefore, the far-field momentum equation is
\begin{equation}
\frac{\partial \mathbf{u}_\infty}{\partial t}+\textbf{ADV}(\mathbf{u}_\infty,\mathbf{u}_p,\boldsymbol{\omega}_p;\mathbf{r})=-\frac{1}{\rho_f}\nabla p_\infty.\label{eq:MomentumConsvFar}
\end{equation}
Hence, the governing mass and momentum equations for the deviation of velocity field and the pressure field from the far-field flow,
\begin{equation}
\widetilde{\mathbf{u}}=\mathbf{u}-\mathbf{u}_\infty,\hspace{0.2in}\widetilde{p}=p-p_\infty.
\end{equation}
are,
\begin{equation}
\nabla \cdot \widetilde{\mathbf{u}}=0,\label{eq:MassConsvpert}
\end{equation}
\begin{equation}
{\rho_f}\big(\frac{\partial \widetilde{\mathbf{u}}}{\partial t}+\widetilde{\textbf{ADV}}(\widetilde{\mathbf{u}},\mathbf{u}_\infty,\boldsymbol{\omega}_p;\mathbf{r})\big)=-\nabla \widetilde{p} +\nabla^2\widetilde{\mathbf{u}}+\nabla \cdot (\boldsymbol{\Pi}-\boldsymbol{\Pi}_\infty),\label{eq:MomentumConsvpert}
\end{equation}
where
\begin{equation}
\widetilde{\textbf{ADV}}(\widetilde{\mathbf{u}},\mathbf{u}_\infty,\boldsymbol{\omega}_p;\mathbf{r})=\widetilde{\mathbf{u}}\cdot(\nabla\widetilde{\mathbf{u}}+\nabla \mathbf{u}_\infty)+\mathbf{u}_\infty\cdot\nabla\widetilde{\mathbf{u}}+2\boldsymbol{\omega}_p\times\widetilde{\mathbf{u}},\label{eq:AdvectionTermspert}
\end{equation}
and the boundary conditions are
\begin{align}\begin{split}
\widetilde{\mathbf{u}}=-\mathbf{u}_\infty(\mathbf{r})=-\mathbf{r}\cdot\boldsymbol{\Gamma}-\mathbf{u}_0+\mathbf{u}_p+\boldsymbol{\omega}_p\times\mathbf{r},\text{ on particle surface },\hspace{0.2in}
\widetilde{\mathbf{u}}=0, \text{ as }|\mathbf{r}|\rightarrow \infty\approx\mathbf{r}_\infty.
\end{split}\label{eq:VelBCspert}\end{align}
Simulating $\widetilde{\mathbf{u}}$ and $\widetilde{p}$ instead of $\mathbf{u}$ and $p$ allows free-stream preservation trivially.  Furthermore, the momentum equation for $\widetilde{\mathbf{u}}$ does not include centrifugal and angular acceleration terms (compare \eqref{eq:AdvectionTerms} with \eqref{eq:AdvectionTermspert}) and hence we do not need to numerically evaluate ${d\boldsymbol{\omega}_p}/{d t}$ at different times during a simulation involving particle rotation.

To model the polymer stress, $\boldsymbol{\Pi}$, we consider various dumbbell models that consider polymer molecules as a spring attached to Brownian beads \cite{bird2016polymer}. These dumbbell-based models define a constitutive equation for the polymer configuration, $\boldsymbol{\Lambda}=\langle\mathbf{qq}\rangle_\text{polymer configuration}$, where $\mathbf{q}$ is the end-to-end vector of the dumbbell and the angle brackets represent the average over polymer configurations. $\boldsymbol{\Lambda}$ is non-dimensionalized with the square of the radius of gyration of the polymer, and the dumbbell models have the following form of the constitutive equation,
\begin{equation}
\frac{\partial \boldsymbol{\Lambda}}{\partial t}+\mathbf{u}\cdot \nabla (\boldsymbol{\Lambda}-\boldsymbol{\Lambda}_\infty)=\nabla \mathbf{u}^\text{T}\cdot\boldsymbol{\Lambda}+\boldsymbol{\Lambda}\cdot\nabla\mathbf{u}{-\frac{1}{\lambda}\mathbf{R}(\boldsymbol{\Lambda})},\label{eq:Constitutive}
\end{equation}
where the polymer convection $\mathbf{u}\cdot \nabla (\boldsymbol{\Lambda}-\boldsymbol{\Lambda}_\infty)=\mathbf{u}\cdot \nabla \boldsymbol{\Lambda} $ is balanced by its stretching ($\nabla \mathbf{u}^\text{T}\cdot\boldsymbol{\Lambda}+\boldsymbol{\Lambda}\cdot\nabla\mathbf{u}$) and relaxation {(${\mathbf{R}(\boldsymbol{\Lambda})}/{\lambda}$) for a polymer solution with relaxation time $\lambda$. Since any valid constitutive equation is materially frame-invariant or objective \cite{graham2018microhydrodynamics}, as long as the components of the $\boldsymbol{\Lambda}$ tensor are expressed in the appropriate reference frame, the constitutive equations retain their form in different reference frames. In other words, unlike the momentum equation governing the fluid velocity (a frame variant quantity), particle rotation does not introduce additional terms in the constitutive equation governing $\boldsymbol{\Lambda}$ (a frame-invariant quantity) in a non-inertial or rotating reference frame. We subtract the spatially constant far-field polymer configuration, $\boldsymbol{\Lambda}_\infty$, before taking the gradient in the convective term to ensure free-stream preservation. 
The polymeric stress, $\boldsymbol{\Pi}$ is directly proportional to the polymer configuration, $\boldsymbol{\Lambda}$ and can be written as}
\begin{equation}
\boldsymbol{\Pi}={\frac{c\mu}{\lambda}\mathbf{F}}(\boldsymbol{\Lambda}),\label{eq:ConstitutiveForce}
\end{equation}
where $c$ is the polymer concentration defined as the ratio of zero shear rate polymer to solvent viscosity ($\mu$).
The exact form of the relaxation tensor, $\mathbf{R}(\boldsymbol{\Lambda})$, and the polymer force-configuration relation, $\mathbf{F}(\boldsymbol{\Lambda})$, depends on the particular model being used to represent the spring force.
If the spring is considered to be Hookean, one obtains the Oldroyd-B model; if the spring force is nonlinear and has finite extensibility, one obtains a FENE (finite extensible nonlinear elastic) model. A closure approximation is needed to obtain continuum-level constitutive equations from the FENE model, and different choices lead to FENE-P and FENE-CR models. The latter does not exhibit shear-thinning. The Giesekus model is also a nonlinear model but has a quadratic nonlinearity. In FENE models, the maximum length of the polymers is $L$ (non-dimensionalized with the radius of gyration). The Giesekus model describes concentrated polymer solutions and melts by including the anisotropic effect of nearby dumbbells through the mobility parameter $\alpha$. See \cite{bird2016polymer} for a review of various constitutive models. $\mathbf{F}(\boldsymbol{\Lambda})$ and $\mathbf{R}(\boldsymbol{\Lambda})$ for the different constitutive models we consider are,
\begin{equation}
\begin{matrix}
\text{Model}:&\text{Oldroyd-B}&\text{FENE-P}&\text{FENE-CR}&\text{Giesekus}\\
{\mathbf{R}(\boldsymbol{\Lambda})}:&{\boldsymbol{\Lambda}-\boldsymbol{\delta}}&{f\boldsymbol{\Lambda}-b\boldsymbol{\delta}}&{f\boldsymbol{\Lambda}-f\boldsymbol{\delta}}&{(\boldsymbol{\Lambda}-\boldsymbol{\delta})-\alpha(\boldsymbol{\Lambda}-\boldsymbol{\delta})^2}\\
\mathbf{F}(\boldsymbol{\Lambda}):&\boldsymbol{\Lambda}-\boldsymbol{\delta}&f\boldsymbol{\Lambda}-b\boldsymbol{\delta}&f\boldsymbol{\Lambda}-f\boldsymbol{\delta}&\boldsymbol{\Lambda}-\boldsymbol{\delta},
\end{matrix}\label{eq:ConstitutiveSpecific}
\end{equation}
where ,
\begin{equation}
f=1/(1-\text{tr}(\boldsymbol{\Lambda})/L^2), \text{ and } b=1/(1-\text{tr}(\boldsymbol{\delta})/L^2).\label{eq:FENEParams}
\end{equation}
The polymer constitutive equation \eqref{eq:Constitutive} (or equation \eqref{eq:ConstitutivePsi} discussed later) is a hyperbolic equation. Therefore, the boundary conditions are required at the locations the streamlines of the velocity field, $\mathbf{u}$, enter the computation domain. At these locations, the boundary condition is a time-dependent, spatially constant polymer configuration tensor, $\boldsymbol{\Lambda}=\boldsymbol{\Lambda}_\infty$, driven by the imposed velocity field and is governed by,
\begin{equation}
\frac{\partial \boldsymbol{\Lambda}_\infty}{\partial t}=\nabla \mathbf{u}_\infty^\text{T}\cdot\boldsymbol{\Lambda}_\infty+\boldsymbol{\Lambda}_\infty\cdot\nabla\mathbf{u}_\infty{-\frac{1}{\lambda}\mathbf{R}(\boldsymbol{\Lambda}_\infty)}.\label{eq:ConstitutiveUndisturbed}
\end{equation}
Using equation \eqref{eq:ConstitutiveForce}, $\boldsymbol{\Pi}_\infty={c\mu/\lambda}\mathbf{F}(\boldsymbol{\Lambda}_\infty)$ and $\mathbf{F}(\boldsymbol{\Lambda}_\infty)$ is evaluated from equation \eqref{eq:ConstitutiveSpecific}.
\subsection{Log Conformal Form of Constitutive equations}
An important property of $\boldsymbol{\Lambda}$ that is preserved by an appropriate constitutive equation such as the ones introduced above is its positive definiteness \cite{fattal2004constitutive}. However, numerical discretization of constitutive equations of the form in equations equation \eqref{eq:Constitutive} and \eqref{eq:ConstitutiveForce} with a model from \eqref{eq:ConstitutiveSpecific} may lead to violation of positive definiteness in the numerical solution of $\boldsymbol{\Lambda}$ at high $De$. This manifests as a numerical instability and is also termed the high Weissenberg number problem (HWNP)-- Weissenberg number is defined as the product of the characteristic flow gradient magnitude and the polymer relaxation time; for steady linear flows Deborah and Weissenberg numbers are equivalent. Fattal and Kupferman \cite{fattal2004constitutive} remedied the HWNP by introducing an equivalent constitutive equation for the matrix logarithm of the conformation tensor, $\boldsymbol{\Lambda}$,
\begin{equation}
\boldsymbol{\Psi}=\log(\boldsymbol{\Lambda}).\label{eq:DefineLog}
\end{equation}
Solving the governing equation for $\boldsymbol{\Psi}$ instead of $\boldsymbol{\Lambda}$ provides a more stable numerical solution as found in numerous numerical studies such as \cite{yang2016numerical,binagia2020swimming,afonso2009log,d2010viscoelasticity,zhong2022modeling} after the seminal work of \cite{fattal2004constitutive}. Fattal and Kupferman \cite{fattal2004constitutive} derived an equation for $\boldsymbol{\Psi}$ based on the eigen-decomposition of the velocity gradient. We use the alternative derivation, provided by Hulsen and the previous authors \cite{hulsen2005flow}, based on the evolution of the principal axes of $\boldsymbol{\Lambda}$ (and hence also $\boldsymbol{\Psi}$ since $\boldsymbol{\Lambda}$ and $\boldsymbol{\Psi}$ have same eigenvectors). We find this form to be simpler in treating cases such as biaxial extensional flow for which the two eigenvalues are identical. We use the Jacobi algorithm provided in \cite{kopp2008efficient} to obtain the eigen-decomposition of $\boldsymbol{\Lambda}$ and $\boldsymbol{\Psi}$,
\begin{equation}
\boldsymbol{\Lambda}=\mathbf{V}\cdot\mathbf{D}_{\boldsymbol{\Lambda}}\cdot\mathbf{V}^T,\hspace{0.2in}\boldsymbol{\Psi}=\mathbf{V}\cdot\mathbf{D}_{\boldsymbol{\Psi}}\cdot\mathbf{V}^T,\label{eq:EigenDecomp}
\end{equation}
where $\mathbf{V}$ is a $3\times 3$ matrix with the eigenvectors, $\mathbf{v}_i,i\in[1,3]$ as its columns and
$\mathbf{D}_{\boldsymbol{\Lambda}}$ and \begin{equation}\mathbf{D}_{\boldsymbol{\Psi}}=\log(\mathbf{D}_{\boldsymbol{\Lambda}}),\label{eq:EigenDecomp2}\end{equation} are diagonal matrices with eigenvalues $\lambda_i$ and $\psi_i=\log(\lambda_i),i\in[1,3]$ as their entries. The governing equation for the matrix logarithm, $\boldsymbol{\Psi}$ is,
\begin{equation}
\frac{\partial \boldsymbol{\Psi}}{\partial t}+\mathbf{u}\cdot \nabla( \boldsymbol{\Psi}- \boldsymbol{\Psi}_\infty)=\textbf{SR}(\boldsymbol{\Psi},\mathbf{u}),\label{eq:ConstitutivePsi}
\end{equation}
where,
\begin{equation}
\textbf{SR}(\boldsymbol{\Psi},\mathbf{u})=2\Sigma_{i=1}^3L_{ii}\mathbf{v}_i\mathbf{v}_i+\Sigma_{i=1}^3\Sigma_{j=1,j\ne i}^3\frac{\psi_i-\psi_j}{\lambda_i-\lambda_j}(\lambda_jL_{ij}+\lambda_iL_{ji})\mathbf{v}_i\mathbf{v}_j{- \exp(-\boldsymbol{\Psi})\cdot\mathbf{R}(\exp(\boldsymbol{\Psi}))},\label{eq:StretchingRelaxation}
\end{equation}
$\mathbf{L}=\nabla\mathbf{u}$ is the velocity gradient tensor, $\exp(\boldsymbol{\Psi})=\mathbf{V}\cdot\mathbf{D}_{\boldsymbol{\Lambda}}\cdot\mathbf{V}^T$ and $\exp(-\boldsymbol{\Psi})=\mathbf{V}\cdot\mathbf{D}_{\boldsymbol{\Lambda}}^{-1}\cdot\mathbf{V}^T$. The three terms of $\textbf{SR}(\boldsymbol{\Psi},\mathbf{u})$ in equation \eqref{eq:StretchingRelaxation} represent the stretching of eigenvectors by $L_{ii}$, rotation of eigenvectors by vorticity and their relaxation respectively \cite{hulsen2005flow}. When two eigenvalues are identical \cite{hulsen2005flow} in the second term of equation \eqref{eq:StretchingRelaxation},
\begin{equation}
\lim\limits_{\lambda_i\rightarrow\lambda_j}\frac{\psi_i-\psi_j}{\lambda_i-\lambda_j}(\lambda_jL_{ij}+\lambda_iL_{ji})\rightarrow L_{ij}+L_{ji}.
\end{equation}
The governing equation for the undisturbed matrix logarithm, $\boldsymbol{\Psi}_\infty$, is obtained by setting $\mathbf{L}=\nabla\mathbf{u}_\infty$ and $\boldsymbol{\Psi}=\boldsymbol{\Psi}_\infty$ in equations \eqref{eq:ConstitutivePsi} and \eqref{eq:StretchingRelaxation}.

\section{Temporal discretization and coupling between equations and boundary conditions}\label{sec:TemporalDiscretization}
The constitutive equation \eqref{eq:ConstitutivePsi} for the matrix logarithm, $\boldsymbol{\Psi}$ is driven by the velocity and velocity gradients. The polymer stress hence generated (see equations \eqref{eq:ConstitutiveForce}, \eqref{eq:ConstitutiveSpecific} and \eqref{eq:DefineLog}) acts as a body force in the momentum equation \eqref{eq:MomentumConsv} and influences the velocity and pressure field. The constraint of divergence-free velocity field defined by equation \eqref{eq:MassConsv} ensures mass conservation. In this section, we describe the temporal discretization and methods that treat the coupling between mass (equation \eqref{eq:MassConsv}) and momentum (equation \eqref{eq:MomentumConsv}) conservation and the polymer constitutive equation \eqref{eq:ConstitutivePsi} . We also describe the separate numerical treatments of the Newton's equations (equations \eqref{eq:ForceOnParticle} and \eqref{eq:TorqueOnParticle}) governing the particle dynamics for the case when particle inertia is negligible and when it is finite. The former case is of particular interest when fluid inertia is also neglected.

We adopt a similar methodology as \cite{cai2014efficient} to treat the time discretization of the momentum and mass conservation equations. The mass conservation equation at time step $n+1$ is
\begin{equation}
\nabla\cdot\widetilde{\mathbf{u}}^{n+1}=0.\label{eq:MassDiscrete}
\end{equation}
Using a backward Euler temporal discretization scheme, at a time step $n+1$, the momentum equation is written as,
\begin{align}\begin{split}
&\rho_f\big(\frac{\widetilde{\mathbf{u}}^{n+1}-\widetilde{\mathbf{u}}^n}{\Delta t}+\frac{3}{2}\widetilde{\textbf{ADV}}(\widetilde{\mathbf{u}}^n,\mathbf{u}_\infty^n,\boldsymbol{\omega}_p^n;\mathbf{r})-\frac{1}{2}\widetilde{\textbf{ADV}}(\widetilde{\mathbf{u}}^{n-1},\mathbf{u}_\infty^{n-1},\boldsymbol{\omega}_p^{n-1};\mathbf{r})\big)=\\&-\nabla \widetilde{p}^{n+1} +\nabla^2\widetilde{\mathbf{u}}^{n+1}+\nabla \cdot (\boldsymbol{\Pi}^{n+1}-\boldsymbol{\Pi}^{n+1}_\infty),
\end{split}\label{eq:MometumDiscrete}
\end{align}
where the non-linear terms i.e. $\widetilde{\textbf{ADV}}(\widetilde{\mathbf{u}},\mathbf{u}_\infty,\boldsymbol{\omega}_p;\mathbf{r})$ from equation \eqref{eq:AdvectionTermspert} are treated explicitly using a second-order Adams–Bashforth scheme after the first time step (in the first time step a first-order explicit Euler scheme is used). Explicit treatment of these non-linear terms dependent upon fluid inertia is a valid strategy because we aim to study the effect of zero to moderate fluid inertia on flows of viscoelastic fluids. After the first time step, we use a second-order implicit Crank-Nicholson scheme where the polymer constitutive equation is temporally discretized as following,
\begin{equation}
\frac{\boldsymbol{\Psi}^{n+1}-\boldsymbol{\Psi}^n}{\Delta t}+\frac{\mathbf{u}^{n+1}\cdot \nabla (\boldsymbol{\Psi}^{n+1}-\boldsymbol{\Psi}^{n+1}_\infty)}{2}+\frac{\mathbf{u}^n\cdot \nabla (\boldsymbol{\Psi}^n-\boldsymbol{\Psi}^n_\infty)}{2}=\frac{\textbf{SR}(\boldsymbol{\Psi}^{n+1},\mathbf{u}^{n+1})}{2}+\frac{\textbf{SR}(\boldsymbol{\Psi}^{n},\mathbf{u}^{n})}{2}.\label{eq:ConstitutiveTimeDiscrete}
\end{equation}
We consider both a fixed time step and a variable time step in our studies. ${\textbf{SR}(\boldsymbol{\Psi}^{n+1},\mathbf{u}^{n+1})}$ is dependent on the unknown values at the current time step. We use a weighted Jacobi method to iteratively solve equation \eqref{eq:ConstitutiveTimeDiscrete} for $\boldsymbol{\Psi}^{n+1}$. In the first time step we solve the polymer constitutive equation with a first order implicit Euler method. To obtain $\boldsymbol{\Pi}^{n+1}$ from $\boldsymbol{\Psi}^{n+1}$, we use the eigen-decomposition of $\boldsymbol{\Psi}^{n+1}$ and then use equations \eqref{eq:EigenDecomp} and \eqref{eq:EigenDecomp2} to obtain $\boldsymbol{\Lambda}^{n+1}$ along with the relevant relation between $\boldsymbol{\Lambda}^{n+1}$ and $\boldsymbol{\Pi}^{n+1}$ from equation \eqref{eq:ConstitutiveForce} and \eqref{eq:ConstitutiveSpecific}.

\subsection{Decoupling mass-momentum system from polymer constitutive equations}\label{eq:DecouplingEqns}
The polymer stress, $\boldsymbol{\Pi}^{n+1}$, is considered constant while solving the mass-momentum system of equations \eqref{eq:MassDiscrete} and \eqref{eq:MometumDiscrete}. Similarly, the velocity field, $\widetilde{\mathbf{u}}^{n+1}$ is considered constant when solving the constitutive equation \eqref{eq:ConstitutiveTimeDiscrete}. This allows us to numerically decouple the mass and momentum equations from the polymer constitutive equations. To properly account for the correct time location of $\boldsymbol{\Pi}^{n+1}$, $\boldsymbol{\psi}^{n+1}$, $\widetilde{p}^{n+1}$ and $\widetilde{\mathbf{u}}^{n+1}$, similar to \cite{richter2010simulations}, we consider $K$ inner iterations. At $k^{th}$ iteration, first the polymer constitutive equations are solved using velocity information from the previous, $1^{st}<(k-1)^{th}\le K^{th}$, step to update the values of $\boldsymbol{\Pi}^{n+1}$ and $\boldsymbol{\psi}^{n+1}$ (at $k=1^{st}$ inner iteration velocity information is taken from the previous time step). Then the mass-momentum system of equations is solved using the latest polymer stress to update the values of $\widetilde{\mathbf{u}}^{n+1}$ and $\widetilde{p}^{n+1}$.  The inner iterations are terminated when a user defined tolerance or $K$ is reached. We obtain accurate results even with $K=1$.

\subsection{Decoupled Schur complement approach to solve mass-momentum system}\label{sec:SchurMethod}
The coupled system of discrete mass and momentum equations \eqref{eq:MassDiscrete} and \eqref{eq:MometumDiscrete} is written in an operator form,
\begin{equation}
\begin{bmatrix}
\boldsymbol{\mathcal{L}}&\boldsymbol{\mathcal{G}}\\\boldsymbol{\mathcal{D}}&\boldsymbol{0}
\end{bmatrix}\begin{bmatrix}
\widetilde{\mathbf{u}}^{n+1}\\\widetilde{p}^{n+1}
\end{bmatrix}=\begin{bmatrix}
{\boldsymbol{m}}\\{0}
\end{bmatrix},\label{eq:MomentumMass}
\end{equation}
{or, using a Schur complement reduction approach similar to Furuichi et al. (2011), \cite{furuichi2011development} as,}
\begin{equation}
\begin{bmatrix}
\boldsymbol{\mathcal{L}}&\boldsymbol{\mathcal{G}}\\\boldsymbol{0}&\boldsymbol{\mathcal{S}}
\end{bmatrix}\begin{bmatrix}
\widetilde{\mathbf{u}}^{n+1}\\\widetilde{p}^{n+1}
\end{bmatrix}=\begin{bmatrix}
{\boldsymbol{m}}\\{h}
\end{bmatrix},\label{eq:MomentumMassSchur}
\end{equation}
{to decouple pressure from velocity. Here, $\boldsymbol{\mathcal{S}}=-\boldsymbol{\mathcal{D}}\boldsymbol{\mathcal{L}}^{-1}\boldsymbol{\mathcal{G}}$ is the Schur complement of the matrix in equation \eqref{eq:MomentumMass} and $h=-\boldsymbol{\mathcal{D}}\boldsymbol{\mathcal{L}}^{-1}{\boldsymbol{m}}$. In these equations, $\boldsymbol{\mathcal{L}}=(\rho_f/\Delta t)\boldsymbol{\delta}-\nabla^2$, $\boldsymbol{\mathcal{G}}=\nabla$, and, $\boldsymbol{\mathcal{D}}=\nabla\cdot$ represent different spatial operators. Here, $\nabla^2$ is a vector Laplacian operator. In the Cartesian basis, the unit vectors are spatially constant. However, due to the spatial dependence of unit vectors in a curvilinear basis, cross-terms exist that involve multiple vector components in a particular component of the Laplacian of that vector. This guides our choice of basis for vectors and tensors in section \ref{sec:ChoiceOfbasis}. The discretization of these spatial operators will be described in more detail in section \ref{sec:SpatialDiscretization}.}
\begin{equation}
{\boldsymbol{m}}=\rho_f\big(\frac{\widetilde{\mathbf{u}}^{n}}{\Delta t}-\frac{3}{2}\widetilde{\textbf{ADV}}(\widetilde{\mathbf{u}}^n,\mathbf{u}_\infty^n,\boldsymbol{\omega}_p^n;\mathbf{r})+\frac{1}{2}\widetilde{\textbf{ADV}}(\widetilde{\mathbf{u}}^{n-1},\mathbf{u}_\infty^{n-1},\boldsymbol{\omega}_p^{n-1};\mathbf{r})\big)+\nabla \cdot (\boldsymbol{\Pi}^{n+1}-\boldsymbol{\Pi}^{n+1}_\infty)\label{eq:GvecinMassMom}
\end{equation} {represents the sum of terms in the momentum equations that are considered constant within a time-step (or within a $k^{th}$ iteration as described in section \ref{eq:DecouplingEqns}), i.e., the explicit terms and the divergence of the polymer stress. On the boundaries of the computational domain, i.e., the particle surface and the outer boundary, the matrix-vector system of equation \eqref{eq:MomentumMass} is appropriately changed to represent the velocity boundary conditions of equation \eqref{eq:VelBCspert}.}

{As highlighted in section \ref{sec:Introduction}, in the context of finite difference spatial discretization, splitting methods (that advance the momentum equation in time and solve a pressure-Poisson equation to obtain an appropriate pressure for ensuring incompressibility) are not suitable for cases in which the fluid inertia is small. This is related to artificial boundary conditions for pressure (since a second-order partial differential equation now governs it) required in the splitting methods \cite{gresho1987pressure,sani2006pressure}. The splitting errors may be ignored compared with more important momentum advection terms when fluid inertia is larger than the viscous forces, making them suitable for studies of turbulent flows. The dominance of viscous terms at lower fluid inertia values in studies of our interest prevents us from using such methods. Fluid inertia is quantified in the above equations by the fluid density, $\rho_f$. Furthermore, due to the time advancement of the momentum equation in splitting methods, studies with $\rho_f=0$ cannot be considered. $\rho_f=0$ studies are useful to isolate viscoelasticity's effect from fluid inertia completely. This Schur complement reduction method was originally developed to study the behavior of fluids with zero inertia and large viscosity variations, such as in long time scale dynamics of the Earth's convecting mantle \cite{furuichi2011development}. We have found it useful in studies of zero to moderate inertia viscoelastic fluids.}

 Similar to \cite{furuichi2011development}, we begin by solving the decoupled pressure equation
\begin{equation}\boldsymbol{\mathcal{S}}\widetilde{p}^{n+1}=h,\label{eq:PressureDecoupled}\end{equation} using a Krylov subspace method that constructs a subspace $K(\boldsymbol{\mathcal{S}},r_0)=\text{Span}(r_0,\boldsymbol{\mathcal{S}}r_0,\boldsymbol{\mathcal{S}}^2r_0,\cdots,\boldsymbol{\mathcal{S}}^{N-1}r_0)$, for $\boldsymbol{\mathcal{S}}\in\mathcal{R}^{N\times N}$, with \begin{equation}
\widetilde{\mathbf{u}}_0=\boldsymbol{\mathcal{L}}^{-1}(\boldsymbol{G}\widetilde{p}_0-{\boldsymbol{m}}),\hspace{0.2in}r_0=h-\boldsymbol{\mathcal{S}}\widetilde{p}_0=\boldsymbol{\mathcal{D}}\widetilde{\mathbf{u}}_0,\label{eq:GMRESInit}
\end{equation} for an initial guess $\widetilde{p}_0$ for $\widetilde{p}^{n+1}$. The generalized minimum residual (GMRES) method \cite{saad1986gmres} is our choice of Krylov subspace method. The matrix vector product $y_i=\boldsymbol{\mathcal{S}}x_i$ consists of three separate operations defined as (similar to \cite{furuichi2011development}),
\begin{align}
\mathbf{a}^*=\boldsymbol{\mathcal{G}}x_i,\hspace{0.2in}
\widetilde{\mathbf{u}}^*_i=\boldsymbol{\mathcal{L}}^{-1}\mathbf{a}^*,\hspace{0.2in}
y_i=\boldsymbol{\mathcal{D}}\widetilde{\mathbf{u}}^*_i.\label{eq:MatVec}
\end{align}
The first and third steps are simple matrix-vector products of a known matrix and vector that can be computed straightforwardly in an efficient manner. The second step requires solution of a matrix equation $\boldsymbol{\mathcal{L}}\widetilde{\mathbf{u}}*=\mathbf{a}^*$ and the operator $\boldsymbol{\mathcal{L}}$ involves the sum of an identity and a Laplacian operator. Therefore, we use the aggregation-based algebraic multigrid (AGMG) method of \cite{notay2010aggregation,napov2012algebraic} to solve this elliptic equation efficiently. We use the implementation provided in \cite{agmg}. Once the GMRES method terminates upon reaching a sufficient user-defined tolerance (see \cite{saad1986gmres,golub2013matrix} for details), say in $M$ iterations, the solution for  $\widetilde{p}^{n+1}$ is formed as,
\begin{equation}
\widetilde{p}^{n+1}=\widetilde{p}_0+\Sigma_{i=1}^My_i/||y_i||_2 l_i,
\end{equation}
where for each GMRES iteration $i$, $y_i$ is a vector defined by equation \eqref{eq:MatVec} and $l_i$ is a scalar defined within the GMRES procedure (see \cite{saad1986gmres,golub2013matrix} for details). If we keep track of different $\widetilde{\mathbf{u}}^*_i$ in the GMRES iteration for solution of the pressure equation \eqref{eq:PressureDecoupled}, by a simple and fast operation we can construct,
\begin{equation}
\delta \widetilde{\mathbf{u}}^{n+1}=\Sigma_{i=1}^M\widetilde{\mathbf{u}}^*_i l_i.
\end{equation}
The velocity field at step n+1 is thus obtained by
\begin{equation}
\widetilde{\mathbf{u}}^{n+1}=\widetilde{\mathbf{u}}_0-\delta \widetilde{\mathbf{u}}^{n+1},
\end{equation}
where $\widetilde{\mathbf{u}}_0$ is already calculated and defined in equation \eqref{eq:GMRESInit}. Our methodology to solve the coupled system of mass and momentum equation closely follows that of \cite{furuichi2011development} for solving the decoupled pressure equation \eqref{eq:PressureDecoupled}. However, careful observation of the GMRES method allows us to obtain the solution for velocity, $\widetilde{\mathbf{u}}^{n+1}$ as an auxiliary product of the same calculation. This avoids the need to solve a velocity equation $\widetilde{\mathbf{u}}^{n+1}=\boldsymbol{\mathcal{L}}^{-1}({\boldsymbol{m}}-\boldsymbol{\mathcal{G}}p^{n+1})$ and saves CPU time. In \cite{furuichi2011development} authors note a similar point but still solve the velocity equation after obtaining the pressure solution. This has so far proven to be unnecessary for our studies.

\subsection{Ensuring stretch limited by maximum polymer extensibility in FENE models}
In dumbbell based models for polymer configuration, as discussed in section \ref{sec:GoverningEquations}, the polymer configuration is $\boldsymbol{\Lambda}=\langle\mathbf{qq}\rangle_\text{polymer configuration}$, where $\mathbf{q}$ is end-to-end vector (non-dimensionalized with polymer's radius of gyration) of the dumbbell and the angle brackets represent average over polymer configuration. Therefore, the polymer stretch is $\sqrt{\text{tr}(\boldsymbol{\Lambda})}$. In FENE models such as the FENE-P and FENE-CR (equation \eqref{eq:ConstitutiveSpecific}) models, the polymers have a maximum extensibility $L$. Following the technique introduced in \cite{richter2010simulations} we numerically impose this constraint by separately evolving the variable $\gamma=1/f=1-\text{tr}(\boldsymbol{\Lambda})/L^2$ (equation \eqref{eq:FENEParams}) used in the FENE models. An evolution equation for $\gamma$ is obtained by taking the trace of the polymer constitutive equation \eqref{eq:Constitutive} with one of the FENE models from equation \eqref{eq:ConstitutiveSpecific},
\begin{equation}
\frac{\partial \gamma}{\partial t}+\mathbf{u}\cdot \nabla \gamma+\frac{2}{L^2}\boldsymbol{\Lambda}:\nabla\mathbf{u}+\frac{1}{De}(\frac{\gamma-1}{\gamma}+\beta)=0,\hspace{0.2in}\beta=\begin{cases}\frac{3}{L^2-3}, &\text{FENE-P}\\
\frac{3}{L^2\gamma}, &\text{FENE-CR}
\end{cases}.\label{eq:ConstitutiveGama}
\end{equation}
Similar to \cite{richter2010simulations}, we temporally discretize equation \eqref{eq:ConstitutiveGama} using a Cranck-Nicholson scheme for the relaxation terms (${1}/{De}((\gamma-1)/{\gamma}+\beta)$) and treating the advection ($\mathbf{u}\cdot \nabla \gamma$) and stretching (${2}/{L^2}\boldsymbol{\Lambda}:\nabla\mathbf{u}$) terms explicitly. This leads to the following quadratic equation for $\gamma^{n+1}$,
\begin{equation}
(\gamma^{n+1})^2+\gamma^{n+1}\Delta t\Big(-\frac{\gamma^n}{\Delta t}+\mathbf{u}^n\cdot \nabla \gamma^n+\frac{2\boldsymbol{\Lambda^n}:\nabla\mathbf{u}^n}{L^2}+\frac{\gamma^n-0.5}{De\gamma^n}+\frac{1}{DeL^2}\alpha_1^n\Big)-\frac{\Delta t}{2 De}\alpha_2=0,\label{eq:FeneStretcheqn}
\end{equation}
where $\alpha_1^n=3/(L^2-3)$ for FENE-P and  $1.5/\gamma^n$ for FENE-CR, and, $\alpha_2=1$ for FENE-P and $1-3/L^2$ for FENE-CR. This equation leads to two real roots with opposite signs \cite{richter2010simulations}. The negative root is unphysical since it implies $\text{tr}(\boldsymbol{\Lambda}^{n+1})>L^2$. Choosing the positive root ensures that the polymer stretch is upper-bounded by the maximum extensibility, $L$. At each time step, $n+1$, we calculate $f^{n+1}=1/\gamma^{n+1}$ from this treatment and use this value of $f^{n+1}$ in the relaxation term of the discretized polymer constitutive equation \eqref{eq:ConstitutiveTimeDiscrete}.

\subsection{Evolution of particle orientation and velocity boundary conditions}\label{sec:Rotations}
As mentioned in section \ref{sec:GoverningEquations}, we solve the governing equations in a particle fixed reference frame. In our simulations, the inertial (or laboratory) frame is defined either from the initial particle orientation (e.g., a particle rotating about its axis in quiescent fluid) or through the geometry of the imposed flow (e.g., a reference frame fixed with the imposed simple shear flow, uniaxial extensional flow or a uniform flow field). The particle orientation is defined using quaternions, $\mathbf{q}=\begin{bmatrix}
q_1&q_2&q_3&q_4
\end{bmatrix}^T$, (see chapter 8 of \cite{rapaport2004art}) that are related to the Euler angles, $\theta,\phi,\psi$, between the particle-fixed and inertial reference frames,
\begin{align}
\begin{split}
q_1=&\sin(\theta/2)\cos((\phi-\psi)/2),\hspace{0.2in}
q_2=\sin(\theta/2)\sin((\phi-\psi)/2),\\
q_3=&\cos(\theta/2)\sin((\phi+\psi)/2),\hspace{0.2in}
q_4=\cos(\theta/2)\cos((\phi+\psi)/2).
\end{split}
\end{align}
The sequence of three rotations that define the Euler angles are described in \cite{rapaport2004art}: $\theta=\sin^{-1}(-X_3)$, $\phi=\sin^{-1}(Y_3/\sqrt{1-X_3^2})$, $\psi=\sin^{-1}(X_2/\sqrt{1-X_3^2})$. Here, $X_2$ and $X_3$ are the projections of the $X$ (or 1) axis of the particle-fixed reference frame on the $2$ and $3$ axis, respectively, of the inertial reference frame. $Y_2$ and $Y_3$ are the same projections of the $Y$ (or 2) axis of the particle-fixed reference frame. The quaternion formulation has previously been used to study fluid flows around prolate spheroids in \cite{d2014bistability,yu2007direct}. The evolution equation for quaternions is related to the particle's angular velocity,
$\boldsymbol{\omega}_p=\begin{bmatrix}
\omega_1&\omega_2&\omega_3
\end{bmatrix}^T$,
\begin{equation}
\frac{d}{dt}\begin{bmatrix}
q_1\\q_2\\q_3\\q_4
\end{bmatrix}=\frac{1}{2}\begin{bmatrix}
q_4&-q_3&q_2&q_1\\
q_3&q_4&-q_1&q_2\\
-q_2&q_1&q_4&q_3\\
-q_1&-q_2&-q_3&q_4
\end{bmatrix}\begin{bmatrix}
\omega_1\\\omega_2\\\omega_3\\0
\end{bmatrix}.\label{eq:QuaternionEvol}
\end{equation}
We consider second and third-order accurate Adams-Bashforth schemes to discretize equation \eqref{eq:QuaternionEvol} temporally.
The transformation matrix from the inertial frame to the particle-fixed frame is,
\begin{equation}
\boldsymbol{A}(t)=2\begin{bmatrix}
q_1^2+q_4^2-1/2&q_1q_2+q_3q_4&q_1q_3-q_2q_4\\
q_1q_2-q_3q_4&q_2^2+q_4^2-1/2&q_2q_3+q_1q_4\\
q_1q_3+q_2q_4&q_2q_3-q_1q_4&q_3^2+q_4^2-1/2	
\end{bmatrix}.\label{eq:TransformationMatrix}
\end{equation}
$\boldsymbol{A}(t)$ is an orthonormal matrix. A vector in the particle-fixed frame, $\mathbf{b}_\text{particle}$ is transformed into the inertial reference frame with $\mathbf{b}_\text{inertial}=\boldsymbol{A}(t)^T\cdot\mathbf{b}_\text{particle}$. The major axis or the center-line of the spheroidal particle is along the $x_3$ axis of the particle-fixed frame. At a particular time, in the inertial reference frame the particle orientation vector i.e. the particle center line is $\mathbf{p}=\begin{bmatrix}q_1q_3+q_2q_4&q_2q_3-q_1q_4&q_3^2+q_4^2-1/2\end{bmatrix}$. At the end of each time step, after updating the quaternions we update the velocity boundary conditions. Consider a general case where the imposed fluid motion consists of a uniform velocity, $\mathbf{u}_{far}^\text{inertial}$ and a velocity gradient, $\boldsymbol{\Gamma}_\text{far}^\text{inertial}$ in the laboratory frame. The particle's translation and angular velocities (numerical methods to evaluate these are discussed in section \ref{sec:ParticleVels}) in the frame aligned with the particle are $\mathbf{u}_p$ and $\boldsymbol{\omega}_p$ respectively. In the frame of reference aligned with the particle coordinates, the appropriately rotated imposed fluid motion at the outer boundary is,
\begin{equation}
\mathbf{u}_\infty(\mathbf{r},t)=\boldsymbol{A}(t)\cdot\mathbf{u}_\text{far}^\text{inertial}+\mathbf{r}\cdot\boldsymbol{A}(t)\cdot\boldsymbol{\Gamma}_\text{far}^\text{inertial}\cdot\boldsymbol{A}(t)^T-\mathbf{u}_p-\boldsymbol{\omega}_p\times \mathbf{r},\hspace{0.2in} \mathbf{r}=\mathbf{r}_\infty.\label{eq:VelBCFinal}
\end{equation}

\subsection{Particle velocities}\label{sec:ParticleVels}
As mentioned in section \ref{sec:GoverningEquations} the particle's angular and translation velocities, $\boldsymbol{\omega}_p$ and $\mathbf{u}_p$ may be a prescribed quantity in a study of interest, so that the procedure mentioned above has all the required quantities to solve the equations. We are also interested in studying the scenarios where the particle is free to move due to the imposed and fluid-generated forces and torques. The case when particle inertia is zero must be dealt with differently than the one with finite particle inertia, and we will consider these two cases next.

\subsubsection{Finite particle inertia}\label{sec:FiniteParticleInertial}
At each time step the particle motion must satisfy the Newton's equations \eqref{eq:ForceOnParticle} and \eqref{eq:TorqueOnParticle}. These equations are discretized using a second order implicit Crank-Nicholson scheme,
\begin{align}
\mathbf{u_p}^{n+1}=&\mathbf{u_p}^{n}+\frac{\Delta t}{2\rho_pV_p}(\mathbf{f}^{n+1}_\text{net}+\mathbf{f}^{n}_\text{net})\\
\boldsymbol{\omega}_p^{n+1}=&\boldsymbol{\omega}_p^{n}+\frac{5\Delta t}{2\rho_pV_pr_\text{minor}^2}\begin{bmatrix}
\frac{1}{1+\kappa^2}&0&0\\
0&\frac{1}{1+\kappa^2}&0\\
0&0&0.5
\end{bmatrix}\cdot( \mathbf{q}_\text{net}^{n+1}+\mathbf{q}_\text{net}^{n}),
\end{align}
where $V_p=4\pi\kappa r_{minor}^3/3$,
\begin{align}
\begin{split}
\mathbf{f}^{n+1}_\text{net}=&{\mathbf{f}_\text{fluid}^{n+1}+\mathbf{f}_\text{ext.}^{n+1},\hspace{0.2in} \mathbf{f}_\text{fluid}^{n+1}=\mathbf{f}(\boldsymbol{\tau})^{n+1}+\mathbf{f}(\boldsymbol{\Pi})^{n+1}}\\
\mathbf{q}_\text{net}^{n+1}=&\mathbf{q}_\text{fluid}^{n+1}+\mathbf{q}_\text{ext.}^{n+1},\hspace{0.2in} \mathbf{q}_\text{fluid}^{n+1}=\mathbf{q}(\boldsymbol{\tau})^{n+1}+\mathbf{q}(\boldsymbol{\Pi})^{n+1}.
\end{split}\label{eq:ParticleMotion}
\end{align}
are the net force and torque acting on the particle at time step $n+1$. {$\mathbf{f}_\text{fluid}^{n+1}$ and $\mathbf{q}_\text{fluid}^{n+1}$ are evaluated from equations \eqref{eq:FluidTorqueandForce} and \eqref{eq:TorqueDecompose} once the fluid stress on the particle surface at time step $n+1$,  $\boldsymbol{\sigma}^{n+1}=\boldsymbol{\tau}^{n+1}+\boldsymbol{\Pi}^{n+1}$, is available. $\mathbf{f}_\text{ext.}^{n+1}$ and $\mathbf{q}_\text{ext.}^{n+1}$ are the externally imposed (possibly time-varying) force and torques in the frame aligned with the particle at time step $n+1$.} We use the first order explicit Euler scheme in the first time step.

\subsubsection{Zero particle inertia} \label{sec:ZeroParticleInertial}
As mentioned earlier in section \ref{sec:Introduction} and \ref{sec:SchurMethod}, we are also interested in the scenario of zero particle inertia or zero $\rho_p$. In our studies, such a case arises when we want to completely remove the inertial effects and study the influence of viscoelasticity, but it could also be used to investigate massless particles in a fluid with finite inertia. In {both of these cases}, Newton's equations governing the particle motion at each time step are,
\begin{align}
\begin{split}
\mathbf{f}^{n+1}_\text{net}=&{\mathbf{f}_\text{fluid}^{n+1}+\mathbf{f}_\text{ext.}^{n+1}}=0,\hspace{0.2in} \mathbf{f}_\text{fluid}^{n+1}=\mathbf{f}(\boldsymbol{\tau})^{n+1}+\mathbf{f}(\boldsymbol{\Pi})^{n+1},\\
\mathbf{q}_\text{net}^{n+1}=&\mathbf{q}_\text{fluid}^{n+1}+\mathbf{q}_\text{ext.}^{n+1}=0,\hspace{0.2in} \mathbf{q}_\text{fluid}^{n+1}=\mathbf{q}(\boldsymbol{\tau})^{n+1}+\mathbf{q}(\boldsymbol{\Pi})^{n+1},
\end{split}\label{eq:ForceTorqueConstraints}
\end{align}
and the time marching used for finite $\rho_p$ in \eqref{eq:ParticleMotion} cannot be employed. The governing equations \eqref{eq:ForceTorqueConstraints}  are viewed as force and torque constraints that velocity, pressure and polymer stress field must satisfy at each time step to yield an appropriate $\mathbf{u_p}^{n+1}$ and $\boldsymbol{\omega}_p^{n+1}$. {Padhy et al. (2013) \cite{padhy2013simulations} used a secant method to iteratively impose the torque-free constraint on a sphere rotating in a cross shear flow. We first consider this method for imposing force- and torque-free constraints for a finite fluid inertia case. Subsequently, we show that a novel decomposition of inertia-less fluid's momentum equation can be used to impose these constraints in a non-iterative and hence computationally efficient manner. These two techniques are discussed next.}

{\textbf{Secant iteration method for finite fluid inertia and zero particle inertia:} At each time step or within each inner $k$ iteration (section \ref{eq:DecouplingEqns}) the polymer stress, $\boldsymbol{\Pi}$ and hence the polymer torque and force, $\mathbf{q}(\boldsymbol{\Pi})^{n+1}$ and $\mathbf{f}(\boldsymbol{\Pi})^{n+1}$, from equations equations \eqref{eq:FluidTorqueandForce} and \eqref{eq:TorqueDecompose}, are fixed. $\mathbf{u_p}^{n+1}$ and $\boldsymbol{\omega}_p^{n+1}$ are iterated along with $\widetilde{\mathbf{u}}^{n+1}$ and $\widetilde{p}^{n+1}$ to generate the appropriate Newtonian solvent torque and force, $\mathbf{q}(\boldsymbol{\tau})^{n+1}$ and $\mathbf{f}(\boldsymbol{\tau})^{n+1}$, that ensures the torque and force balance in equation \eqref{eq:ForceTorqueConstraints}. In practice, once the polymer constitutive equation \eqref{eq:ConstitutiveTimeDiscrete} is solved, we first obtain the polymeric torque and force, $\mathbf{q}(\boldsymbol{\Pi})^{n+1}$ and $\mathbf{f}(\boldsymbol{\Pi})^{n+1}$, from equation \eqref{eq:FluidTorqueandForce}. The mass-momentum system described by equation \eqref{eq:MomentumMassSchur} depends on the particle's angular velocity $\boldsymbol{\omega}_p$ (see equations \eqref{eq:AdvectionTermspert}, \eqref{eq:VelBCspert}, and \eqref{eq:GvecinMassMom}). In the secant iterations method, iterations proceed by solving the mass-momentum system by the Schur complement method described in section \ref{sec:SchurMethod} and obtaining the Newtonian torque and force, $\mathbf{q}(\boldsymbol{\tau})^{n+1}$ and $\mathbf{f}(\boldsymbol{\tau})^{n+1}$, from equation \eqref{eq:FluidTorqueandForce}. After each secant iteration, $s$, the particle's angular and translational velocities $\boldsymbol{\omega}_p$ and $\mathbf{u}_p$ are updated component-wise,}
\begin{equation}
\omega_{p,i}^{s+1}=\omega_{p,i}^s-q_{net,i}^s\frac{\omega_{p,i}^{s}-\omega_{p,i}^{s-1}}{q_{net,i}^s-q_{net,i}^{s-1}},\hspace{0.2in}u_{p,i}^{s+1}=u_{p,i}^s-f_{net,i}^s\frac{u_{p,i}^{s}-u_{p,i}^{s-1}}{f_{net,i}^s-f_{net,i}^{s-1}},\hspace{0.2in}i\in[1,3].
\end{equation}
$\boldsymbol{\omega}_p^{s+1}$ and $\mathbf{u}_p^{s+1}$ are used in the velocity boundary condition (equation \eqref{eq:VelBCFinal}) for the next secant iteration. The secant iterations are stopped once the magnitudes of all components of the net torque and force on the particle, $\mathbf{q}_\text{net}^{n+1}$ and $\mathbf{f}_\text{net}^{n+1}$, are below  prescribed tolerances. The numerical solution at this point consists of the velocity, pressure, and polymer stress fields and the particle's required angular and translational velocity that satisfy the force and torque constraints of equation \eqref{eq:ForceTorqueConstraints}.

{\textbf{Novel resistivity formulation for zero fluid and particle inertia in non-Newtonian fluids}: The system of mass and momentum equations and the torque- and force- free constraint is quasi-steady when both fluid and particle inertia are neglected. All the variables from here until the end of this section are taken at time step $n+1$ and we will supress the superscript for clarity. Upon neglecting fluid inertia, the momentum equation \eqref{eq:MomentumConsv} becomes
\begin{equation}
\nabla \cdot \boldsymbol{\sigma}=0,
\end{equation}
subject to the boundary conditions mentioned in equation \eqref{eq:VelBCs} and the stress tensor, $\boldsymbol{\sigma}$, in equation \eqref{eq:StressDefinition}.  In this case, the momentum and mass \eqref{eq:MassConsv} conservation equations are linear in the velocity and pressure. Therefore, using
\begin{equation}
\mathbf{u}=\mathbf{u}^\text{M}+\mathbf{u}^\text{P},\hspace{0.2in} p=p^\text{M}+p^\text{P},
\end{equation}
the system of mass and momentum equations along with the associated boundary conditions is linearly decomposed into two parts. $\mathbf{u}^\text{M}$ and $p^\text{M}$ represent the motion-induced velocity and pressure fields and $\mathbf{u}^\text{P}$ and $p^\text{P}$ the polymer-induced fields. The equations governing the motion-induced part are,
\begin{equation}
\nabla \cdot \mathbf{u}^\text{M}=0,\hspace{0.2in}
\nabla \cdot \boldsymbol{\sigma}^\text{M}=0,
\end{equation}
with,
\begin{equation}
\boldsymbol{\sigma}^\text{M}=\boldsymbol{\tau}^\text{M}=-p^\text{M}\boldsymbol{\delta}+\mu(\nabla\mathbf{u}^\text{M}+(\nabla\mathbf{u}^\text{M})^T)
\end{equation}
and the boundary conditions,
\begin{align}\begin{split}
\mathbf{u}^\text{M}=\mathbf{0}, \text{ on particle surface },\hspace{0.2in}
\mathbf{u}^\text{M}=\mathbf{u}_\infty(\mathbf{r})=\mathbf{r}\cdot\boldsymbol{\Gamma}+\mathbf{u}_0-\mathbf{u}_p-\boldsymbol{\omega}_p\times\mathbf{r}, \text{ as }|\mathbf{r}|\rightarrow \infty\approx\mathbf{r}_\infty.
\end{split}\label{eq:VelBCsMotion}\end{align}
The polymer-induced part is governed by,
\begin{align}
\nabla \cdot \mathbf{u}^\text{P}=0,\hspace{0.2in}\nabla \cdot \boldsymbol{\sigma}^\text{P}=0,
\end{align}
with,
\begin{equation}
\boldsymbol{\sigma}^\text{P}=\boldsymbol{\tau}^\text{P}+\boldsymbol{\Pi}=-p^\text{P}\boldsymbol{\delta}+\mu(\nabla\mathbf{u}^\text{P}+(\nabla\mathbf{u}^\text{P})^T+\boldsymbol{\Pi})
\end{equation}
and the boundary conditions,
\begin{equation}
\mathbf{u}^\text{P}=\mathbf{0}, \text{ on particle surface },\hspace{0.2in} \mathbf{u}^\text{P}=\mathbf{0}, \text{ as }|\mathbf{r}|\rightarrow \infty\approx\mathbf{r}_\infty.\label{eq:VelBCsPoly}\end{equation}
The hydrodynamic force and torque acting on the particle are also decomposed into a motion- and polymer-induced part,
\begin{equation}
\mathbf{f}_\text{fluid}=\mathbf{f}_\text{fluid}^\text{M}+\mathbf{f}_\text{fluid}^\text{P},\hspace{0.1in}\mathbf{q}_\text{fluid}=\mathbf{q}_\text{fluid}^\text{M}+\mathbf{q}_\text{fluid}^\text{P},\label{eq:TorqueDecomposeMotionPolymer}
\end{equation}
with,
\begin{equation}
\mathbf{f}_\text{fluid}^\text{M}=\mathbf{f}(\boldsymbol{\sigma}^\text{M}),\hspace{0.05in} \mathbf{f}_\text{fluid}^\text{P}=\mathbf{f}(\boldsymbol{\sigma}^\text{P}),\hspace{0.05in}
\mathbf{q}_\text{fluid}^\text{M}=\mathbf{q}(\boldsymbol{\sigma}^\text{M}),\hspace{0.05in} \text{and},\hspace{0.05in} \mathbf{q}_\text{fluid}^\text{P}=\mathbf{q}(\boldsymbol{\sigma}^\text{P}),
\end{equation}
where, $\mathbf{f}(\boldsymbol{\sigma}) $ and $\mathbf{q}(\boldsymbol{\sigma}) $ are defined in equation \eqref{eq:FluidTorqueandForce}. This way of decomposing the momentum equation attributes the entire effect of particle motion (via the boundary conditions in equation \eqref{eq:VelBCsMotion}) to the motion-induced part of the governing equations. This part has no influence of the polymer stress. The polymer-induced pressure and velocity fields are forced by the effect of the polymer stress, but are not explicitly affected by the particle's motion. A key observation allowing us to circumvent an iterative procedure in calculating the particle's motion is that the motion-induced equations represent a Stokes flow (inertia-less Newtonian flow). Hence, the motion induced force and torque are simply,
\begin{align}
\begin{split}
\mathbf{f}_{\text{fluid}}^M=\boldsymbol{F}_p\cdot\mathbf{u}_p+\mathbf{f}_\infty,\hspace{0.05in} \mathbf{q}_{\text{fluid}}^M=\boldsymbol{Q}_p\cdot\boldsymbol{\omega}_p+\mathbf{q}_\infty.
\end{split}
\end{align}
The tensors $\boldsymbol{F}_p$ and $\boldsymbol{Q}_p$ depend upon only on the particle shape, and the vectors $\mathbf{f}_\infty$ and $\mathbf{q}_\infty$ depend upon both the particle shape and the imposed flow. These are either evaluated analytically for simple particle shapes and imposed flows or calculated by considering only the motion-induced mass and momentum equations with appropriate boundary conditions. For example, the 21 component of $\boldsymbol{F}_p$ is simply the second component of $\mathbf{f}(\boldsymbol{\sigma}^\text{M})$ for a boundary condition $\mathbf{u}^\text{M}=\mathbf{0}, \text{ on particle surface and }\mathbf{u}^\text{M}=[
1\hspace{0.05in} 0\hspace{0.05in} 0]^T, \text{ as }|\mathbf{r}|\rightarrow \infty\approx\mathbf{r}_\infty$. $\mathbf{f}_\infty$ and $\mathbf{q}_\infty$ are simply the hydrodynamic force and torque on a fixed particle in the imposed flow ($\mathbf{r}\cdot\boldsymbol{\Gamma}+\mathbf{u}_0$) of an inertia-less Newtonian fluid. From the force- and torque-free constraints, we obtain
\begin{equation}
\mathbf{u}_p=-\boldsymbol{F}_p^{-1}\cdot(\mathbf{f}_\infty+\mathbf{f}_{\text{fluid}}^\text{P}+\mathbf{f}_{\text{ext}}),\text{ and, }
\boldsymbol{\omega}_p=-\boldsymbol{Q}_p^{-1}\cdot(\mathbf{q}_\infty+\mathbf{q}_{\text{fluid}}^\text{P}+\mathbf{q}_{\text{ext}})
\end{equation}
in a non-iterative way. This method of obtaining a particle's motion has long been used in micro-hydrodynamics of Newtonian fluids and is called a resistivity formulation \cite{kim2013microhydrodynamics}. However, with the novel decomposition of the inertia-less momentum equation we have shown that a similar principal can be applied in a computationally useful manner for non-Newtonian problems.}

{Unlike the polymer induced pressure and velocity fields, the determination of the motion induced velocity and pressure fields does not require application of the Schur complement approach described in section \ref{sec:SchurMethod} at each time step of the simulation. Once the particle's translational and angular velocity are available, these fields are either taken from the analytical solution of Stokes flow around the particle or from a linear superposition of pre-calculated numerical solutions of a few fundamental incompressible flows around the particle. The boundary conditions at a given time instant can be represented by a superposition of the boundary conditions in these fundamental flows. In the most general case when the velocity at the outer boundary in equation \eqref{eq:VelBCsMotion} has non-zero components for all the components of the effective (incompressible) imposed velocity gradient ($\boldsymbol{\Gamma+\epsilon\cdot\omega}_p$) and the effective imposed velocity ($\mathbf{u}_0-\mathbf{u}_p$), 11 fundamental Stokes flows have to be pre-calculated. While we are primarily considering linear velocity boundary conditions, this method can be extended to other types of boundary conditions where the Stokes flows required to be pre-calculated will be different. Using this resistivity formulation, our simulations become significantly faster relative to the secant iteration method.}

{The two most computationally intensive components of our method are the weighted Jacobi method to iteratively solve equation \eqref{eq:ConstitutiveTimeDiscrete} for $\boldsymbol{\Psi}^{n+1}$ and the algebraic multigrid method to invert $\boldsymbol{\mathcal{L}}$. The domain decomposition method implemented in Message Passing Interface (MPI) is used to parallelize the code in the $\xi_1$ and $\xi_2$ directions (the $\xi_3$ direction could also be parallelized in future). The weighted Jacobi method is implemented completely in-house and shows good strong and weak scaling for 100s of processing units. The parallel scaling of the algebraic multigrid method is dependent on the external vendor's capability. We use the academic version of AGMG \cite{agmg} for which good scaling was only obtained up to about 20 processing units. The examples considered in section \ref{sec:Examples} are completed within a reasonable time of up to four days using 20 processing units. However, for more stringent cases, AGMG can be replaced with other open source algebraic multigrid solvers available in libraries such as the BoomerAMG of HYPRE \cite{falgout2002hypre} that scales well up to a large number of processing units, but requires more parameter tuning.}

In this section \ref{sec:TemporalDiscretization}, we have described the algorithmic structure of our numerical method in a coordinate system free manner. In the next section, we will delve into the specific choice of the prolate spheroidal coordinate system and the finite difference schemes used to discretize the spatial gradients in the governing equations. This will allow us to discretize the computational space and various operators mentioned above.  

\section{Spatial discretization}\label{sec:SpatialDiscretization}
One of the benefits of using body-fitted grids and expressing equations in the particle reference frame is that the discrete space or mesh remains fixed as the particle rotates, and the discretized spatial operators defined here need to be calculated only once before the equations are evolved in time.
\subsection{Prolate spheroidal coordinate system for spatial discretization of computational domain}\label{sec:GridDefinition}
\begin{figure}[h!]
\centering
\subfloat{\includegraphics[width=0.48\textwidth]{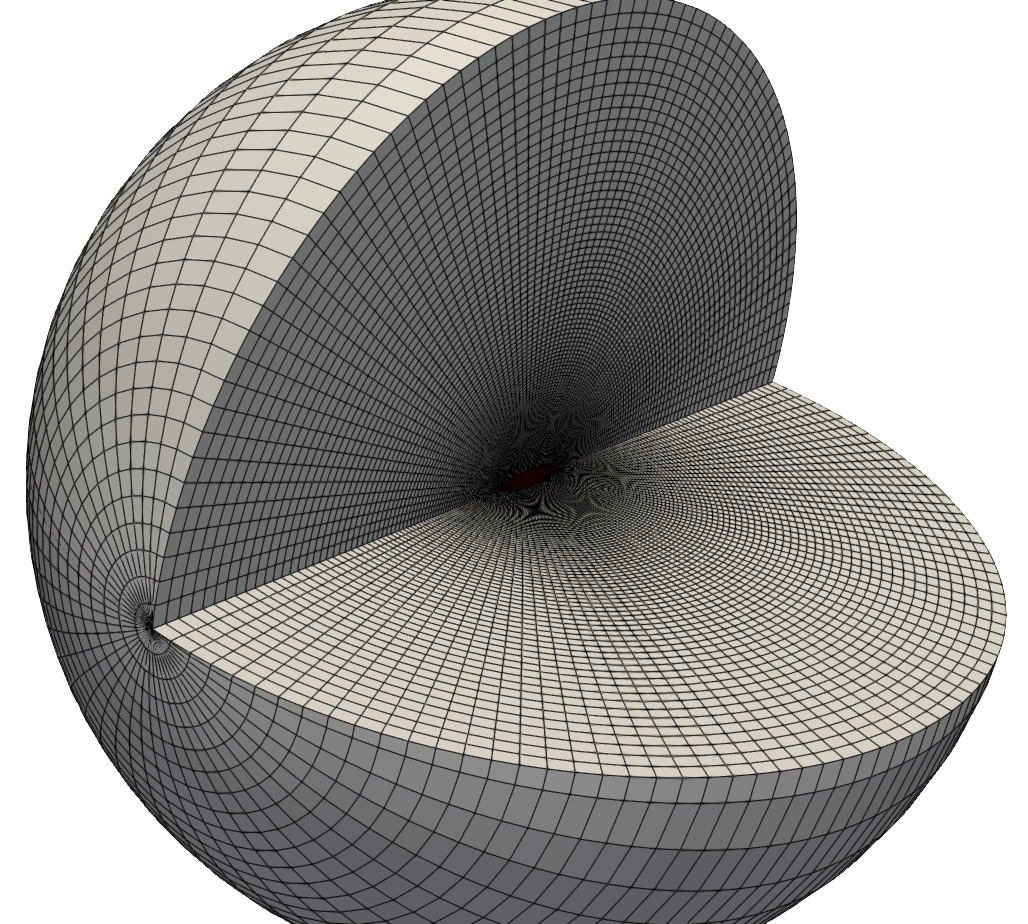}}\hfill
\subfloat{\includegraphics[width=0.48\textwidth]{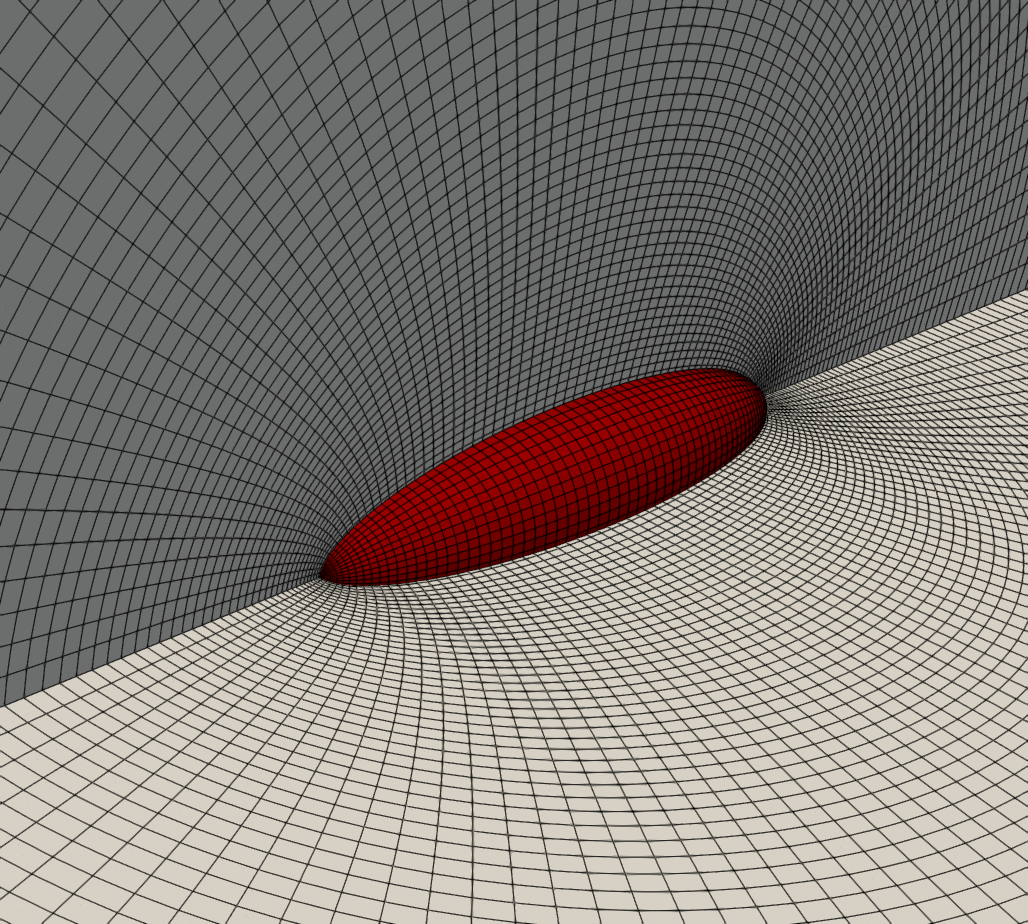}}
\caption {Discretized computational domain ($\kappa=4$, $r_\text{minor}=1$, $||\mathbf{r}_\infty||_2\approx||\mathbf{r}^\text{minor}_\infty||_2=80$, $N_1=90$, $N_2=N_3=71$ and $c_1=c_2=0$): {left panel shows the full domain, and right panel shows a zoomed region near particle surface (red).} Pressure grid is shown here, velocity grid is staggered relative to this according to definitions in equation \eqref{eq:Grid}.\label{fig:discretizeddomainfull}}
\end{figure}

We use a prolate spheroidal coordinate system to spatially discretize the particle surface and the fluid region around it. The transformation between the prolate spheroidal ($\boldsymbol{\xi}$) and cartesian ($\mathbf{x}$) coordinates, defined for the particle-fixed reference frame, are
\begin{equation}
x_1=f\sinh(\xi_1)\sin(\xi_2)\sin(\xi_3),\hspace{0.2in}x_2=f\sinh(\xi_1)\sin(\xi_2)\cos(\xi_3),\hspace{0.2in}x_3=f\cosh(\xi_1)\cos(\xi_2),\label{eq:CoordinateTransform}
\end{equation}
where $\xi_2\in[0,\pi]$ and $\xi_3\in[0,2\pi]$. $f$ is the focal length of the prolate spheroidal particle. Here, $\xi_1$, $\xi_2$, and $\xi_3$ are similar to the radial, polar and azimuthal directions respectively in a spherical coordinate system. The surface of the particle with an aspect ratio, $\kappa$, is exactly modeled as one of the coordinate surface i.e. $\xi_1=\xi_1^\text{surface}$. We consider a prolate spheroid spheroid with minor radius, $r_\text{minor}$. The surface, $\xi_1^\text{surface}$ and focal length, $f$, are,
\begin{equation}
\xi_1^\text{surface}=\frac{1}{2}\log\Big(\frac{\kappa+1}{\kappa-1}\Big),\hspace{0.2in}f=\frac{r_\text{minor}}{\sinh(\xi_1^\text{surface})}.
\end{equation}
The outer surface of the computational domain, $\mathbf{r}_\infty$, is also a prolate spheroidal surface that has a constant $\xi_1=\xi_1^\infty$,
\begin{equation}
\xi_1^\infty=\sinh^{-1}(||\mathbf{r}^\text{minor}_\infty||_2/f),
\end{equation}
where $||\mathbf{r}^\text{minor}_\infty||_2$ (2-norm or the Euclidean norm of $\mathbf{r}^\text{minor}$) is the minor radius of the outer surface. {It is a user prescribed parameter. The Euclidean distance of any point on a prolate spheroid $||\mathbf{r}_\infty||_2$ is related to its minor axis through $||\mathbf{r}_\infty||_2^2=||\mathbf{r}^\text{minor}_\infty||_2^2+f^2\cos^2(\xi_2)$. If this surface is placed far from the particle, such as the outer surface, $||\mathbf{r}_\infty||\approx||\mathbf{r}^\text{minor}_\infty||_2$. In other words, the outer surface is a nearly spherical surface. This is shown schematically in figure \ref{fig:computationaldomain} and for an actual discretized example in the left panel of figure \ref{fig:discretizeddomainfull}.} The computational domain is defined within the limits: $\xi_1\in[\xi_1^\text{surface},\xi_1^\infty]$, $\xi_2\in[0,\pi]$, $\xi_3\in[0,2\pi]$.

In order to prevent pressure aliasing that leads to the checkerboard effect causing spurious pressure oscillations, we use a staggered grid arrangement \cite{chung2002computational}. The three velocity components are stored at the same location that we term velocity grid. Pressure and the components of the polymer configuration tensor $\boldsymbol{\Lambda}$ and the latter's matrix logarithm $\boldsymbol{\Psi}$ are stored at a location that is staggered relative to the velocity grid. These staggered locations form the pressure grid. The velocity and pressure grids are
\begin{align}
\text{Velocity Grid: } \xi_1^\text{dist.,vel}\cup\xi_2^\text{dist.,vel}\cup\xi_3^\text{dist.,vel},\hspace{0.2in}
\text{Pressure Grid: } \xi_1^\text{dist.,pres}\cup\xi_2^\text{dist.,pres}\cup\xi_3^\text{dist.,pres},
\end{align}
where
\begin{align}\begin{split}
\xi_{1,i}^\text{dist.,vel}=&\xi_1^\text{surface}+\frac{\xi_1^\infty-\xi_1^\text{surface}}{N_1-1}(i-1)[c_1\frac{(i-1)(i-0.5)}{(N_1-1)^2}+1],\hspace{0.1in}i=[1,2,\cdots,N_1],\\
\xi_{1,1}^\text{dist.,pres}=&\xi_1^\text{surface},\hspace{0.1in}\xi_{1,i}^\text{dist.,pres}=\frac{\xi_{1,i}^\text{dist.,vel}+\xi_{1,i-1}^\text{dist.,vel}}{2},\hspace{0.1in}i=[2,3,\cdots,N_1],\hspace{0.1in}\xi_{1,N_1+1}^\text{dist.,pres}=\xi_1^\infty,\\
\xi_{2,j}^\text{dist.,pres}=&\pi\frac{j-1}{N_2}[c_2\frac{(j-1)(j-0.5)}{(N_2-1)^2}+1],\hspace{0.1in}j=[2,3,\cdots,N_2+1],\\
\xi_{2,j}^\text{dist.,vel}=&\frac{\xi_{2,j}^\text{dist.,pres}+\xi_{2,j+1}^\text{dist.,pres}}{2},\hspace{0.1in}j=[1,2,\cdots,N_2],\\
\xi_{3,k}^\text{dist.,vel}=&2\pi\frac{k-1}{N_3-1},\hspace{0.1in}k=[1,2,\cdots,N_3],\hspace{0.1in}
\xi_{3,k}^\text{dist.,pres}=\frac{\xi_{3,k}^\text{dist.,vel}+\xi_{3,k+1}^\text{dist.,vel}}{2},\hspace{0.1in}k=[1,2,\cdots,N_3-1],\\
\end{split}\label{eq:Grid}\end{align}
when $c_1$ and $c_2$ are zero, we obtain a uniform grid in spheroidal coordinates. A uniform spheroidal grid is naturally more clustered near the particle surface in the Euclidean sense than a uniform Cartesian grid. {When $c_1<0$, further grid clustering towards the particle and the outer surface is obtained. Another useful grid for this purpose to obtain grid clustering near the particle surface only (and appropriate when $c_1\ne0$) is $\xi_{1,i}^\text{dist.,vel}=\xi_1^\text{surface}+({\xi_1^\infty-\xi_1^\text{surface}})\{\exp[-c_1(i-1)/(N_1-1)]-1\}/\{\exp(-c_1)-1\},\hspace{0.1in}i=[1,2,\cdots,N_1]$.} Setting $c_2<0$ allows us to cluster the grids (in $\xi_2$ coordinate) towards the major axis of the particle (which is useful in studies such as extensional flow around the particle with its major axis aligned with the extensional axis). We only consider a uniform grid in the azimuthal, $\xi_3$ direction as the particle is axisymmetric, and there is no a priori preference of gradients at a particular $\xi_3$ location for a general linear flow around the particle. Any other function to describe a non-uniform grid can be used with our method as the finite difference and interpolation schemes described in section \ref{sec:FDSchemes} do not assume a grid type. Figure \ref{fig:discretizeddomainfull} shows the pressure grid for a spheroid with $\kappa=4$, $||\mathbf{r}_\infty||_2\approx||\mathbf{r}^\text{minor}_\infty||_2=80$, $N_1=90$, $N_2=N_3=71$ and $c_1=c_2=0$.

\subsection{Caretsian basis for vectors and tensors and discrete representation of spatial operators}\label{sec:ChoiceOfbasis}
We use the Cartesian basis on this prolate spheroidal grid to represent the velocity vector or the polymer configuration tensor at each location. This is done to prevent the coupling of different velocity vector components in the vector Laplacian operator (also see equation \eqref{eq:MomentumMass} and following discussion) appearing in the momentum equations. The momentum equation written in Cartesian basis on the prolate spheroidal grid is,
\begin{equation}
{\rho_f}\big(\frac{\partial \widetilde{u}_i}{\partial t}+\widetilde{\text{ADV}}(\widetilde{\mathbf{u}},\mathbf{u}_\infty,\boldsymbol{\omega}_p;\mathbf{r})_i\big)=-\frac{\partial \xi_k}{\partial x_i}\frac{\partial \widetilde{p}}{\partial \xi_k} +\nabla^2_\text{SphinCart}\widetilde{u}_i+\frac{\partial \xi_k}{\partial x_j}\frac{\partial (\Pi_{ji}-\Pi_{ji,\infty})}{\partial \xi_k},
\end{equation}
where,
\begin{equation}
\nabla^2_\text{SphinCart}=\frac{1}{h_1^2}\Big(\frac{1}{\tanh(\xi_1)}\frac{\partial}{\partial \xi_1}+\frac{\partial^2}{\partial\xi_1^2}\Big)+\frac{1}{h_2^2}\Big(\frac{1}{\tan(\xi_2)}\frac{\partial}{\partial \xi_2}+\frac{\partial^2}{\partial\xi_2^2}\Big)+\frac{1}{h_3^2}\frac{\partial^2}{\partial\xi_3^2},
\end{equation}
is a Laplacian operator that acts on individual Cartesian velocity component, $u_i$ with derivatives defined in prolate spheroidal coordinates. In the above equation
\begin{equation}
h_1=h_2=f\sqrt{\sinh^2(\xi_1)+\sin^2(\xi_2)}, h_3=f\sinh(\xi_1)\sin(\xi_2).\label{eq:ShapeFactors}
\end{equation}
An advantage of this operator is that $\text{SphinCart}\widetilde{u}_i$ has no cross-terms with other velocity components.
This decoupling allows us to solve the mass-momentum system of equations more efficiently. As mentioned in section \ref{sec:SchurMethod} the second step of the matrix-vector product in the Krylov subspace procedure (equation \eqref{eq:MatVec}) within the Schur complement method involves inversion of a discrete Laplacian operator acting on velocity vector. Unlike in curvilinear basis, the vector Laplacian in Cartesian basis can be treated as a set of three independent scalar Laplacians acting on three velocity components. Hence, it allows us to use the algebraic multi-grid methods such as AGMG \cite{agmg}, developed for scalar elliptic equations, to obtain the efficient inversion of the vector Laplacian operator. In other words, this choice of basis allows us to have a smaller bandwidth and number of off-diagonal components of the discrete matrix operators in equation \eqref{eq:MomentumMass} as compared to using a more natural choice of spheroidal basis.

As mentioned earlier in section \ref{sec:GridDefinition}, we use a staggered grid arrangement to represent different variables: $\widetilde{\mathbf{u}}$ is stored on the velocity and $\widetilde{p}$, $\boldsymbol{\Psi}$ and $\gamma$ on the pressure grid. The momentum equation \eqref{eq:MometumDiscrete} is discretized on the velocity grid. ${\mathbf{u}}_\infty$ is an analytically known function that can be evaluated at any location. The mass conservation equation \eqref{eq:MassDiscrete}, polymer constitutive equation \eqref{eq:ConstitutiveTimeDiscrete} and $\gamma$ equation \eqref{eq:FeneStretcheqn} for FENE models are discretized on the pressure grid. Similar to the finite difference representation of the viscous terms by \cite{desjardins2008high}, we use a straightforward local Lagrange polynomial representation of the spatial operators on a discretized quantity. This allows flexibility in choosing the order of accuracy of finite difference schemes in different directions and for different interpolation and spatial derivative operators and is described in section \ref{sec:FDSchemes}. The spatially and temporally discretized governing equations at each grid point and at time step $n+1$ are,
\begin{align}
&\widehat{\frac{\partial \xi_k}{\partial x_i}}\widehat{\frac{\delta \widetilde{u}_i^{n+1}}{\delta \xi_k}}=0,\\
& \Big(\frac{\rho_f}{\Delta t}-\nabla^2_\text{SphinCart}\Big|^\text{disc.}\Big)\widetilde{u}^{n+1}_i+\overline{\frac{\partial \xi_k}{\partial x_i}}\overline{\frac{\delta \widetilde{p}^{n+1}}{\delta \xi_k}}=g_i^\text{disc.},\label{eq:DiscMom}\\
&\frac{\Psi^{n+1}_{ij}}{\Delta t}+\frac{1}{2}\hat{u}_l^{n+1}\widehat{\frac{\partial \xi_k}{\partial x_l}}\frac{\delta {\Psi}_{ij}^{n+1}-\delta {\Psi}_{ij,\infty}^{n+1}}{\delta \xi_k}-\frac{\text{SR}(\boldsymbol{\Psi}^{n+1},\hat{\mathbf{u}}^{n+1})_{ij}^\text{disc.}}{2}=\frac{\text{SR}(\boldsymbol{\Psi}^{n},\hat{\mathbf{u}}^{n})_{ij}^\text{disc.}}{2}-\frac{1}{2}\hat{u}_l^{n}\widehat{\frac{\partial \xi_k}{\partial x_l}}\frac{\delta {\Psi}_{ij}^{n}}{\delta \xi_k}\label{eq:DiscreteConst},\\
&(\gamma^{n+1})^2+\gamma^{n+1}\Delta t\Big(-\frac{\gamma^n}{\Delta t}+\hat{u}_j^{n}\widehat{\frac{\partial \xi_k}{\partial x_j}}\frac{\delta \gamma^n}{\delta \xi_k}+\frac{2{\Lambda^n_{ij}}}{L^2}\widehat{\frac{\partial \xi_k}{\partial x_i}}\widehat{\frac{\partial u_j}{\partial \xi_k}}+\frac{\gamma^n-0.5}{De\gamma^n}+\frac{1}{DeL^2}\alpha_1^n\Big)-\frac{\Delta t}{2 De}\alpha_2=0\label{eq:DiscreteGamma},
\end{align}
$\mathbf{u}^{n+1}$ in the polymer equations \eqref{eq:DiscreteConst} and \eqref{eq:DiscreteGamma} is the total velocity i.e. the sum of velocity deviation $\widetilde{\mathbf{u}}^{n+1}$ and undisturbed velocity $\mathbf{u}_\infty^{n+1}$ evaluated on the pressure grid.
In these equations the hat, $\hat{}$, represents the scenario when the velocity or its spatial gradients are evaluated on the pressure grid. Similarly, the overbar, $\bar{}$, represents the case when a pressure grid variable or its gradient is evaluated on the velocity grid. Various terms appearing in these equations are now defined. $\alpha_1^n=3/(L^2-3)$ for FENE-P and  $1.5/\gamma^n$ for FENE-CR model, and, $\alpha_2=1$ for FENE-P and $1-3/L^2$ for FENE-CR model. These variables are irrelevant for Oldroyd-B or Giesekus models as the $\gamma$ equation is unnecessary.
\begin{equation}
\nabla_\mathbf{x}\boldsymbol{\xi}=\begin{bmatrix}
\frac{\partial \xi_1}{\partial x_1}&\frac{\partial \xi_2}{\partial x_1}&\frac{\partial \xi_3}{\partial x_1}\\
\frac{\partial \xi_1}{\partial x_2}&\frac{\partial \xi_2}{\partial x_2}&\frac{\partial \xi_3}{\partial x_2}\\
\frac{\partial \xi_1}{\partial x_3}&\frac{\partial \xi_2}{\partial x_3}&\frac{\partial \xi_3}{\partial x_3}
\end{bmatrix}=\begin{bmatrix}
\frac{f\cosh(\xi_1)\sin(\xi_2)\cos(\xi_3)}{h_1^2}&\frac{f\sinh(\xi_1)\cos(\xi_2)\cos(\xi_3)}{h_2^2}&-\frac{\sin(\xi_3)}{h_3}\\
\frac{f\cosh(\xi_1)\sin(\xi_2)\sin(\xi_3)}{h_1^2}&\frac{f\sinh(\xi_1)\cos(\xi_2)\sin(\xi_3)}{h_2^2}&\frac{\cos(\xi_3)}{h_3}\\
\frac{f\sinh(\xi_1)\cos(\xi_2)}{h_1^2}&-\frac{f\cosh(\xi_1)\sin(\xi_2)}{h_2^2}&0
\end{bmatrix},\label{eq:ShapeGradients}
\end{equation}
are spatial functions of the coordinate system that remain constant with time. $\overline{\frac{\partial \xi_i}{\partial x_j}}$ and $\widehat{\frac{\partial \xi_k}{\partial x_i}}$ are temporally constant analytical functions evaluated at the velocity and pressure grid respectively.
\begin{equation}
\nabla^2_\text{SphinCart}|^\text{disc.}\widetilde{u}^{n+1}_i=\frac{1}{h_1^2}\Big(\frac{1}{\tanh(\xi_1)}\frac{\delta \widetilde{u}^{n+1}_i}{\delta \xi_1}+\frac{\delta^2\widetilde{u}^{n+1}_i}{\delta\xi_1^2}\Big)+\frac{1}{h_2^2}\Big(\frac{1}{\tan(\xi_2)}\frac{\delta\widetilde{u}^{n+1}_i}{\delta \xi_2}+\frac{\delta^2\widetilde{u}^{n+1}_i}{\delta\xi_2^2}\Big)+\frac{1}{h_3^2}\frac{\delta^2\widetilde{u}^{n+1}_i}{\delta\xi_3^2},
\end{equation}
is discrete Laplacian operator acting at the velocity grid ($\xi_1$, $\xi_2$, $h_1$, $h_2$ and $h_3$ are evaluated at each grid point on the velocity grid).
\begin{equation}
{g}_i^\text{disc.}=\rho_f\big(\frac{{u}^{n}_i}{\Delta t}-\frac{3}{2}\widetilde{\text{ADV}}(\widetilde{\mathbf{u}}^n,\mathbf{u}^{n}_\infty,\boldsymbol{\omega}_p^n;\mathbf{r})_i+\frac{1}{2}\widetilde{\text{ADV}}(\widetilde{\mathbf{u}}^{n-1},\mathbf{u}^{n-1}_\infty,\boldsymbol{\omega}_p^{n-1};\mathbf{r})\big)_i+\overline{\frac{\partial \xi_k}{\partial x_j}}\overline{\frac{\delta(\Pi_{ji}^{n+1}-\Pi_{ji,\infty}^{n+1})}{\delta \xi_k}},\label{eq:DiscMom2}
\end{equation}
\begin{equation}
\widetilde{\text{ADV}}(\widetilde{\mathbf{u}}^n,\mathbf{u}^{n}_\infty,\boldsymbol{\omega}_p^n;\mathbf{r})_i=(\widetilde{u}_j^{n}+{u}_{\infty,j}^{n})\overline{\frac{\partial \xi_k}{\partial x_j}}\frac{\delta \widetilde{u}_i^{n}}{\delta \xi_k}+\widetilde{u}_j^{n}(\nabla \mathbf{u}_\infty)_{ji}+ 2\epsilon_{ijk}{\omega_j^{n}}\widetilde{u}_k^{n},
\end{equation}
$\nabla \mathbf{u}_\infty$ is a spatially constant tensor, that is analytically determined from the imposed flow and particle velocities along with the particle orientation (equation \eqref{eq:VelBCFinal}). $\text{SR}(\boldsymbol{\Psi}^n,\hat{\mathbf{u}}^n)_{ij}|^\text{disc.},$ is defined in equation \eqref{eq:StretchingRelaxation}  with $\boldsymbol{\Psi}=\boldsymbol{\Psi}^n$ and $\mathbf{u}=\hat{\mathbf{u}}^n$ where the velocity, $\hat{\mathbf{u}}^n$, dependence appears through  $\mathbf{L}=\nabla\mathbf{u}^n$ defined as,
\begin{equation}
{L}_{ij}={L}_{ij}^n=\widehat{\frac{\partial \xi_k}{\partial x_i}}\widehat{\frac{\delta u_j^n}{\delta \xi_k}}.
\end{equation}
\subsection{Finite Difference Schemes for spatial derivatives and interpolations}\label{sec:FDSchemes}
Values of various spatial derivatives and interpolated quantities appearing in the discretized equations, $\widehat{\frac{\delta \widetilde{u}_i^{n}}{\delta \xi_j}}$, $\overline{\frac{\delta \widetilde{p}^{n}}{\delta \xi_j}}$, $\frac{\delta ({\Psi}_{ij}^{n+1}-{\Psi}_{ij,\infty}^{n+1})}{\delta \xi_k}$, $\overline{\frac{\delta (\Pi_{ji}^{n+1}-\Pi_{ji,\infty}^{n+1})}{\delta \xi_k}}$, $\frac{\delta \gamma^n}{\delta \xi_j}$,  $\hat{u}_i^{n}$, $\frac{\partial \widetilde{u}_i^{n}}{\partial \xi_j}$ and $\frac{\delta^2\widetilde{u}^{n}_i}{\delta\xi_j^2}$, are approximated by higher order finite difference or interpolation schemes. Lagrange polynomials provide a convenient and flexible way to generate finite difference weights, irrespective of the grid spacing and order of accuracy \cite{fornberg1988generation}. Each variable, $\widetilde{u}_i$, $\widetilde{p}$, $\Psi_{ij}$, $\Pi_{ij}$ and $\gamma$ is locally represented by an $(M-1)^{th}$ order Lagrange polynomial by fitting the variable's data on $M$ neighboring grid points on which it is stored. Consider any of these variables to be represented by $\phi(\xi_1,\xi_2,\xi_3)$ which is stored at locations $[\xi_{1,i}^\text{dist.},\xi_{2,j}^\text{dist.},\xi_{3,k}^\text{dist.}]$, $i,j,k,\in\mathbb{N}$ (which represent either the pressure or the velocity grid defined in equation \eqref{eq:Grid}).
Interpolation along coordinate $\xi_i,i=[1,3]$, to a location $\widetilde{\xi}_{i}^\text{dist.}$, in Lagrange basis polynomials is,
\begin{align}\begin{split}
\phi(\widetilde{\xi}_{1}^\text{dist.},\xi_{2,j}^\text{dist.},\xi_{3,k}^\text{dist.})&=\Sigma_{q=n_1}^{M-n_1}l_q^{(0)}(\widetilde{\xi}_{1}^\text{dist.},{\xi}_{1,(.)}^\text{dist.})\phi(\xi_{1,q}^\text{dist.},\xi_{2,j}^\text{dist.},\xi_{3,k}^\text{dist.}),\\
\phi({\xi}_{1,i}^\text{dist.},\widetilde{\xi}_{2}^\text{dist.},\xi_{3,k}^\text{dist.})&=\Sigma_{q=n_1}^{M-n_1}l_q^{(0)}(\widetilde{\xi}_{2}^\text{dist.},{\xi}_{2,(.)}^\text{dist.})\phi(\xi_{1,i}^\text{dist.},\xi_{2,q}^\text{dist.},\xi_{3,k}^\text{dist.}),\\
\phi({\xi}_{1,i}^\text{dist.},{\xi}_{2,j}^\text{dist.},\widetilde{\xi}_{3}^\text{dist.})&=\Sigma_{q=n_1}^{M-n_1}l_q^{(0)}(\widetilde{\xi}_{3}^\text{dist.},{\xi}_{3,(.)}^\text{dist.})\phi(\xi_{1,i}^\text{dist.},\xi_{2,j}^\text{dist.},\xi_{3,q}^\text{dist.}).\\
&l_q^{(0)}(\widetilde{\xi}_{i}^\text{dist.},{\xi}_{i,(.)}^\text{dist.})=\prod\limits_{j=n_1,j\ne q}^{M-n_1}\frac{\widetilde{\xi}_{i}^\text{dist.}-\xi_{i,j}^\text{dist.}}{\xi_{i,q}^\text{dist.}-\xi_{i,j}^\text{dist.}},
\end{split}\label{eq:LagrangeInterp}\end{align}
where the set of discrete points $\{\xi_{i,q}^\text{dist.}, q=n_1,n_1+1,\cdots,M+n_1, n_1\in\mathbb{N}\}$ form an M point stencil around $\widetilde{\xi}_{i}^\text{dist.}$. The value of the variable $\phi$ is thus interpolated from its representation at these $M$ locations to a location represented by the point $\widetilde{\xi}_{i}^\text{dist.}$. As expected, if $\widetilde{\xi}_{i}^\text{dist.}$ is one of the locations in the $M$ point stencil, $\xi_{i,q}^\text{dist.}$, the original value of $\phi$ is recovered. Equation \eqref{eq:LagrangeInterp} is used to interpolate a variable from the pressure or velocity grid to a point $\widetilde{\xi}_{i}^\text{dist.}$  which is not one of $\xi_{i,q}^\text{dist.}$ locations but lies somewhere between these in the continuous space.  One can then choose to have $\widetilde{\xi}_{i}^\text{dist.}$ represent a point on the other grid.
The discrete first and second derivatives in $\xi_1$ are,
\begin{align}\begin{split}
\frac{\delta \phi}{\delta \xi_1}(\widetilde{\xi}_{1}^\text{dist.},\xi_{2,j}^\text{dist.},\xi_{3,k}^\text{dist.})&=\Sigma_{q=n_1}^{M+1-n_1}l_q^{(1)}(\widetilde{\xi}_{1}^\text{dist.},{\xi}_{1,(.)}^\text{dist.})\phi(\xi_{1,q}^\text{dist.},\xi_{2,j}^\text{dist.},\xi_{3,k}^\text{dist.})\\
\frac{\delta^2 \phi}{\delta \xi_1^2}(\widetilde{\xi}_{1}^\text{dist.},\xi_{2,j}^\text{dist.},\xi_{3,k}^\text{dist.})&=\Sigma_{q=n_1}^{M+1-n_1}l_q^{(2)}(\widetilde{\xi}_{1}^\text{dist.},{\xi}_{1,(.)}^\text{dist.})\phi(\xi_{1,q}^\text{dist.},\xi_{2,j}^\text{dist.},\xi_{3,k}^\text{dist.}),
\end{split}\end{align}
where,
\begin{align}\begin{split}
&l_q^{(1)}(\widetilde{\xi}_{i}^\text{dist.},{\xi}_{i,(.)}^\text{dist.})=\Sigma_{p=n_1,p\ne q}^{M+1-n_1}\frac{1}{\xi_{i,q}^\text{dist.}-\xi_{i,p}^\text{dist.}}\prod\limits_{j=n_1,j\ne (q,p)}^{M+1-n_1}\frac{\widetilde{\xi}_{i}^\text{dist.}-\xi_{i,j}^\text{dist.}}{\xi_{i,q}^\text{dist.}-\xi_{i,j}^\text{dist.}},\\
&l_q^{(2)}(\widetilde{\xi}_{i}^\text{dist.},{\xi}_{i,(.)}^\text{dist.})=\Sigma_{r=n_1,r\ne q}^{M+1-n_1}\frac{1}{\xi_{i,q}^\text{dist.}-\xi_{i,r}^\text{dist.}}\Sigma_{p=n_1,p\ne {(q,r)}}^{M+1-n_1}\frac{1}{\xi_{i,q}^\text{dist.}-\xi_{i,p}^\text{dist.}}\prod\limits_{j=n_1,j\ne (q,p,r)}^{M+1-n_1}\frac{\widetilde{\xi}_{i}^\text{dist.}-\xi_{i,j}^\text{dist.}}{\xi_{i,q}^\text{dist.}-\xi_{i,j}^\text{dist.}},\\
\end{split}\label{eq:Lagrange1st2nd}\end{align}
The derivatives in $\xi_2$ and $\xi_3$ are similarly defined. In the first and second order derivative operators of equation \eqref{eq:Lagrange1st2nd} $\widetilde{\xi}_{i}^\text{dist.}$  may or may not be one of $\xi_{i,q}^\text{dist.}$ locations. In the latter case it is somewhere within these $\xi_{i,q}^\text{dist.}$ locations in the continuous space. The interpolation and discrete differentiation in the above equations is expressed as a discrete operator acting on the variable $\phi$.

\textbf{Terms including variables on the other grid:}
$\widehat{\frac{\delta \widetilde{u}_i^{n}}{\delta \xi_j}}$ is obtained by the multiplying the first order discrete differential operator, $l_q^{(1)}({\xi}_{j}^\text{dist.,pres},{\xi}_{j,(.)}^\text{dist.,vel})$ and two interpolation operators $l_q^{(0)}({\xi}_{f}^\text{dist.,pres},{\xi}_{f,(.)}^\text{dist.,vel})$ for $\{f=1,2,3,f\ne j\}$ with $u_i^{n}$ in any order. Each operator in this case acts on the velocity grid and produces an output on the pressure grid in a particular dimension. Thus, $\widehat{\frac{\delta \widetilde{u}_i^{n}}{\delta \xi_j}}$ is the derivative of the velocity component $u_i$ at time step $n$ with respect to the $\xi_j$  coordinate, evaluated on the pressure grid. Similarly, $\overline{\frac{\delta \widetilde{p}^{n}}{\delta \xi_j}}$ is obtained by multiplication of $l_q^{(1)}({\xi}_{j}^\text{dist.,vel},{\xi}_{j,(.)}^\text{dist.,pres})$, and $l_q^{(0)}({\xi}_{f}^\text{dist.,vel},{\xi}_{f,(.)}^\text{dist.,pres})$ for $\{f=1,2,3,f\ne j\}$ with $\widetilde{p}^{n}$ in any order to give the pressure
derivative in $\xi_j$ direction, evaluated on the velocity grid. $\overline{\frac{\delta (\Pi_{ji}^{n+1}-\Pi_{ji,\infty}^{n+1})}{\delta \xi_k}}$ is obtained in a way similar to pressure derivative after the polymer stress $\Pi_{ji}^{n+1}$ and $\Pi_{ji,\infty}^{n+1}$ are obtained from the matrix logarithm of the polymer configurations $\Psi_{ji}^{n+1}$ and $\Psi_{ji,\infty}^{n+1}$ respectively using the method described just before section \ref{eq:DecouplingEqns}. $\hat{u}_i^{n}$, the velocity component in $i^{th}$ coordinate evaluated on the pressure grid, is obtained by three successive interpolations of velocity deviation from the undisturbed value in each direction from velocity to pressure grid and then adding the interpolated value to the undisturbed velocity evaluated at the pressure grid. Therefore, the interpolation errors are only incurred on the velocity deviation from the far field. Since centered schemes introduce no numerical diffusion the stencils in these cases consist of an even number of grid points centered (equal on each side) around the output location, ${\xi}_{j}^\text{dist.,vel}$ or ${\xi}_{j}^\text{dist.,pres}$, i.e. $M$ is an even integer. Near the boundaries of the computational domain, a centered stencil is not possible. In this case we still use an $M$ point stencil, but with more points on one side of the output location so as to keep all the stencil points within the computation domain and avoid using any ghost nodes. This issue arises only in the $\xi_1$ direction. $\xi_3$ coordinate is periodic about $\xi_3=0$ or $2\pi$ and we utilize this periodicity. The $\xi_2$ coordinate also has no external boundaries in the computational domain and we will discuss the special treatment required near the coordinate system dependent internal boundaries at $\xi_2=0$ and $\pi$ in section \ref{sec:OtherSide}.

\textbf{Non-convective terms including variables on the same grid:} The terms $\frac{\partial \widetilde{u}_i^{n}}{\partial \xi_j}$ and $\frac{\delta^2\widetilde{u}^{n}_i}{\delta\xi_j^2}$ appearing in the Laplacian operator, $\nabla^2_\text{SphinCart}|^\text{disc.}{u}^{n+1}_i$, require a single multiplication of the discrete derivative operator (first or second order) with the respective variable. In this case, the input and output grid are the same, i.e., the velocity grid. For $\frac{\partial \widetilde{u}_i^{n}}{\partial \xi_j}$ and $\frac{\delta^2\widetilde{u}^{n}_i}{\delta\xi_j^2}$ we use a stencil of odd length as the output location is one of the stencil points and there are equal number of points on either side of it. We deal with the $\xi_1$ boundaries in the similar way as described above for terms including variables on the other grid.

\textbf{Convective terms (HOUC schemes of Nourgaliev et al. (2007) \cite{nourgaliev2007high}):} The spatial gradients appearing in the convective terms must be treated carefully to prevent numerical instability. These include $\frac{\delta ({\Psi}_{ij}^{n+1}-{\Psi}_{ij,\infty}^{n+1})}{\delta \xi_k}$ and $\frac{\delta \gamma^n}{\delta \xi_j}$ in equations \eqref{eq:DiscreteConst} and \eqref{eq:DiscreteGamma}, and, $\frac{\delta \widetilde{u}_i^{n}}{\delta \xi_k}$ within $\widetilde{\text{ADV}}(\widetilde{\mathbf{u}}^{n},\mathbf{u}^{n}_\infty,\boldsymbol{\omega}_p^{n};\mathbf{r})_i$ of the discrete momentum equation (equations \eqref{eq:DiscMom} and \eqref{eq:DiscMom2}). Using centered difference schemes such as the one noted above for non-convective terms lead to a numerical instability. A slight upwinding, where one more point upstream is used relative to the downstream direction, allows one to maintain high spatial accuracy while maintaining numerically smooth and stable solution \cite{desjardins2008high,nourgaliev2007high}. This is termed as a higher order upwind central scheme (HOUC). It is straightforward to implement and introduces much less diffusion error compared to WENO schemes used to discretize the convective term \cite{nourgaliev2007high}. Near the boundaries of the computational domain ($\mathbf{r}_p$ and $\mathbf{r}_\infty$) we reduce the length of the stencil while maintaining upwinding. The direction of the velocity term multiplying the convective spatial gradient, $\tilde{u}_j^{n}\overline{\frac{\partial \xi_k}{\partial x_j}}$, ${u}_{j,\infty}^{n}\overline{\frac{\partial \xi_k}{\partial x_j}}$ or $\hat{u}_j^{n}\widehat{\frac{\partial \xi_k}{\partial x_j}}$, determines the shape of the stencil used at the time step, $n$. This term is equivalent to the velocity components in the spheroidal basis for $k=1$ and 2 i.e. the velocity along $\xi_1$ and $\xi_2$ directions respectively. For $k=3$ this term differs from velocity in the azimuthal, $\xi_3$, direction by a factor $h_3$ from equation \eqref{eq:ShapeFactors} and instead we use the direction of velocity in $\xi_3$ to determine the stencil shape. Therefore, similar to the original HOUC proposition of \cite{nourgaliev2007high} we choose the stencil shape based on the direction of convection in each direction. HOUC was originally \cite{nourgaliev2007high} proposed for interface tracking in multiphase flows, but we have found it to improve numerical stability of flow of viscoelastic fluid around a prolate spheroid as well.

\subsubsection{Internal boundaries in the $\xi_2$ direction}\label{sec:OtherSide}
The only physical boundaries of the computational domain are the two $\xi_1$ surfaces at $\xi_1^\text{surface}$ and $\xi_1^\infty$. Due to the choice of the coordinate system, $\xi_2$ is bounded by 0 and $\pi$ and $\xi_3$ by 0 and $2\pi$, which requires one to consider appropriate boundary conditions at these computational boundaries. As mentioned above, $\xi_3$ is a periodic coordinate, and the points $\xi_3=0$ and $2\pi$ represent the same physical location. Therefore periodic boundary condition in $\xi_3$ allows the interpolation mentioned above and discrete derivative operators in the $\xi_3$ direction to be implemented near the boundaries in a way similar to internal $\xi_3$ points. The boundary treatment of $\xi_1$ operators is already described above. The $\xi_2$ coordinate has non-periodic internal boundaries. Such a scenario appears at $r=0$ in the cylindrical coordinate system and has previously been treated in the context of finite difference schemes by several researchers \cite{verzicco1996finite,morinishi2004fully,desjardins2008high,constantinescu2002highly}. In this situation scenarios the finite difference stencil near $r=0$ requires information at an unphysical location $r<0$. In prolate spheroidal coordinates, the information is required in the unphysical regions $\xi_2<0$ and $\xi_2>\pi$. Similar to these previous works, a variable $\phi$ (representing $p$, $\gamma$ or a component of $\mathbf{u}$ or $\mathbf{\Psi}$) at these locations is defined as,
\begin{align}\begin{split}
\phi(\xi_1,\xi_2,\xi_3)&=\phi(\xi_1,-\xi_2,\xi_3+\pi),\hspace{0.2in}\xi_2\le 0\\
\phi(\xi_1,\xi_2,\xi_3)&=\phi(\xi_1,2\pi-\xi_2,\xi_3+\pi),\hspace{0.2in}\xi_2\ge\pi.\label{eq:Otherside}
\end{split}\end{align}
In previous works, the coordinate system and choice of basis for the vectors and tensors are the same, and care must be taken in the sign of certain components of vectors or tensors when using these transformations around the internal points/ axes \cite{desjardins2008high}. However, in a Cartesian basis the transformations of equation \eqref{eq:Otherside} are valid for any component of $\mathbf{u}$ or $\mathbf{\Psi}$ (as well as $p$ or $\gamma$).

{The issue of non-physical internal boundaries also appears in the spherical coordinate system near the origin at $r=0$ and at the polar axis ($\theta=0$ and $\pi$). This has been treated in a recent work of Santelli et al. \cite{santelli2021finite} by writing momentum equations for a transformed vector $[u_\phi\hspace{0.05in}u_rr^2\hspace{0.05in}u_\theta \sin(\theta)]^T$ obtained from the original velocity vector $[u_\phi\hspace{0.05in}u_r\hspace{0.05in}u_\theta]^T$. Zero transformed vector boundary conditions are imposed on the internal boundaries for equations governing the transformed vector by Santelli et al. \cite{santelli2021finite}.}
\subsubsection{Singular axis and boundary conditions}
Another non-physical issue that arises due to the choice of the coordinate system is the appearance of singular terms. $1/h_3$ (defined in equation \eqref{eq:ShapeFactors}), and $\frac{\partial \xi_3}{\partial x_1}$ and $\frac{\partial \xi_3}{\partial x_2}$ (defined in equation \eqref{eq:ShapeGradients}) have $\sin(\xi_2)$ in the denominator. These terms appear as a factor in front of several terms in the governing equations. As expressed so far, these terms lead to a coordinate system generated singularity at the axis on which $\xi_2=0$ or $\pi$. However, these factors appear along with a spatial gradient in the $\xi_3$ direction. The singular part of these terms can be expressed as
\begin{equation}
\text{Sing}=\frac{1}{\sin(\xi_2)}\frac{\partial \phi}{\partial \xi_3}.\label{eq:SingTerms}
\end{equation}
On the singular axes $\phi$, representing either a scalar $p$ or $\gamma$ or a Cartesian component of $\mathbf{u}$ or $\mathbf{\Psi}$, is unique. Therefore,
\begin{equation}
\frac{\partial \phi}{\partial \xi_3}\Big|_{\xi_2=0}=\frac{\partial \phi}{\partial \xi_3}\Big|_{\xi_2=\pi}=0.
\end{equation}
It can be checked from the governing equations that wherever terms of the type in equation \eqref{eq:SingTerms} appear there is no other $\xi_2$ dependence in the multiplicative factors. Therefore, using L'Hopital's theorem
\begin{equation}
\lim\limits_{\xi_2\rightarrow(0,\pi)}\text{Sing}= \frac{1}{\cos(\xi_2)}\frac{\partial^2 \phi}{\partial \xi_2  \partial \xi_3},
\end{equation}
and the non-physical coordinate system generated singularity is removed at the expense of introducing additional derivatives in the $\xi_3$ (azimuthal) direction. This treatment of coordinate system generated axis singularities is motivated by the treatment of the $1/r$ singularity at $r=0$ in cylindrical coordinate system by \cite{morinishi2004fully} and \cite{desjardins2008high}. The axis singularity  appears in the pressure grid where the points $\xi_{2,1}^\text{dist.,pres}=0$ and $\xi_{2,N_2}^\text{dist.,pres}=\pi$ are defined. Therefore, the singularity axis treatment is implemented in the discretization of the mass conservation and polymer constitutive equations. Such singularity does not appear on the velocity grid as it is staggered relative to the pressure grid. This treatment is not required to discretize the momentum equation.

{The previous methods of \cite{morinishi2004fully} and \cite{desjardins2008high} that used L'Hopital's theorem to account for the coordinate system generated singularity used a curvilinear (cylindrical) basis for the vectors. These vector components are multi-valued at the singularity ($r=0$) and as a result the axis values of these components were obtained using the relation between curvilinear and Cartesian components on the axis. The Cartesian components were averaged/ interpolated using the points off the singular axis. This leads to a difference between the analytical equation obtained after L'Hopital's theorem and that obtained using the Taylor series expansion of the actual discretized equations (compare equations (37) and (51) of \cite{desjardins2008high}). As mentioned in those studies, strict energy conservation cannot be obtained. Here, we use Cartesian components for the vectors and tensors and do not encounter this problem.}

\vspace{0.2in} This completes the description of the numerical method. In the next section, we implement this method on several flows of inertia-less Newtonian and viscoelastic fluids, along with examples of Newtonian fluids with inertia.

\section{Numerical Tests}\label{sec:Examples}
In this section, we use the numerical method described above to compute the flow field around particles such as a sphere and a prolate spheroid in a Newtonian fluid with and without (Stokes flow) inertia and an inertia-less viscoelastic fluid. Several fluid-particle interaction problems are considered for each class by changing the imposed velocity boundary conditions and torque constraint on the particle. These include fixed and freely rotating spheres and spheroids (including high aspect ratios) in uniform flow, simple shear flow, extensional flow, and combinations of the first two. These studies show forces, torques and stresslets, fluid streamlines, and the orientational trajectories of freely suspended spheroids.

As mentioned in section \ref{sec:FDSchemes} using Lagrange polynomials to represent finite difference spatial discretization and interpolation allows us great flexibility in choosing the order of spatial discretization schemes. In all the examples presented, we have used a four-point stencil in $\xi_1$, a six-point stencil in $\xi_2$ and an eight-point stencil in the $\xi_3$ direction. The higher-order discretization in $\xi_2$ and $\xi_3$ is useful because for most cases (excluding section \ref{sec:rheology}) we use less mesh points in these directions as compared to $\xi_1$. After initial testing of different orders of accuracy (not shown), we have found these orders of accuracy to be adequate for validating the cases we present while obtaining numerically stable simulations. In all the examples presented $r_\text{minor}=1$.
\subsection{Stokes flow: Motion of particles in inertia-less Newtonian fluid}
One of our primary objectives in developing the numerical method described in the previous sections is to study flows with zero to moderate inertia. The interest in the former is in the presence of viscoelasticity. As discussed in section \ref{sec:Introduction}, \ref{sec:SchurMethod} and \ref{sec:ZeroParticleInertial} zero inertia studies are important in isolating the effect of viscoelastic fluids and our numerical method is suitable for studying flow with zero particle and fluid inertia. In this section, we compare our numerical method against the analytically available solution of Stokes flow ($\rho_f=\rho_p=0$) of a Newtonian fluid ($c=0$) around prolate spheroids.
\subsubsection{Jeffery Orbits: A spheroid rotating in an inertia-less Newtonian fluid}\label{sec:Jeffery}
\begin{figure}[h!]
\centering
\subfloat{\includegraphics[width=0.51\textwidth]{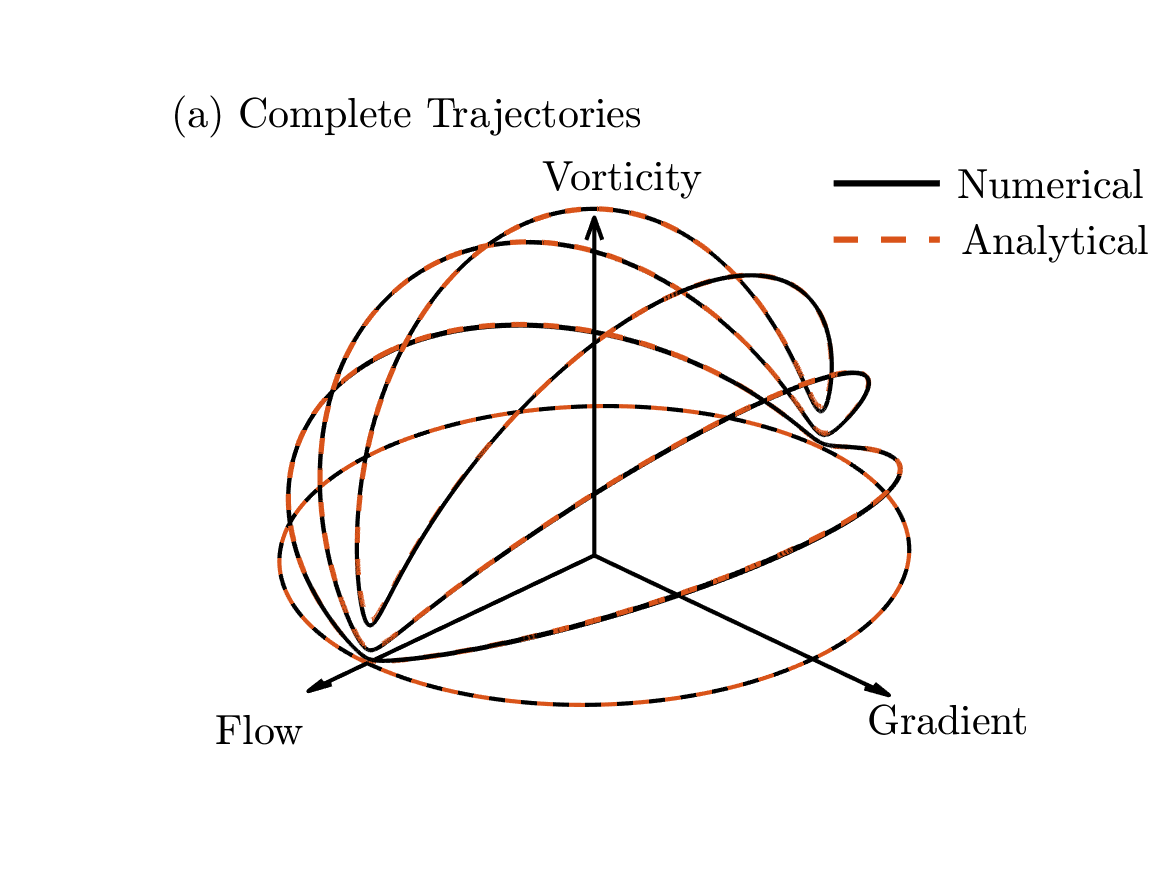}\label{fig:JefferyOrbitsa}}\hfill
\subfloat{\includegraphics[width=0.48\textwidth]{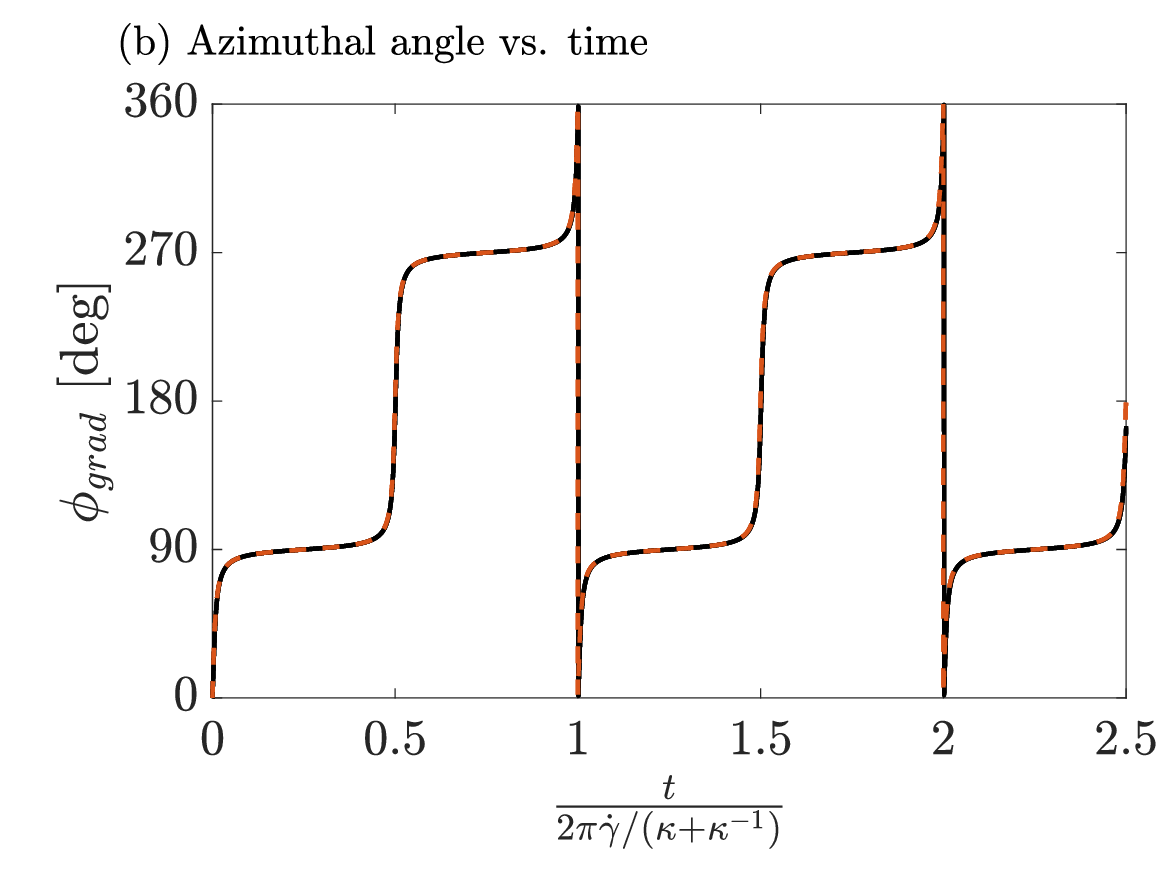}\label{fig:JefferyOrbitsab}}
\caption {Jeffery Orbits: Prolate spheroid of aspect ratio $\kappa=20$ rotating in simple shear flow of Newtonian fluid. Four different starting orientations are plotted in (a), and the orientation trajectory follows one of the degenerate periodic orbits depending upon the starting orientation. (b) shares the same legend as (a) and shows the time evolution of the polar angle $\phi_{grad}$ for the orbits of (a) (curves are indistinguishable as $\phi_{grad}$ is independent of the orbit constant). We obtain good agreement with the analytical prediction of Jeffery \cite{jeffery1922motion} for all orientations.\label{fig:JefferyOrbits}}
\end{figure}

The orientation trajectory of a freely rotating prolate spheroid in a simple shear flow of an inertia-less Newtonian fluid in the absence of particle inertia is available from the work of Jeffery (1922) \cite{jeffery1922motion}. At a given time, $t$, the polar angle, $\theta_{vort}$, with the vorticity axis of the imposed shear flow, and the azimuthal angle, $\phi_{grad}$, with the gradient direction in the flow-gradient plane, subtended by a prolate spheroid of aspect ratio $\kappa$ are given by
\begin{equation}
\tan(\phi_{grad})=\kappa\tan\Big(\frac{\dot{\gamma}t}{\kappa+\kappa^{-1}}\Big),\hspace{ 0.2in}\tan(\theta_{vort})=\frac{C\kappa}{\sqrt{\kappa^2\cos^2(\phi_{grad})+\sin^2(\phi_{grad})}},\label{eq:Jeffery}
\end{equation}
where $\dot{\gamma}$ is the shear rate of the imposed flow and $C$ is an orbital constant that depends on initial orientation. Hence, a prolate spheroid undergoes three-dimensional periodic motion in a shear flow of a Newtonian fluid, and the orbit's shape is determined by initial orientation. In figures \ref{fig:JefferyOrbitsa} and \ref{fig:JefferyOrbitsab}, we show these Jeffery orbits and the evolution of the polar angle, $\phi_{grad}$, respectively, computed from our code along with the analytical expressions from equation \eqref{eq:Jeffery} for $\kappa=20$ at four starting orientations. In the numerical solution, the domain size used is 10$\kappa$ and a uniform mesh is used with $N_1=120$, $N_2=N_3=71$ (these are defined in equation \eqref{eq:Grid}). Terms multiplying fluid and particle density are ignored. At each time step, the {resistivity formulation} method described in section \ref{sec:ZeroParticleInertial} is used to determine the particle's angular velocity, which leads to zero net torque on the particle. {The necessary Stokes flows (motion induced part of the momentum equation) required in this formulation (as mentioned in section \ref{sec:ZeroParticleInertial}) are pre-calculated numerically using the Schur complement approach of section \ref{sec:SchurMethod} before time evolving the particle's orientation. The numerical and the analytical curves shown in figure \ref{fig:JefferyOrbits} are indistinguishable.}

\subsubsection{Forces on a fixed spheroid in a uniform flow}\label{sec:UniformStokes}
The drag, $C_D$, and the lift coefficient, $C_L$, of a prolate spheroid with aspect ratio, $\kappa$, fixed in a uniform flow are given by \cite{oberbeck1876ueber,happel2012low,siewert2014orientation},
\begin{equation}
C_L=\frac{\text{Lift}}{\mu U_0 r_\text{minor}}=16\pi\kappa((K_{zz}-K_{xx})\cos^2(\theta)+K_{xx}),\hspace{0.2in}C_D=\frac{\text{Drag}}{\mu U_0 r_\text{minor}}=16\pi\kappa(K_{xx}-K_{zz})\cos(\theta)\sin(\theta),\label{eq:CoeffDefs}
\end{equation}
where drag and lift are the hydrodynamic forces acting parallel and perpendicular to the imposed flow in the plane of the imposed flow and the particle center line (major axis). $\mu$ and $U_0$ are the fluid viscosity and imposed speed, and $\theta$ is the angle of imposed flow relative to the major axis of the particle.
\begin{equation}
\begin{split}
&K_{xx}=\frac{1}{\xi+\alpha},\hspace{0.2in}
K_{zz}=\frac{1}{\xi+\kappa^2\gamma},\\
\alpha=\frac{\kappa^2}{\kappa^2-1}+\frac{\kappa}{2(\kappa^2-1)^{1.5}}\eta,&\hspace{0.2in}
\gamma=-\frac{2}{\kappa^2-1}-\frac{\kappa}{(\kappa^2-1)^{1.5}}\eta,\hspace{0.2in}
\eta=\log\Big(\frac{\kappa-\sqrt{\kappa^2-1}}{\kappa+\sqrt{\kappa^2-1}}\Big).
\end{split}\label{eq:ForceCoefficients}\end{equation}
\begin{figure}[h!]
\centering
\subfloat{\includegraphics[width=0.49\textwidth]{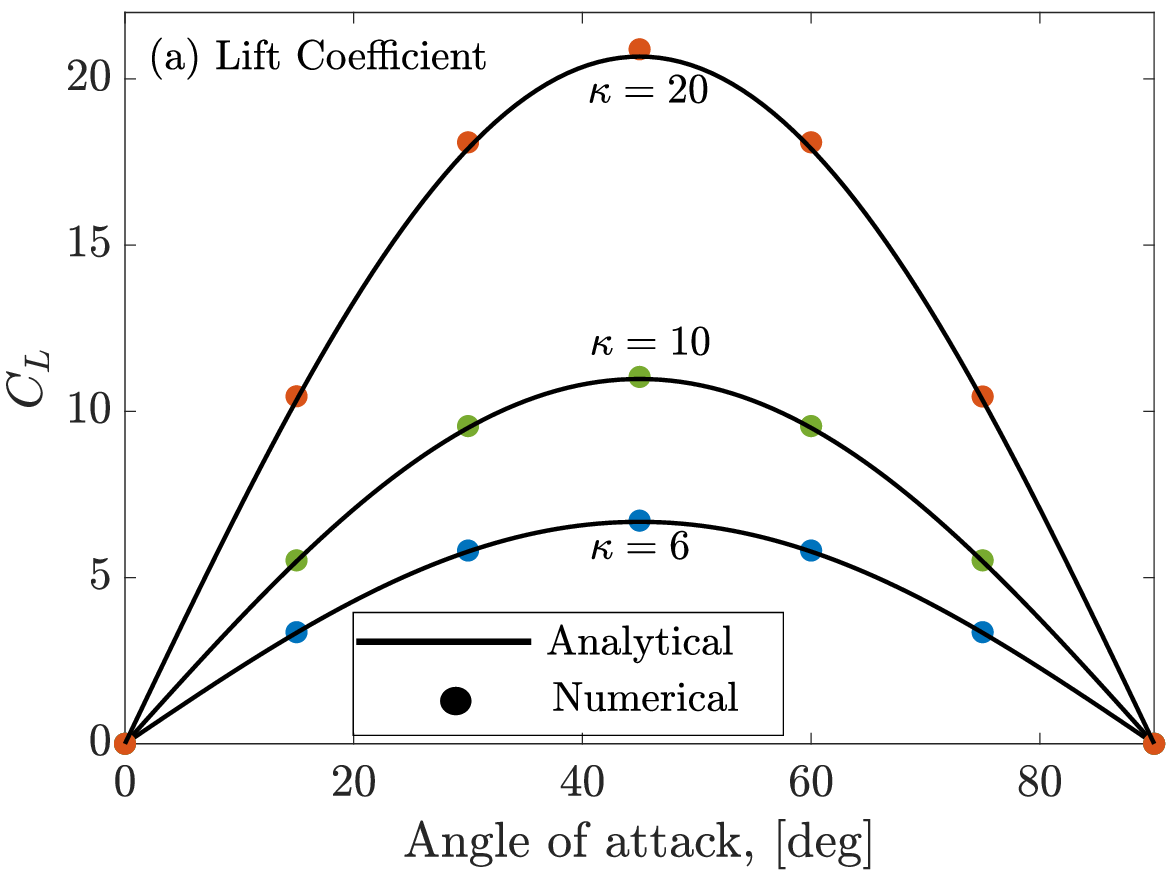}}\hfill
\subfloat{\includegraphics[width=0.49\textwidth]{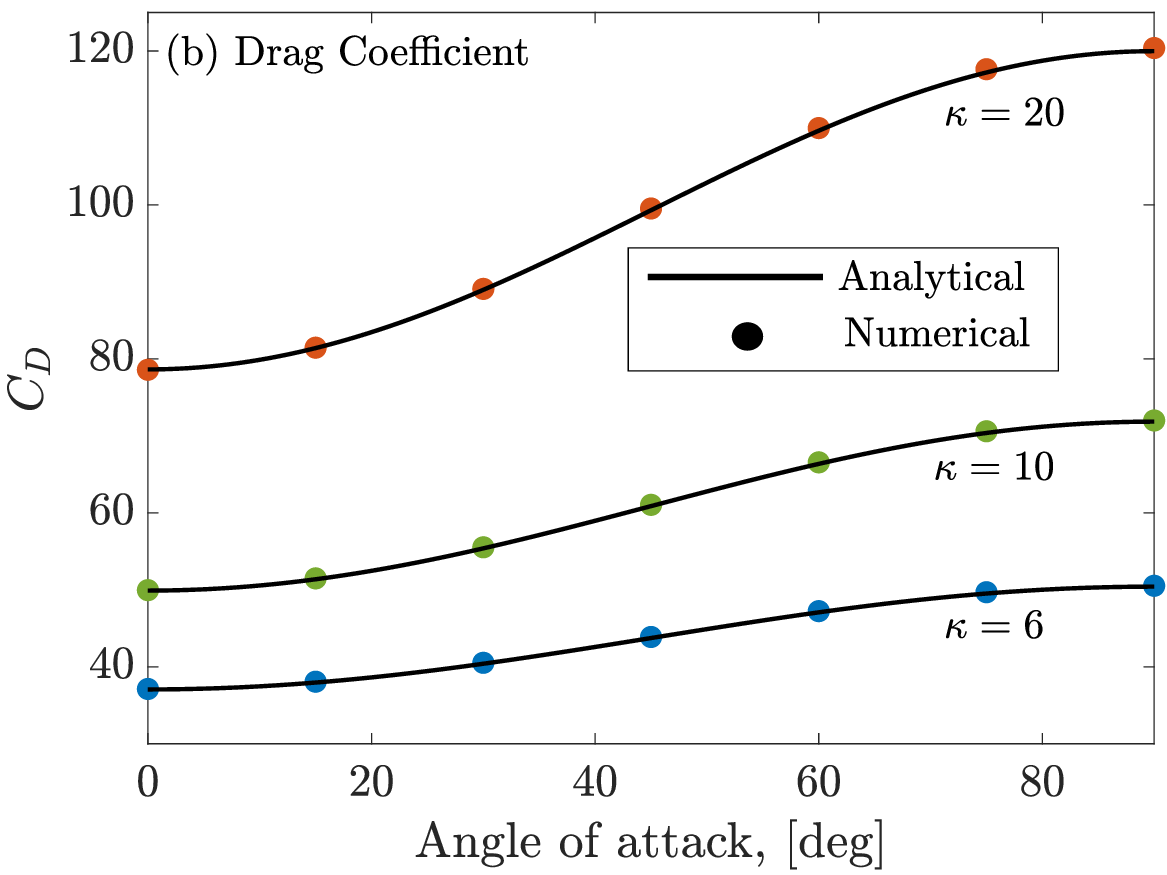}}\hfill
\caption {Lift and drag coefficients \eqref{eq:ForceCoefficients} on a prolate spheroid fixed in a uniform flow at various angle of attacks ($\theta$) and particle aspect ratios ($\kappa$) in the inertia-less limit ($\rho_f=0$). We obtain good agreement with the analytical prediction available in \cite{oberbeck1876ueber,happel2012low,siewert2014orientation}  for each $\kappa$ and $\theta$.\label{fig:StokesUniformFlow}}
\end{figure}
We show the lift and drag coefficient for the flows past spheroids of three different aspect ratios, $\kappa=6$, 10 and 20 in figure \ref{fig:StokesUniformFlow} for seven different angles of attack between the imposed uniform flow and the particle's major axis. Good quantitative agreement between our numerical results and the analytical expressions from equations \eqref{eq:CoeffDefs} and \eqref{eq:ForceCoefficients} is obtained. The domain size used is 100$\kappa$. A large computational domain is required for low inertia flows past a solid particle because the velocity disturbance created by the particle decays as $1/r$ at large distances ($r$) from the particle \cite{kim2013microhydrodynamics,graham2018microhydrodynamics}. This issue has marred previous computations of Andersson et al. \cite{andersson2019forces} who mentioned the requirement of a large computational domain as the Reynolds number in their study of $\kappa=6$ particle was reduced to about $0.1$. They used a rectangular computational domain of size $64\kappa\times64\kappa\times42.7\kappa$. Depending upon the angle of attack, they reported a deviation of 0.6 to 15\% from the analytical $C_D$ and $10.3-24.9$ \% from the analytical $C_L$ values for zero inertia. {In contrast, using a very large computational domain of size 100$\kappa$ with $N_1=250$, $N_2=80$ and $N_3=51$ we obtain highly accurate results for $\kappa=6$, 10 and 20 shown in figure \ref{fig:StokesUniformFlow}. The largest absolute error is at $\theta=45^\circ$ for $\kappa=20$. It is only 1.09\% and 0.23\% in $C_L$ and $C_D$ respectively. An accurate simulation of these cases with a reasonable number of mesh points is possible because while the required size of the computational domain is large, the grid can be relatively sparse in the far-field and the prolate spheroidal grid is naturally more clustered near the particle surface.} Therefore, prolate spheroidal coordinates employed in our simulations are an appropriate choice for such studies, particularly when we test theories designed to be a perturbation from Stokes flow. This example will be shown in section \ref{sec:InertiaUniformFlow}.

\subsection{Particles in Newtonian fluid with finite inertia}
\subsubsection{Uniform flow past a fixed prolate spheroid}\label{sec:InertiaUniformFlow}
\begin{figure}
\centering
\subfloat{\includegraphics[width=0.49\textwidth]{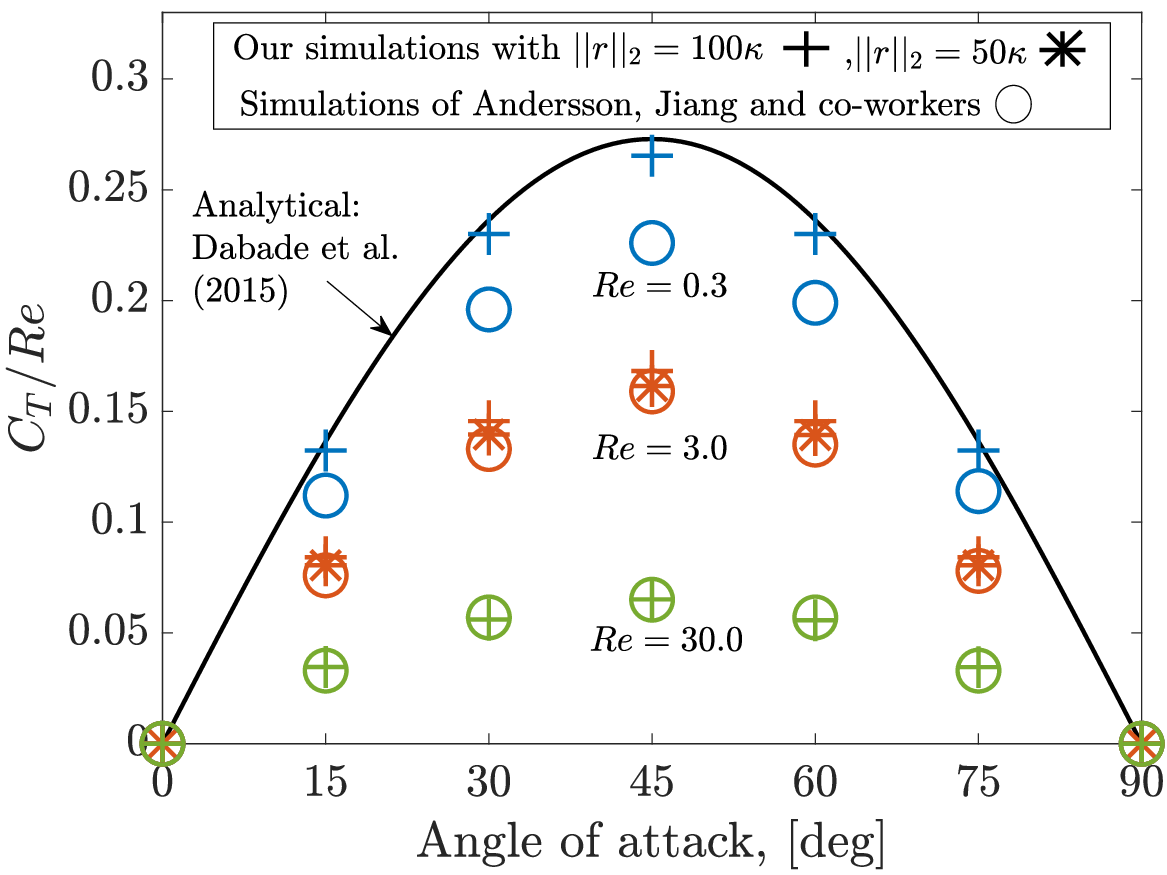}}
\caption{Uniform flow past a fixed spheroid: Steady-state torque coefficient, $C_T$ normalized with Reynolds number, $Re$, from our numerical simulations and those of {Andersson and Jiang (2019) \cite{andersson2019forces} and Jiang et al. (2021) \cite{jiang2021inertial}} at different angles of attack, $\theta$ and Reynolds numbers, $Re=0.3$, 3.0 and 30 along with the analytical calculations of Dabade et al. \cite{dabade2015effects} is also shown. {Our results agree with that of \cite{andersson2019forces,jiang2021inertial} at larger $Re$ and with \cite{dabade2015effects} at  $Re=0.3$.}}
\label{fig:comparemomentsre5}
\end{figure}

A prolate spheroid sediments in an inertial-less Newtonian fluid without any change in orientation. Equivalently, a uniform flow of inertial-less Newtonian fluid past a fixed spheroid, i.e., the cases of section \ref{sec:UniformStokes}, exerts no hydrodynamic torque on the particle. Inertia, however, leads to a finite torque on a fixed spheroid. In the low inertia limit, at the steady-state, Dabade et al. (2015) \cite{dabade2015effects} calculated the coefficient of inertial torque, $C_T$, on a particle with aspect ratio, $\kappa$, fixed in a Newtonian fluid. When the angle of attack of the oncoming flow relative to the particle axis is $\theta$,  they find,
\begin{equation}
C_T=\frac{Re}{2}F(\kappa)\sin(2\theta),
\end{equation}
where, $C_T$ and the Reynolds number, $Re$, are,
\begin{equation}
C_T=\frac{\text{Torque}}{\mu U_0 r_\text{minor}^2\kappa^2},\hspace{0.2in} Re=\frac{\rho_f U_0 r_\text{minor}\kappa}{\mu},\label{eq:TransRe}
\end{equation}
and $U_0$ is the imposed velocity. $F(\kappa)$ is a non-dimensional parameter that depends on the particle aspect ratio and $F(6)=0.5458$ \cite{dabade2015effects}. In figure \ref{fig:comparemomentsre5}, we show the steady-state values of $C_T/Re$ at various $\theta$ for $\kappa=6$ and three different $Re=0.3$, 3.0 and 30.0 from our numerical results along with those of {Andersson and Jiang (2019) \cite{andersson2019forces} and Jiang et al. (2021) \cite{jiang2021inertial}.} We also show the analytical expression $F(6)\sin(2\theta)/2$ of Dabade et al. (2015) \cite{dabade2015effects}.

Due to the inertial screening (see \cite{brenner1999screening,segre2007inertial,bergougnoux2021dilute} for numerical and experimental evidence of this mechanism in a dilute suspension of sedimenting spheres), the velocity disturbance due to the particle in the presence of fluid inertia {decays} at a rate faster than the $1/r$ ($r$ is the distance from the particle) decay characteristic of the inertia-less or Stokes limit (see section \ref{sec:UniformStokes} and \cite{kim2013microhydrodynamics,graham2018microhydrodynamics}). However, the inertial screening length may still be large if Reynolds number is small and a large computational domain may be required to quantitatively assess the validity of theories developed for small inertial corrections. 
In figure \ref{fig:comparemomentsre5} we observe that our numerical results, performed with a large computational domain size of {$||\mathbf{r}_\infty||_2\approx||\mathbf{r}^\text{minor}_\infty||_2=100\kappa$} are closer to the analytical prediction of Dabade et al. (2015) \cite{dabade2015effects} for $Re=0.3$ than the numerical results of Andersson, Jiang and co-workers (2019,2021) \cite{andersson2019forces,jiang2021inertial} performed with a rectangular domain of size $64\kappa\times64\kappa\times42.7\kappa$ \cite{andersson2019forces} or $34\kappa\times34\kappa\times34\kappa$ \cite{jiang2021inertial}. For 45$^\circ$, simulations of \cite{andersson2019forces,jiang2021inertial}  have a deviation of $17\%$ from the analytical prediction of \cite{dabade2015effects} at $Re=0.3$. We have a deviation of $2.7\%$. In addition to larger domain size, we use a straightforward boundary condition of a constant uniform velocity on the outer boundary.  However, the discussion in \cite{andersson2019forces} points to a more involved boundary condition treatment. {At higher $Re=3.0$ and 30.0 our computations agree with that of \cite{andersson2019forces,jiang2021inertial} as shown in figure \ref{fig:comparemomentsre5}. For $Re=3.0$ we show another simulation result at $||\mathbf{r}_\infty||_2\approx||\mathbf{r}^\text{minor}_\infty||_2=50\kappa$. The resolution of our computational grid is $N_1=200$, $N_2=131$ and $N_3=65$ for both computational domains. The agreement with the results of \cite{andersson2019forces,jiang2021inertial} is better for the smaller of our computational domains at $Re=3.0$, a further indication of the importance of domain size at small to moderate inertia.} For $\theta=45^\circ$ at $Re=3.0$ and 30.0 we simulated additional cases with increased resolution $N_1=300$, $N_2=201$ and $N_3=91$ (not shown) and found similar results as with our lower resolution. Due to the nature of spheroidal coordinates (see equation \eqref{eq:CoordinateTransform} and more detailed discussion in section \ref{sec:Introduction} and \ref{sec:GridDefinition} about Euclidean spacing of the spheroidal grid), the domain size can be greatly increased without decreasing the resolution near the particle surface significantly. Therefore, our method is well equipped to study flow around a sedimenting particle or a particle fixed in uniform flow in the limit of small to moderate inertia.

\begin{figure}[h!]
\centering
\subfloat{\includegraphics[width=0.49\textwidth]{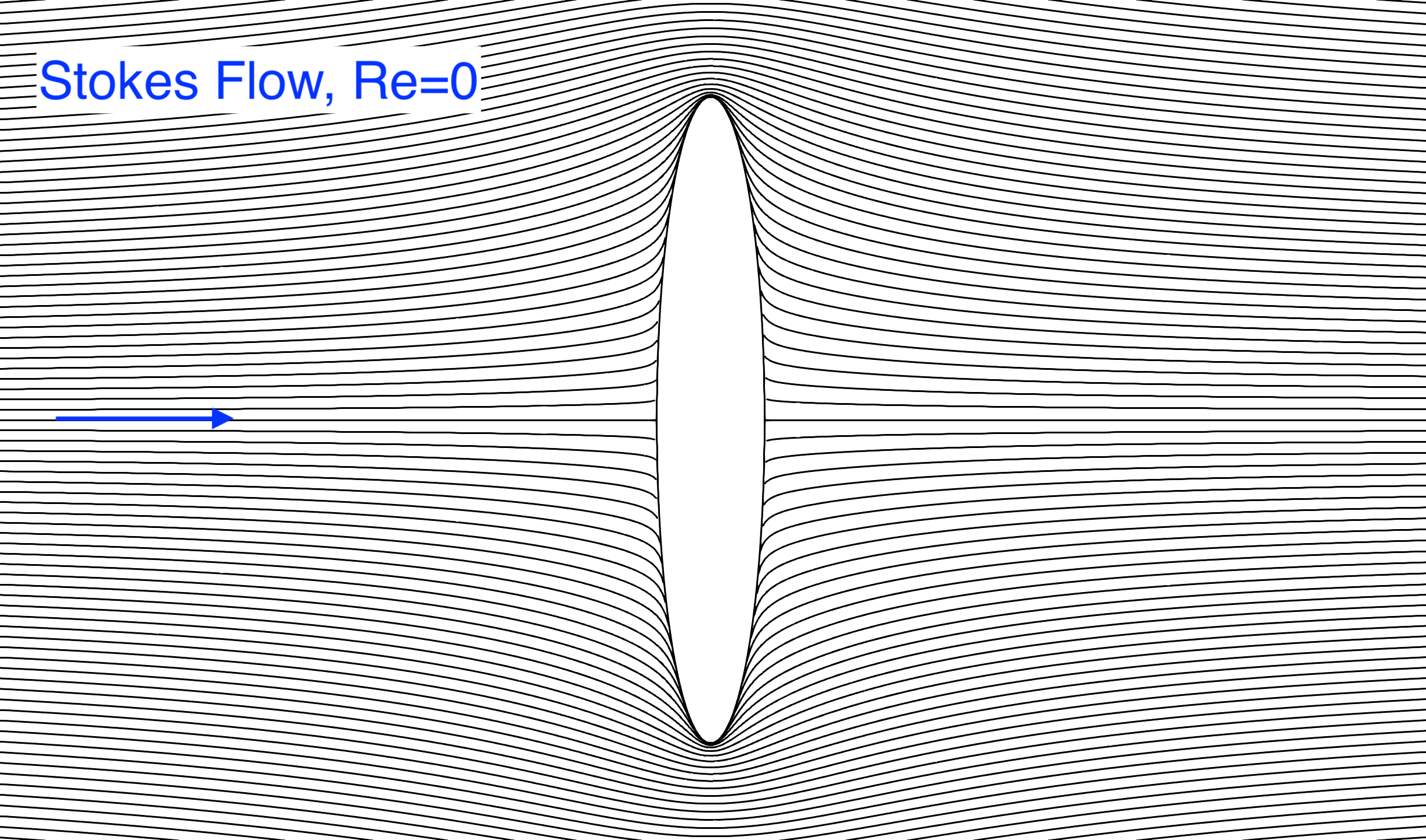}}\hfill
\subfloat{\includegraphics[width=0.49\textwidth]{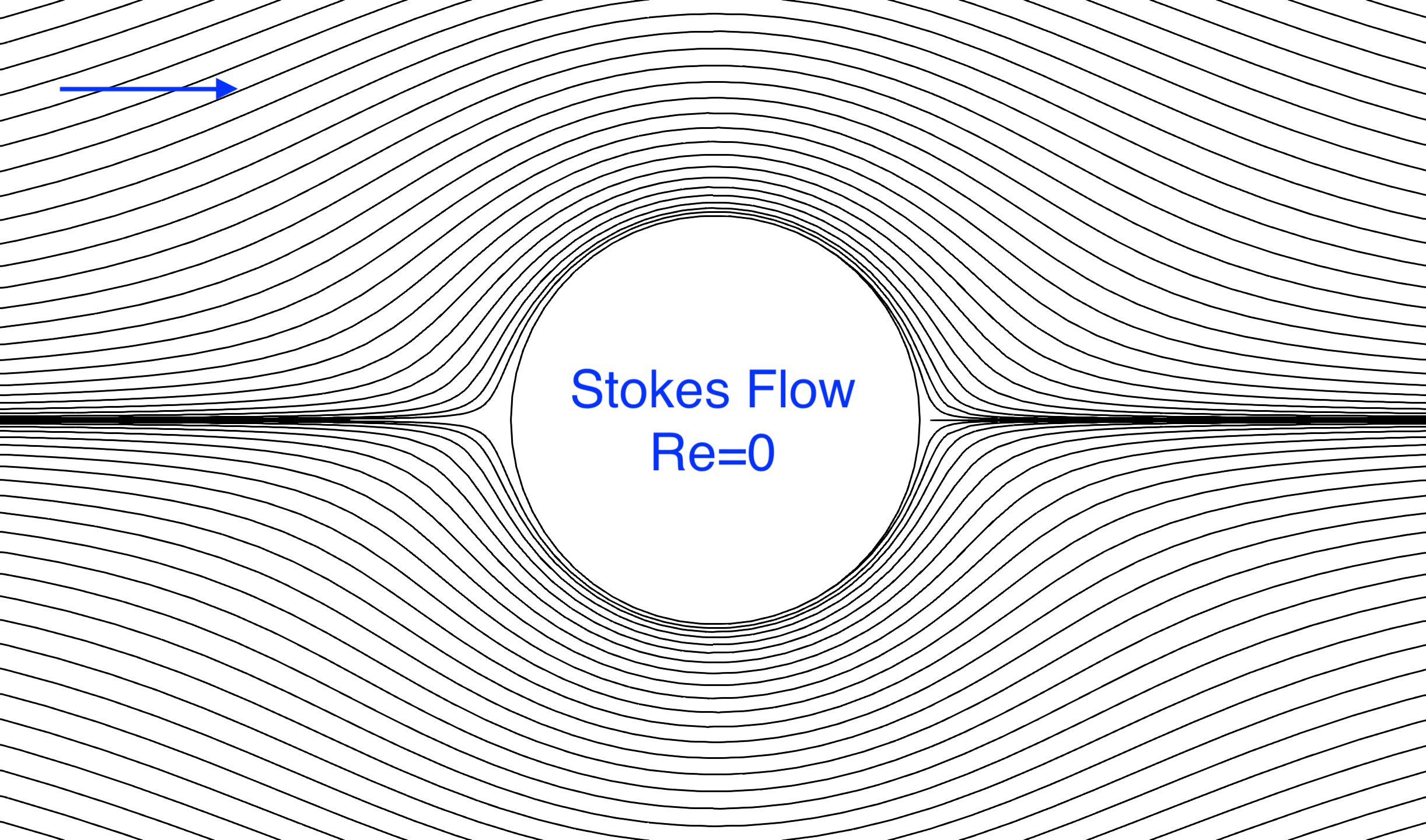}}\hfill
\subfloat{\includegraphics[width=0.49\textwidth]{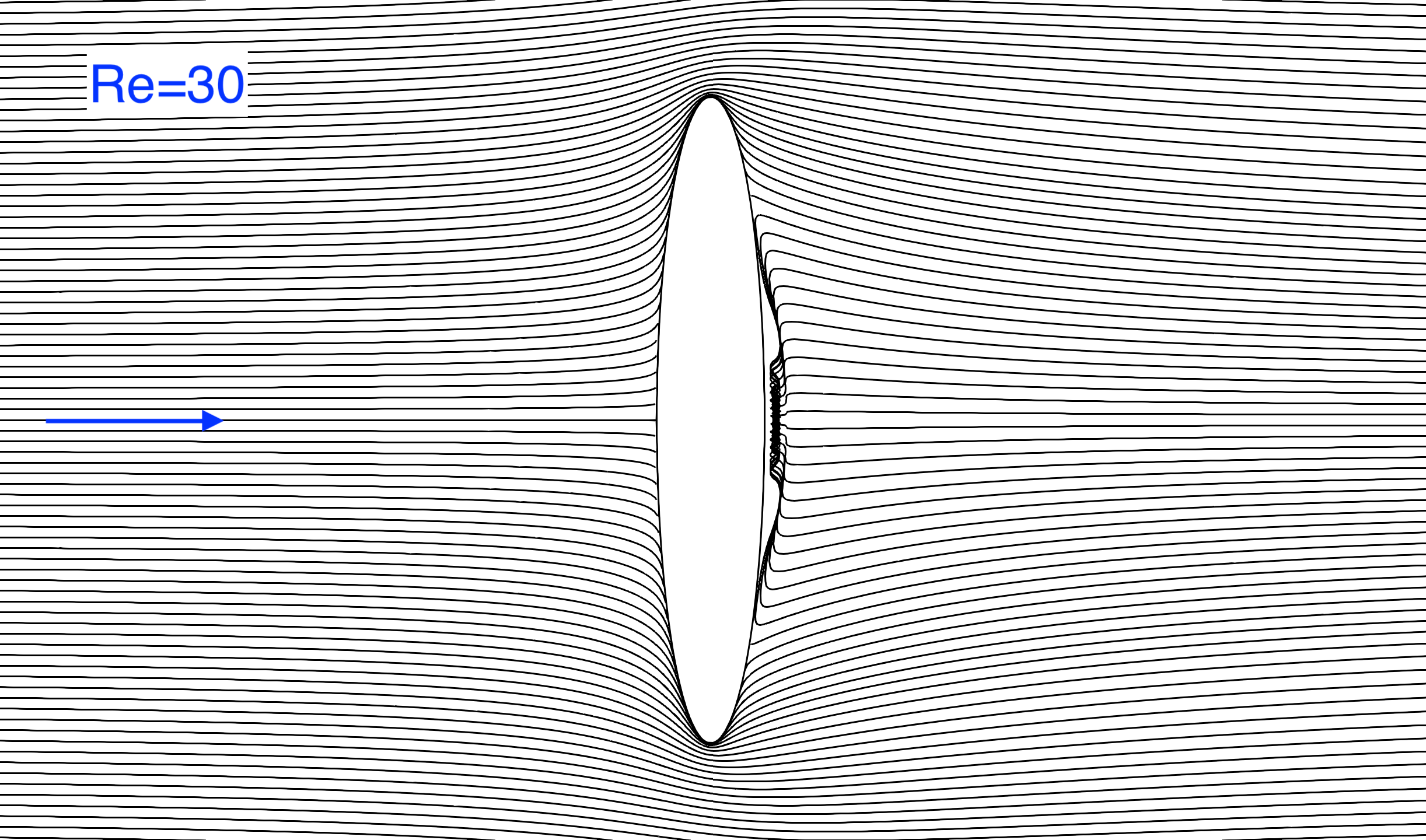}}\hfill
\subfloat{\includegraphics[width=0.49\textwidth]{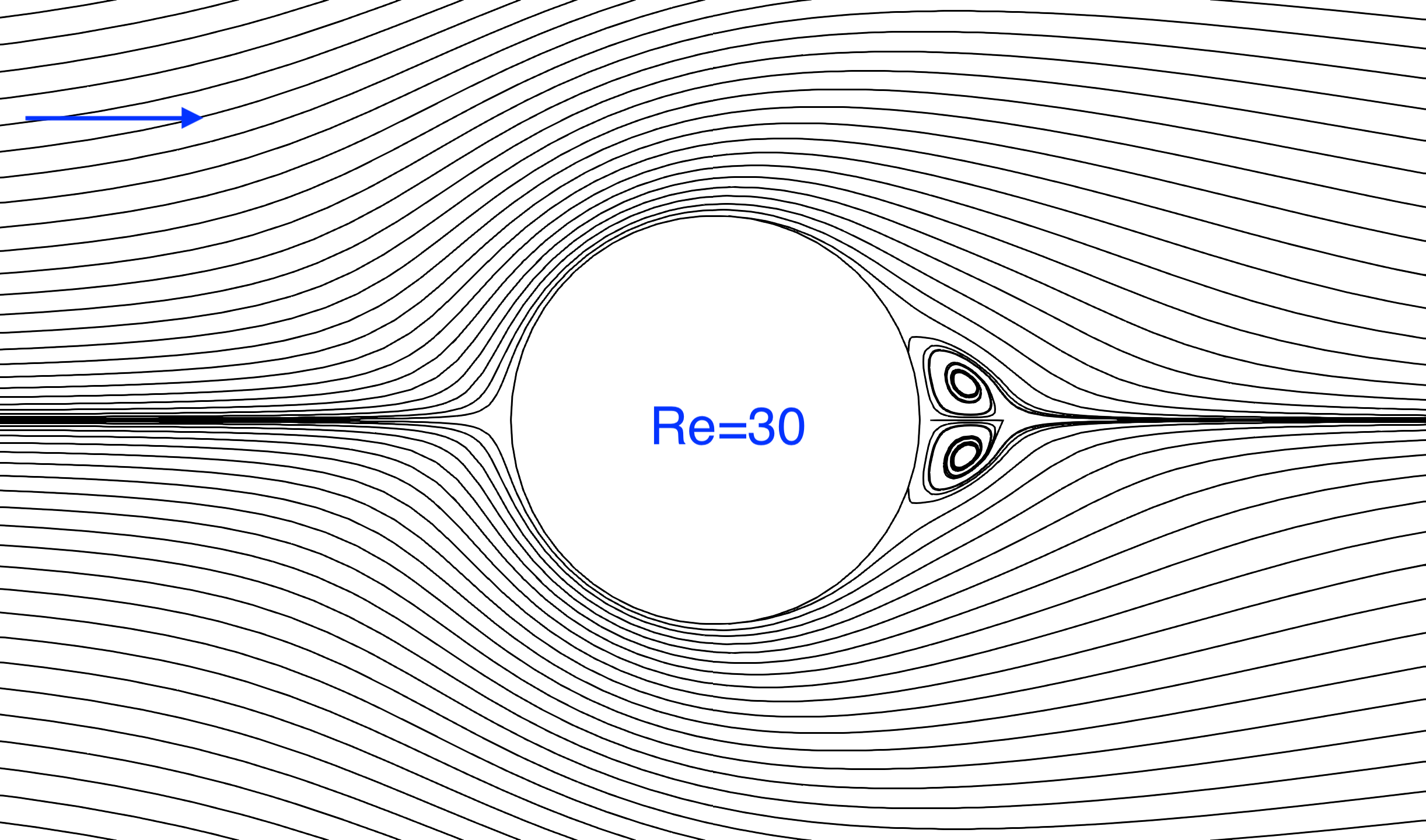}}\hfill
\caption {Streamlines of the Newtonian fluid flow around a fixed prolate spheroid of aspect ratio, $\kappa=6$ at Reynolds number $Re=0$ (Stokes flow) and $Re=30$ with major axis at 90$^\circ$ to the oncoming flow (left to right). For each $Re$, two different views are shown. The $Re=30$ streamlines are the same as those observed i in figures 5(d) and 6(d) of \cite{andersson2019forces}.\label{fig:StreamlinesUniform}}
\end{figure}

We also show the steady-state streamlines of the flow for $Re=0$ and $Re=30$ in figure \ref{fig:StreamlinesUniform} for $\theta=90^\circ$. The presence of inertia leads to the formation of trailing edge vortices. The streamline pictures for $Re=30$ are qualitatively similar to those shown by Andersson et al. \cite{andersson2019forces}.

\subsubsection{Effect of inertia on a freely rotating sphere in simple shear flow}\label{sec:InertialSlowDown}
\begin{figure}[h!]
\centering
\subfloat{\includegraphics[width=0.49\textwidth]{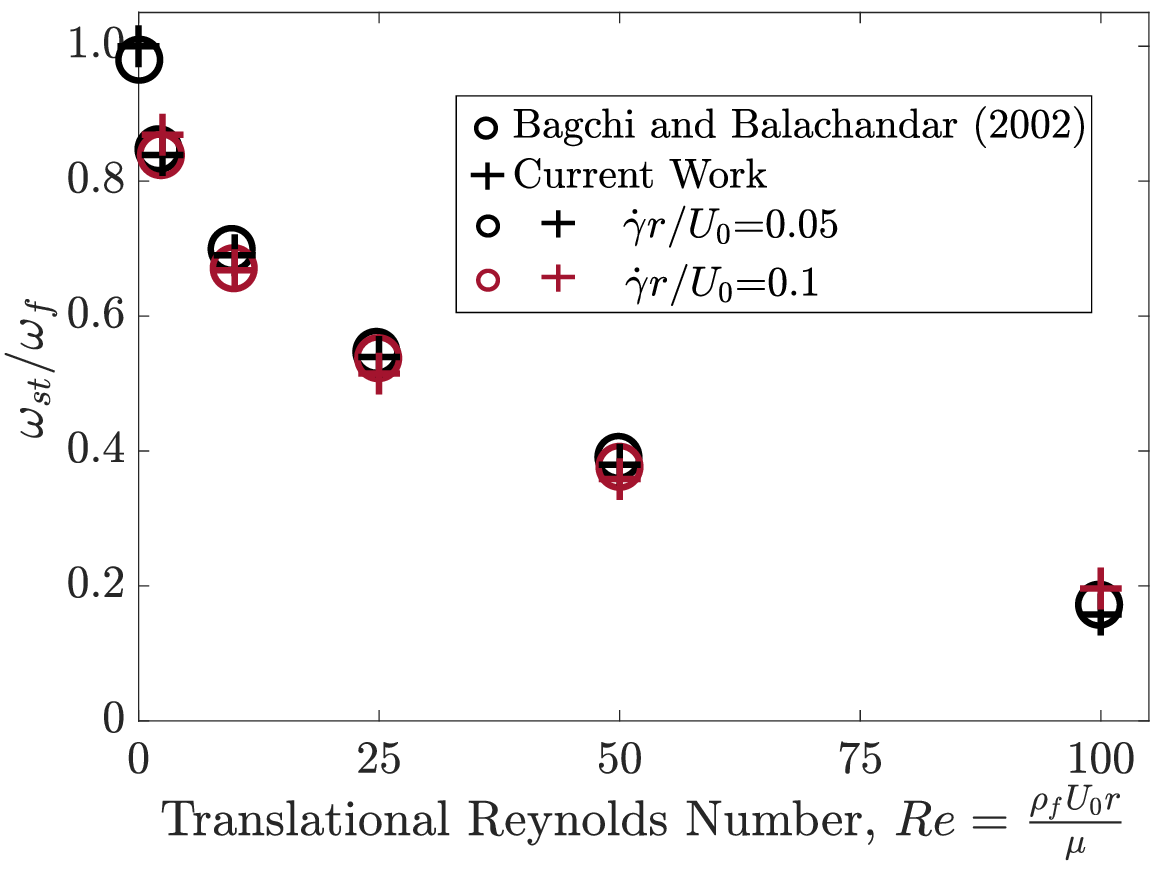}\label{fig:InertialSlowdownVals}}\hfill
\subfloat{\includegraphics[width=0.49\textwidth]{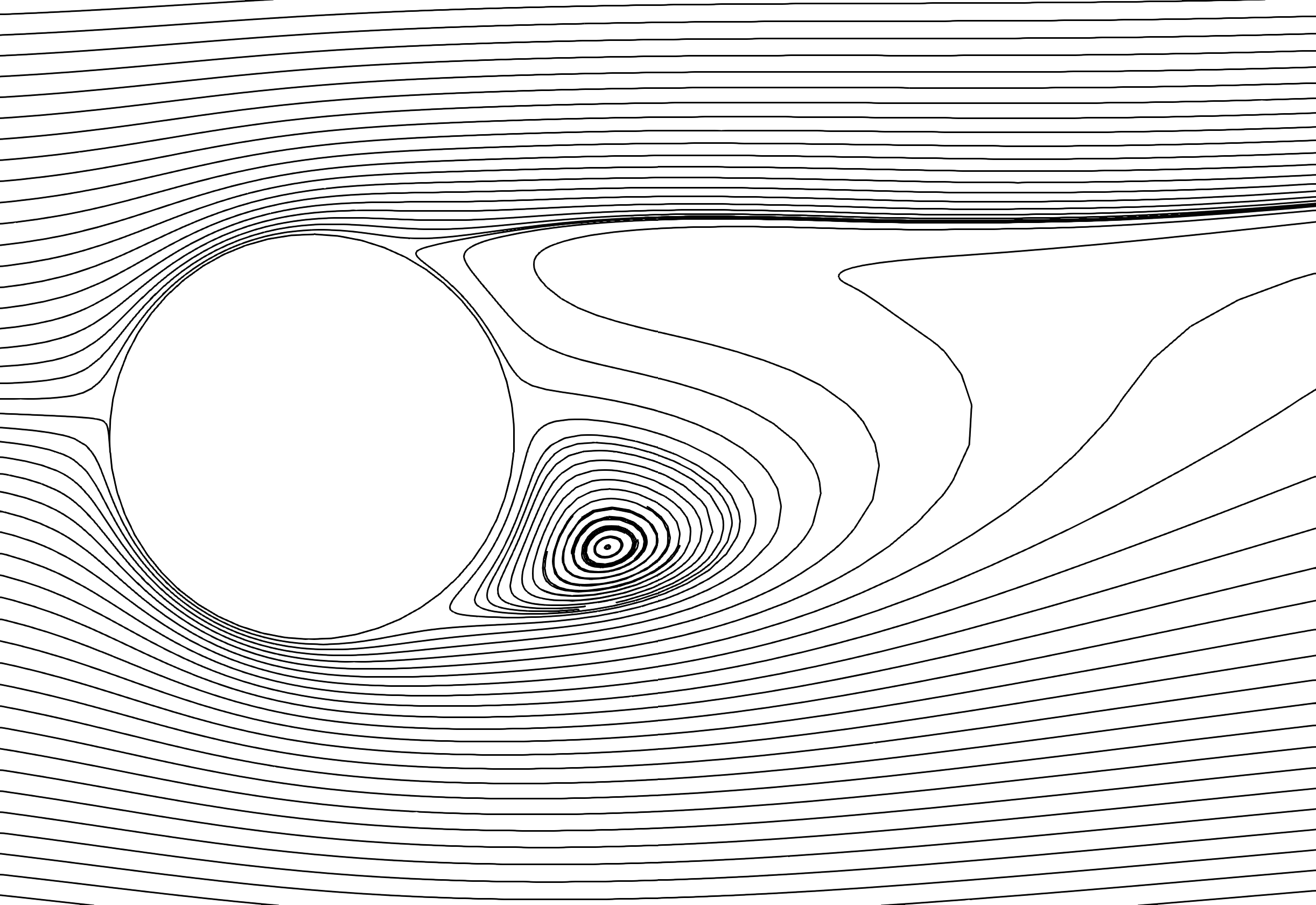}\label{fig:InertialSlowdownStreamlines}}\hfill
\caption {Comparison of the inertial slow down of a freely rotating sphere in a flow $\mathbf{u}=[U_0+\dot{\gamma} y\hspace{0.1in} 0\hspace{0.1in} 0]^T$ from our simulations and those of Bagchi and Balachandar (2002) \cite{bagchi2002effect} ($U_0$ and $\dot{\gamma}$ are constants). The sphere is allowed to rotate freely but not translate. The left figure shows the ratio of steady state rotation rate of a sphere, $\omega_{st}$ to the fluid rotation rate, $\omega_f=\dot{\gamma}/2$ at various translations Reynolds numbers and at two different $\dot{\gamma}r/U_0=0.05$ and 0.1. {Quantitative agreement in $\omega_{st}/\omega_f$ between our simulations and those of  \cite{bagchi2002effect} is found at the $Re$ and $\dot{\gamma}r/U_0$ shown. The right figure shows the streamlines for Re=100 and $\dot{\gamma}r/U_0=0.1$ that are qualitatively similar to the respective streamlines in figure 8(b) of \cite{bagchi2002effect}.} \label{fig:BagchiComparison}}
\end{figure}

In the previous section, we demonstrated the validity of our method to capture the inertial effect on a fixed particle. In this section, we show an example where particle and fluid inertia are moderate, and the particle is allowed to rotate due to the hydrodynamic stresses acting on the particle surface as treated by the method described in section \ref{sec:FiniteParticleInertial}. For this purpose, we validate our code with the simulations of Bagchi and Balachandar (2002) \cite{bagchi2002effect} where the particle is a sphere. We use $\kappa=1.001$ to represent a sphere in the prolate spheroidal coordinate system. The sphere is not allowed to translate but can rotate freely under the influence of the hydrodynamic torque. The imposed fluid velocity is,
\begin{equation}
\mathbf{u}_\text{imposed}^\text{inertial}=[U_0+\dot{\gamma} y\hspace{0.1in} 0\hspace{0.1in} 0]^T,
\end{equation}
in the laboratory frame. Two different Reynolds numbers based on the uniform flow speed, $U_0$, or the shear rate, $\dot{\gamma}$ can be defined in this case. One of them i.e. the translational Reynolds number ($Re$) is the same as defined earlier in equation \eqref{eq:TransRe} and the shear Reynolds number is,
\begin{equation}
Re_{\dot{\gamma}}=\frac{\rho_f\dot{\gamma}r^2}{\mu},
\end{equation}
where 
$r$ ($=r_{major}$) is the radius of the sphere. $Re$ and $Re_{\dot{\gamma}}$ are related by a non-dimensional factor, $\dot{\gamma}r/U_0$. In Stokes flow ($Re=Re_{\dot{\gamma}}=0$) the sphere rotates at the angular velocity of the fluid, $\omega_f=\dot{\gamma}/2$. The presence of inertia lowers the rotation rate of the sphere. In figure  \ref{fig:InertialSlowdownVals} we show the steady state rotation rate of the sphere, $\omega_{st}$, normalized with $\omega_f$ for a range of $Re$ at two different $\dot{\gamma}r/U_0=0.05$ and 0.1 from our simulations and those of Bagchi and Balachandar  (2002) \cite{bagchi2002effect}. We find a good, albeit not exact, agreement of $\omega_{st}/\omega_f$ between the two simulations for all the cases shown up to $Re=100$. Bagchi and Balachandar  (2002) \cite{bagchi2002effect} found the inertial slow-down, $\omega_{st}/\omega_f$, of a sphere's rotation to be independent of {$\dot{\gamma}r/U_0$} for a constant $Re$. This is also captured by our results in figure \ref{fig:InertialSlowdownVals} as the results for the two $\dot{\gamma}r/U_0$ shown nearly collapse. In figure \ref{fig:InertialSlowdownStreamlines} we show the streamlines of the fluid flow around the sphere for $Re=100$ and $\dot{\gamma}r/U+0=0.1$. These are qualitatively similar to the streamlines shown in figure 8(b) of Bagchi and Balachandar  (2002) \cite{bagchi2002effect}.

This example shows the capability of our numerical methodology to handle moving particles in the presence of inertia.

\subsection{Motion of particles in inertia-less viscoelastic fluid}\label{sec:ViscoelasticExamples}
In this section, we compare our numerical method with other numerical and semi-analytical results of the flow of viscoelastic fluids around spheres and prolate spheroids in various linear flows.
\subsubsection{Torque-free rotation in simple shear flow}
\begin{figure}[h!]
\centering
\includegraphics[width=0.49\textwidth]{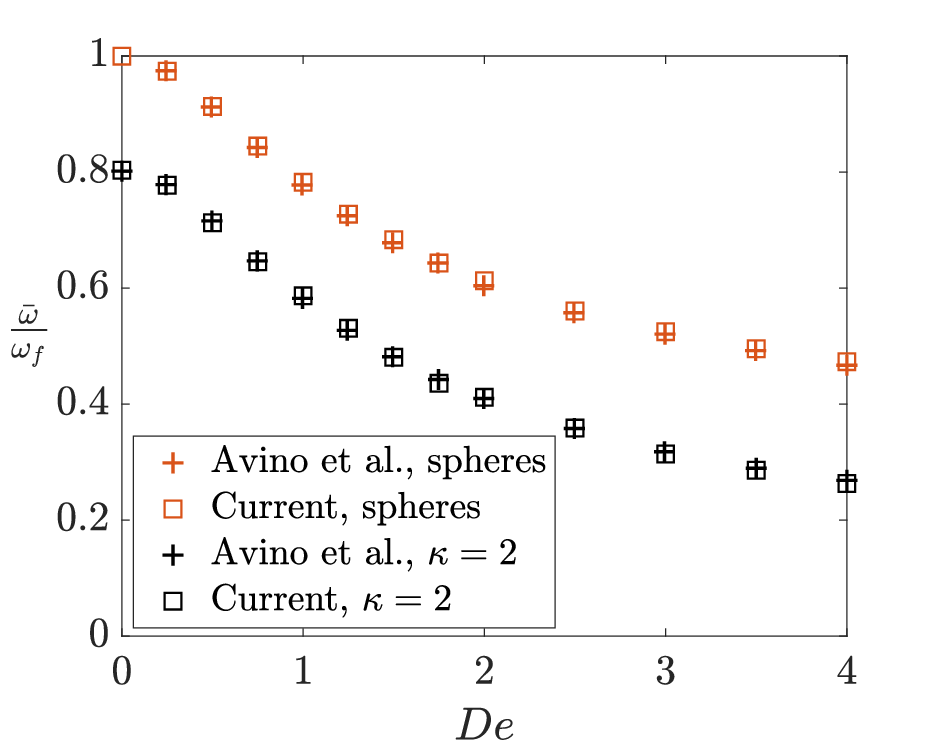}
\caption {Slower particle rotation due to viscoelasticity. Period average angular velocity, $\bar{\omega}$, (normalized with the fluid rotation rate, $\omega_f=\dot{\gamma}/2$) of a torque-free sphere and a prolate spheroid with aspect ratio, $\kappa=2.0$ in the flow-gradient plane of a simple shear flow of Giesekus viscoelastic fluid ($c=10$ and $\alpha=0.2$) at various $De$. {Our results are quantitatively similar to that of Avino et al. (2014) \cite{d2014bistability} at all $De$.}\label{fig:AverageAngularVelocityinFG}}
\end{figure}

\begin{figure}[h!]
\centering
\subfloat{\includegraphics[width=0.33\textwidth]{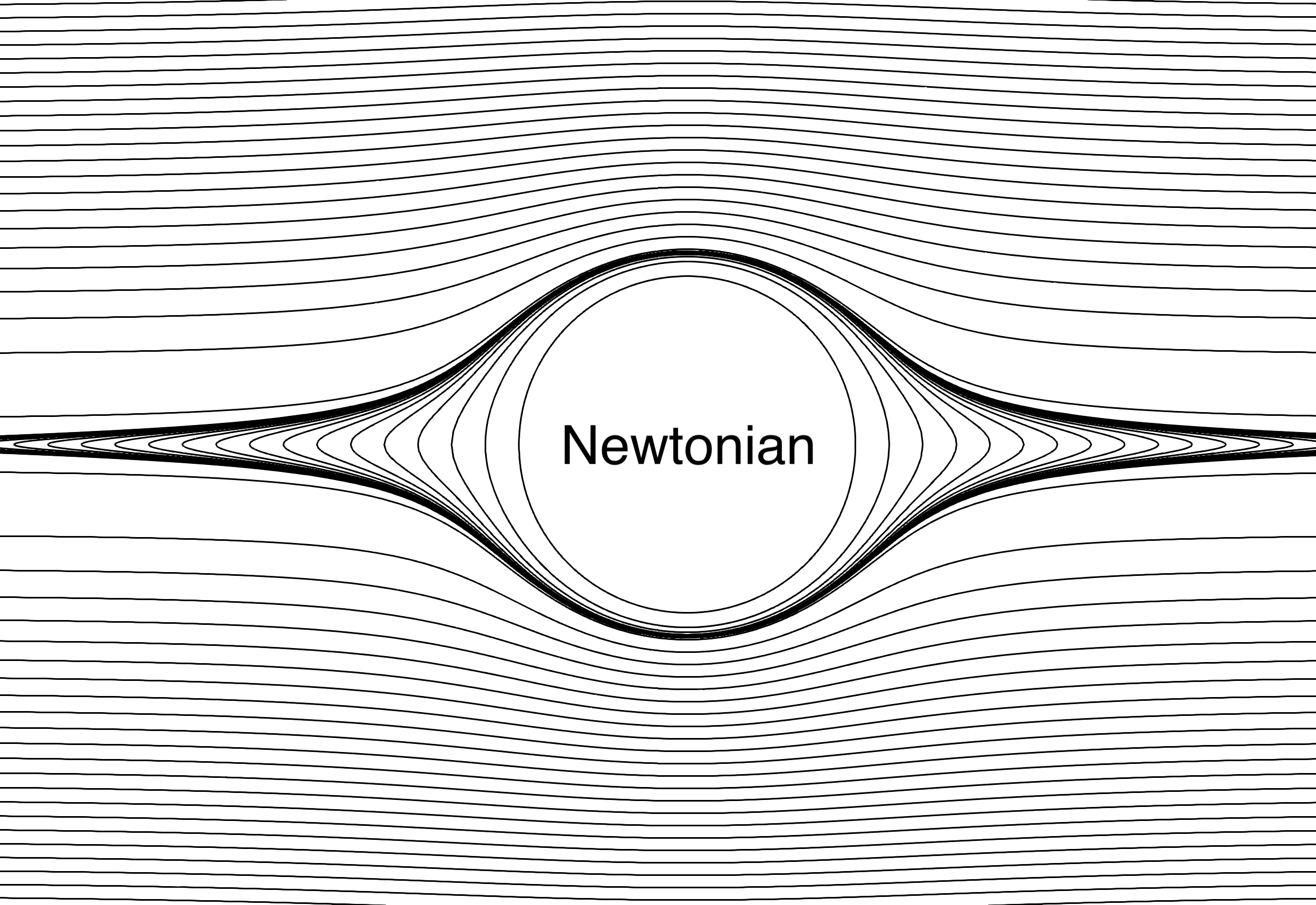}\label{fig:NewtFreeRotSphere}}\hfill
\subfloat{\includegraphics[width=0.33\textwidth]{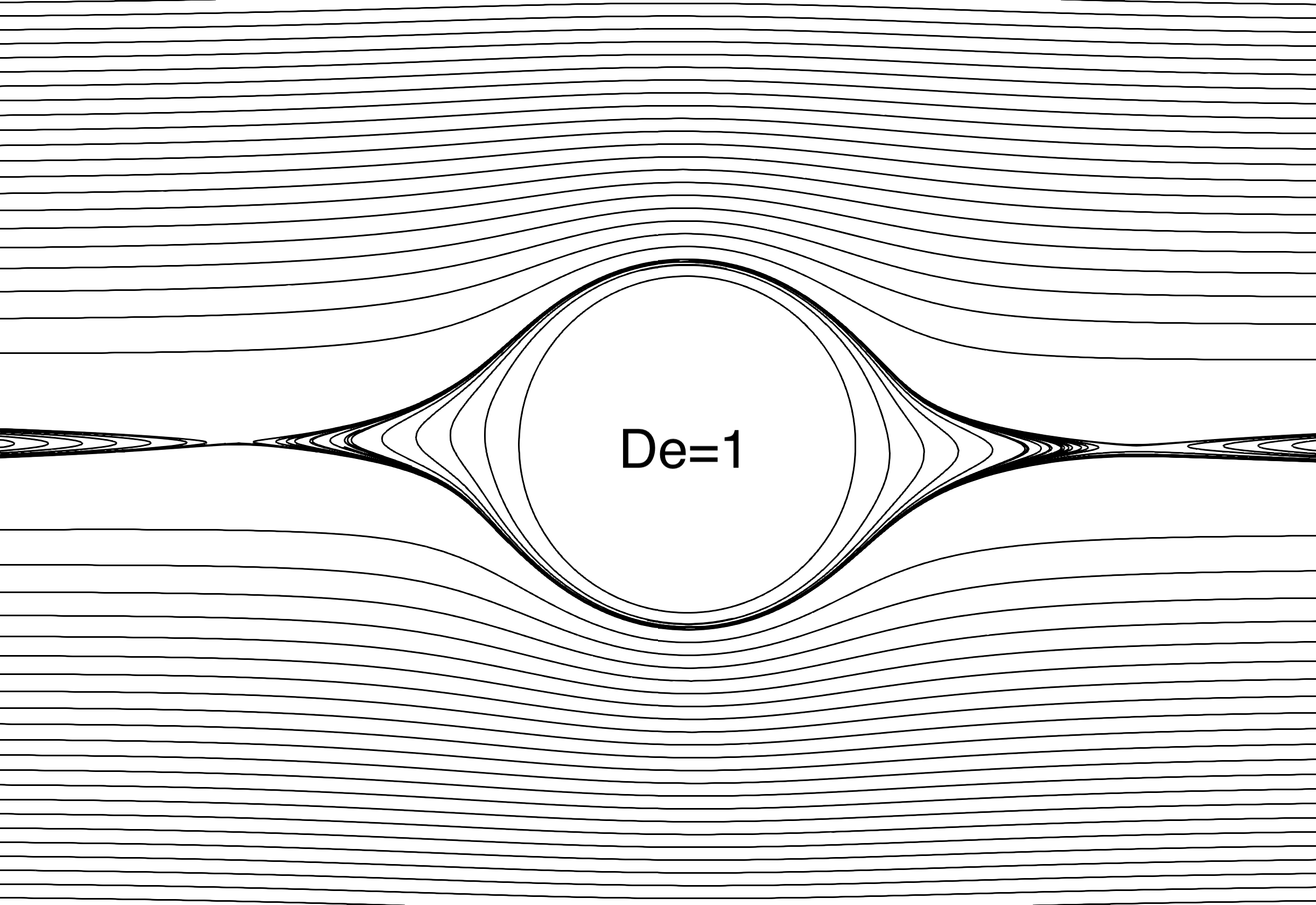}\label{fig:De1FreeRotSphere}}\hfill
\subfloat{\includegraphics[width=0.33\textwidth]{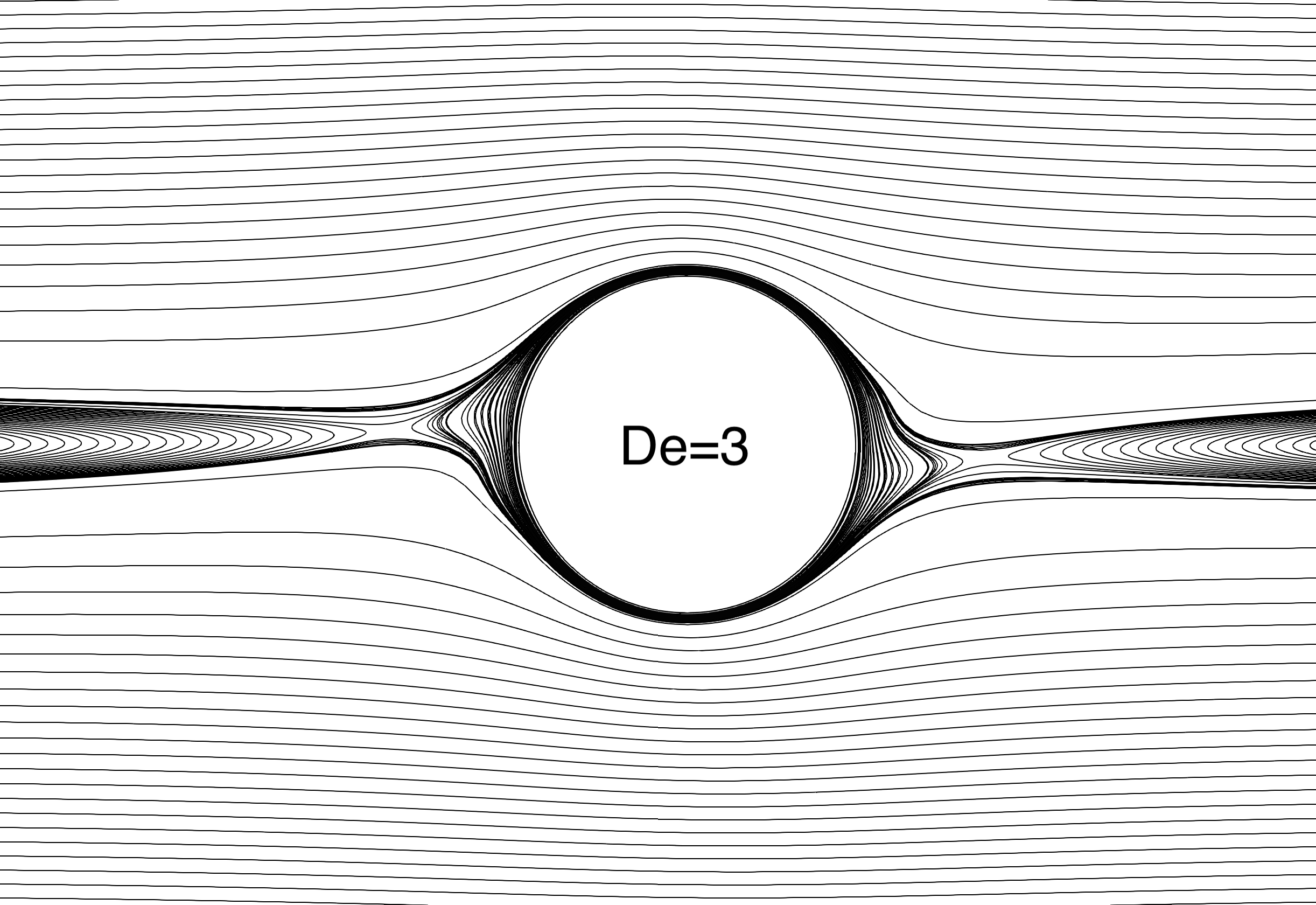}\label{fig:De3FreeRotSphere}}
\caption {Streamlines around a torque-free sphere rotating in a simple shear flow of a Newtonian fluid (left) and of a Giesekus viscoelastic fluid (right) with $c=10.0$, $\alpha=0.2$ and $De=$1.0. The changes in streamlines due to viscoelasticity are consistent with the observations of Avino et al. (2008)  \cite{d2008rotation} (figures 9(a) and 9(d) of \cite{d2008rotation}). Streamlines for a Giesekus fluid with $De=3.0$ are also shown that extend the conclusions of Avino et al. (2008)  \cite{d2008rotation} to higher $De$ than explored in their study. \label{fig:Streamlines}}
\end{figure}

In section \ref{sec:InertialSlowDown}, we discussed the influence of inertia in slowing down the rotation of a sphere in simple shear flow from its value in the Stokes limit. Avino et al. (2008)  \cite{d2008rotation} reported a similar effect due to viscoelasticity where increasing $De$ in an inertia-less viscoelastic fluid lowers the rotation rate of a spherical particle. A subset of these authors reported the steady-state angular velocity of the sphere in Avino et al. (2014) \cite{d2014bistability}. Here $De$ is the non-dimensional parameter known as the Deborah number, representing the product of the imposed shear rate and the particle relaxation time. The fluid and particle inertia are ignored, and the net torque due to the fluid stresses acting on the surface of a freely moving particle is zero. Hence the particle motion in the cases considered in this section is obtained through the method described in section \ref{sec:ZeroParticleInertial}. The Giesekus constitutive relation (equations \eqref{eq:ConstitutiveForce} and \eqref{eq:ConstitutiveSpecific}) with $c=10.0$ and $\alpha=0.2$ is used to model the polymer stress in the viscoelastic fluid. {In figure \ref{fig:AverageAngularVelocityinFG} we show that the results from our simulations for a sphere rotating in a simple shear flow of this fluid are almost identical to that of Avino et al. (2014) \cite{d2014bistability}.} In our simulations, a prolate spheroid with aspect ratio $\kappa=1.001$ represents the sphere. Avino et al. (2014) \cite{d2014bistability} also shown the average angular velocity for the case when the major axis of a prolate spheroid rotates in the flow-gradient plane of a simple shear flow of the same viscoelastic fluid. We compare the results for $\kappa=2.0$ from our simulations with that of Avino et al. (2014) \cite{d2014bistability} in figure \ref{fig:AverageAngularVelocityinFG} and find an excellent agreement at all $De$ shown. For a spheroid in a Newtonian fluid, i.e. at $De=0.0$, according to Jeffery (1922) \cite{jeffery1922motion} $\bar{\omega}/\omega_f=2.0\kappa/(1+\kappa^2)=$ 1.0 and 0.8 for $\kappa=1.0$ (spheres) and 2.0 respectively. {These analytical estimates are the numerical values shown in figure \ref{fig:AverageAngularVelocityinFG} at $De=0.0$}.  We show the streamlines around a freely rotating sphere in a Newtonian fluid and viscoelastic fluid with $De=1.0$ in figures \ref{fig:NewtFreeRotSphere} and \ref{fig:De1FreeRotSphere}. These are qualitatively similar to those shown in figures 9(a) and 9(d) of Avino et al. (2008)  \cite{d2008rotation}. The effect of viscoelasticity is to distort the region of closed streamlines around the sphere that extends to infinity for Newtonian fluid. At $De=1.0$, both the simulations find stagnation points on either side of the sphere in the flow direction, which marks the end of the closed streamline region. By showing streamlines for $De=$ 0.1, 0.3 and 1.0 Avino et al. (2008) \cite{d2008rotation} established the movement of hte stagnation points closer to the sphere upon increasing $De$. We also show the continuation of this trend to $De=3.0$.  At this higher $De$, an asymmetry of the region of closed streamlines and a reverse wake about the flow (horizontal) direction is also observed.

\begin{figure}[h!]
\centering
\subfloat{\includegraphics[width=0.49\textwidth]{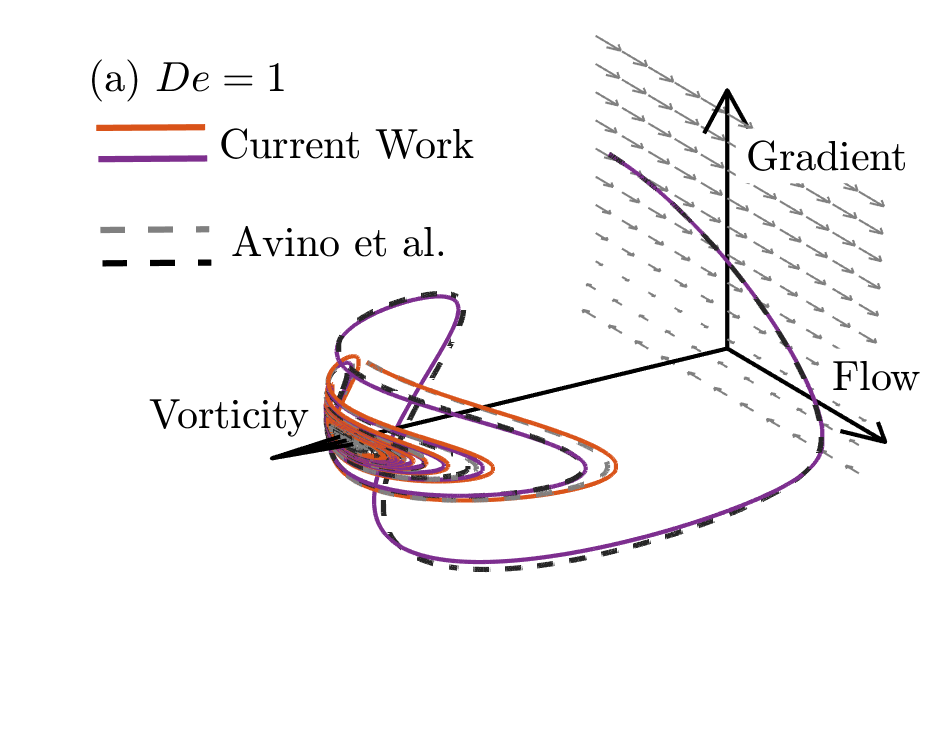}\label{fig:De1Orient1Giesekus}}\hfill
\subfloat{\includegraphics[width=0.49\textwidth]{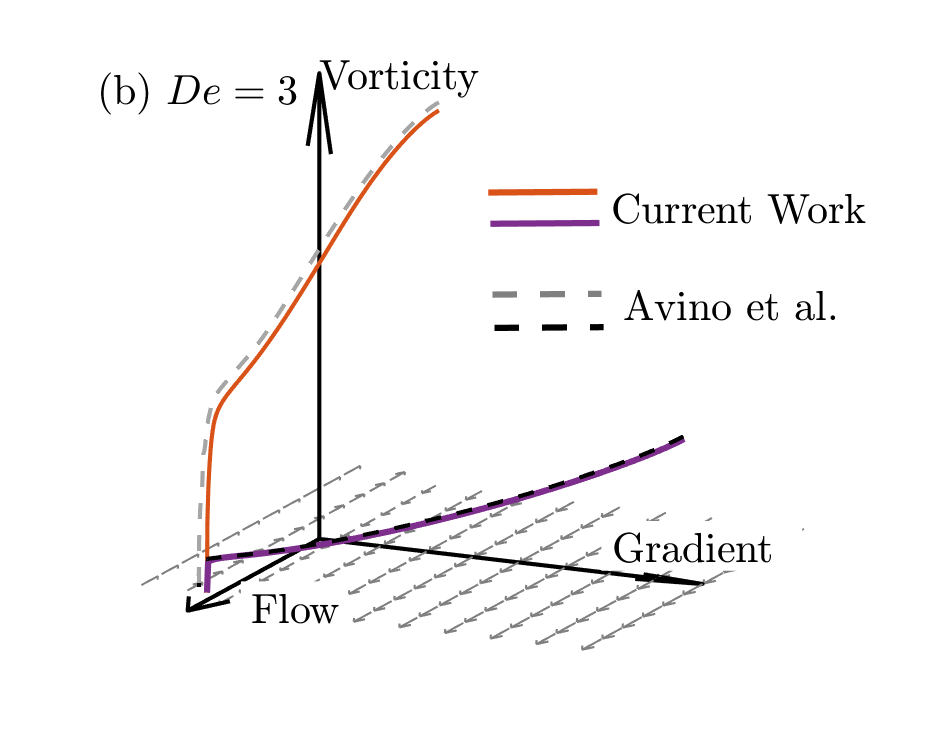}\label{fig:De1Orient2Giesekus}}\vspace{-0.5in}
\caption {{Orientation trajectory of a torque-free prolate spheroid with aspect ratio, $\kappa=4$ released in a simple shear flow of an inertia-less Giesekus viscoelastic fluid ($c=10$ and $\alpha=0.2$) at two different initial orientations in the gradient-vorticity plane at $De=1.0$ and 3.0. Solid orange and purple lines are from our simulations and the dashed grey and black are from Avino et al. (2014) \cite{d2014bistability}. Grey arrows indicate the imposed flow in the shearing plane. Good qualitative agreement is obtained between the two results. The orientation trajectories of Avino et al. (2014) \cite{d2014bistability} were obtained through personal communication with the authors.}\label{fig:AR43DOrientation}}
\end{figure}

Now we consider three dimensional rotation of a $\kappa=4.0$ prolate spheroid in shear flow of the same viscoelastic fluid. As shown by the numerical simulations of Avino et al. (2014) \cite{d2014bistability}, in the presence of viscoelasticity the orientation of a prolate spheroid no longer follows the Jeffery orbits as in Newtonian Stokes flow \cite{jeffery1922motion} (section  \ref{sec:Jeffery}). In figure \ref{fig:AR43DOrientation} we compare the orientation trajectory of a $\kappa=4.0$ prolate spheroid rotating in an inertia-less viscoelastic fluid from our simulations with that of Avino et al. (2014) \cite{d2014bistability}. {We show two different starting orientations for $De=1.0$ and 3.0 in figure \ref{fig:AR43DOrientation}. The resolution used in these simulations is $N_1=150$, $N_2=71$ and $N_3=57$ (equation \eqref{eq:Grid}).}
The size of the computational domain used is $||\mathbf{r}_\infty||_2\approx||\mathbf{r}^\text{minor}_\infty||_2=20\kappa$. Our simulation results qualitatively agree with those of  Avino et al. (2014) \cite{d2014bistability} with small, subtle differences. Similar to Avino et al. (2014) \cite{d2014bistability}, we find the final orientation behavior of the particle at $De=1$ to be spiraling towards the vorticity direction of the imposed simple shear flow, irrespective of the initial orientation. At $De=3.0$, the particle in both simulations settles to a location very close to the flow direction, irrespective of the starting orientation.  

\subsubsection{Rheology of dilute suspensions of spheres}\label{sec:rheology}
As mentioned in section \ref{sec:Introduction}, fluid flow around a particle in an unbounded fluid is useful in studying the rheology of a dilute suspension of particles where the inter-particle interaction is rare. In Newtonian fluids, the presence of particles leads to an additional stress $n\boldsymbol{S}$ in the suspension, where $n$ is the number of particles per unit volume and $\boldsymbol{S}$ is termed stresslet. In rheology studies of incompressible suspensions, the deviatoric or traceless part of the suspension stress is most interesting as the trace of the stress can be absorbed in the modified pressure. The deviatoric part of the stresslet, $\hat{\boldsymbol{S}}$, is an area integral over the particle surface, $\mathbf{r}_p$, of a tensor product of the fluid stress \cite{batchelor1970stress},
\begin{align}
\hat{\text{\textbf{S}}}(\boldsymbol{\sigma})=\int_{\mathbf{r}=\mathbf{r}_p}\text{d}A\big\{\frac{1}{2}[{\text{\textbf{nn}}}\cdot  \boldsymbol{\sigma}+\text{\textbf{n}}\cdot  \boldsymbol{\sigma}\text{\textbf{n}} ]-\frac{1}{3}\boldsymbol{\delta}\text{\textbf{n}}\cdot  \boldsymbol{\sigma}\cdot \text{\textbf{n}}\big\},
\end{align}
where $\textbf{n}$ is the unit surface normal pointing into the fluid. The stresslet also appears in the particle suspension of viscoelastic fluids where $ \boldsymbol{\sigma}$ is the sum of Newtonian solvent and polymer stress. The deviation of the polymeric stress $\boldsymbol{\Pi}$ (equation \eqref{eq:ConstitutiveForce}) from its undisturbed value $\boldsymbol{\Pi}_\infty$ leads to an additional component of the suspension stress known as the ensemble average polymeric stress. This requires more careful treatment, as pointed out by Koch et al. (2016) \cite{koch2016stress}. In this section, we will compare the stresslet due to a sphere (represented by a prolate spheroid with $\kappa=1.001$) in a simple shear flow and a uni-axial extensional flow of viscoelastic fluid with the previously available results.

\begin{figure}
\centering
\includegraphics[width=0.49\linewidth]{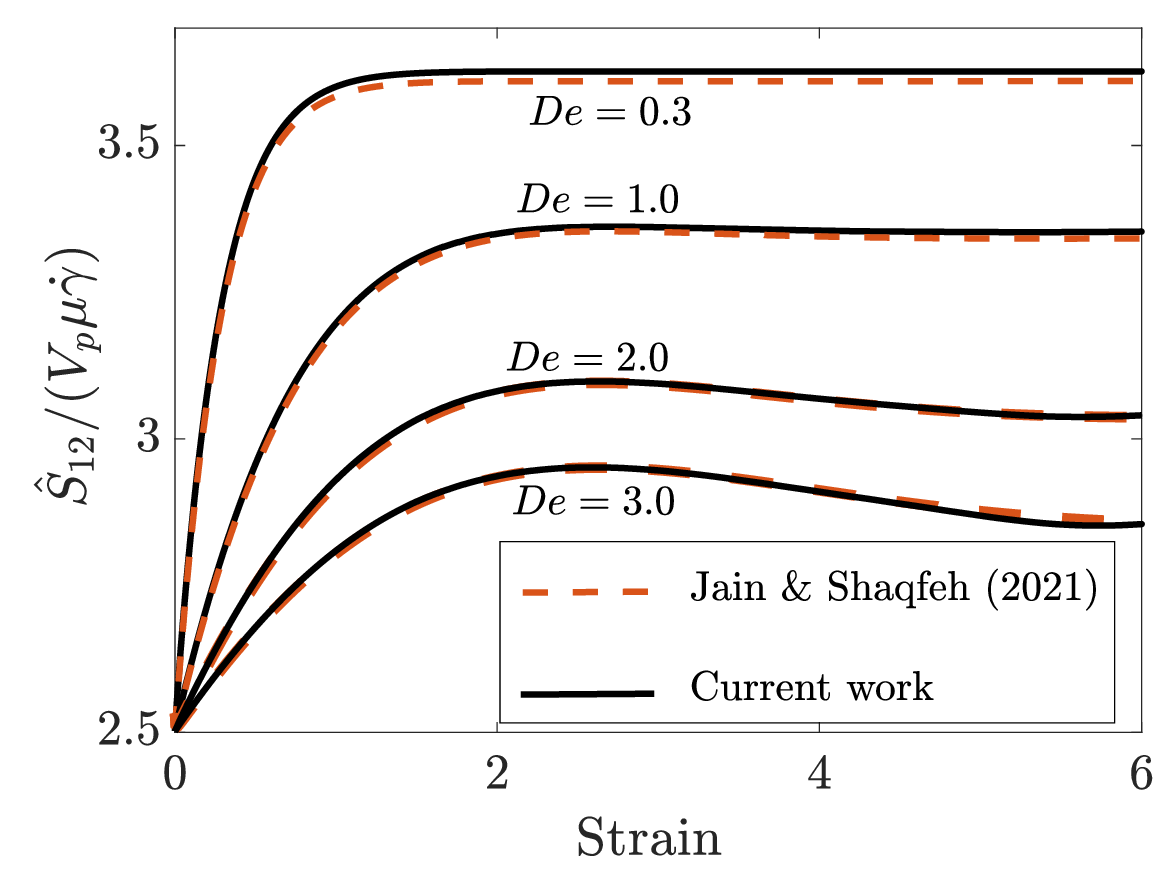}
\caption{Comparison of the stresslet due to a sphere in simple shear flow of a Giesekus viscoelastic fluid with $c=0.471$ and $\alpha=0.0039$ from our simulations and that of Jain \& Shaqfeh (2021) \cite{jain2021transient}. The curves of Jain \& Shaqfeh's (2021) study are obtained via personal communication with the authors.}
\label{fig:validationshearstressletshaq}
\end{figure}

Jain \& Shaqfeh (2021) \cite{jain2021transient} considered the shear rheology of a suspension of spheres in a Giesekus viscoelastic fluid (equations \eqref{eq:ConstitutiveForce} and \eqref{eq:ConstitutiveSpecific}) with $c=0.471$ and $\alpha=0.0039$. In the dilute particle limit, this is obtained by studying simple shear flow around a sphere. {The evolution of the shear or $\hat{\text{{S}}}_{12}$ component (where $1$ and $2$ represent the flow and gradient directions of the imposed shear flow) of the stresslet (normalized with the product of the particle volume, $V_p$, solvent's dynamic viscosity, $\mu$, and imposed shear rate $\dot{\gamma}$) with strain, from our numerical simulations for this case is compared with that of \cite{jain2021transient}  in figure \ref{fig:validationshearstressletshaq} for four different $De$. Here, 
$De$ or Deborah number is the non-dimensional product of the polymer relaxation time and the imposed shear rate. The simulations reported in \cite{jain2021transient} were not converged with mesh size. Through personal communication with the authors, we obtained results from their numerical method at a finer grid and used these refined values to compare with our results here. At the first instant of time, the normalized stresslet value is 2.5, representing the constant stresslet in a Newtonian fluid originally calculated by Einstein (1906) \cite{einsteinoriginal}, as initially, the polymers in the viscoelastic fluid are in equilibrium and have zero stress. The normalized particle stresslet increases with the strain in  the suspension. Stresslet values from the two simulations are in good quantitative agreement at all $De$ and strain. Upon increasing the mesh resolution, their evolution of the normalized stresslet remained qualitatively similar but increased in magnitude by about 0.06 for all $De$ values. 
We have found that our values converged with the mesh size near their most refined values. In the results presented we use 1.72 million mesh points ($N_1=150$, $N_2=201$ and $N_3=57$) and the refined results of Jain \& Shaqfeh (2021) \cite{jain2021transient}  (computed using a finite volume method) are from simulations using 1.75 million volume elements.}

\begin{figure}[h!]
\centering
\subfloat{\includegraphics[width=0.49\textwidth]{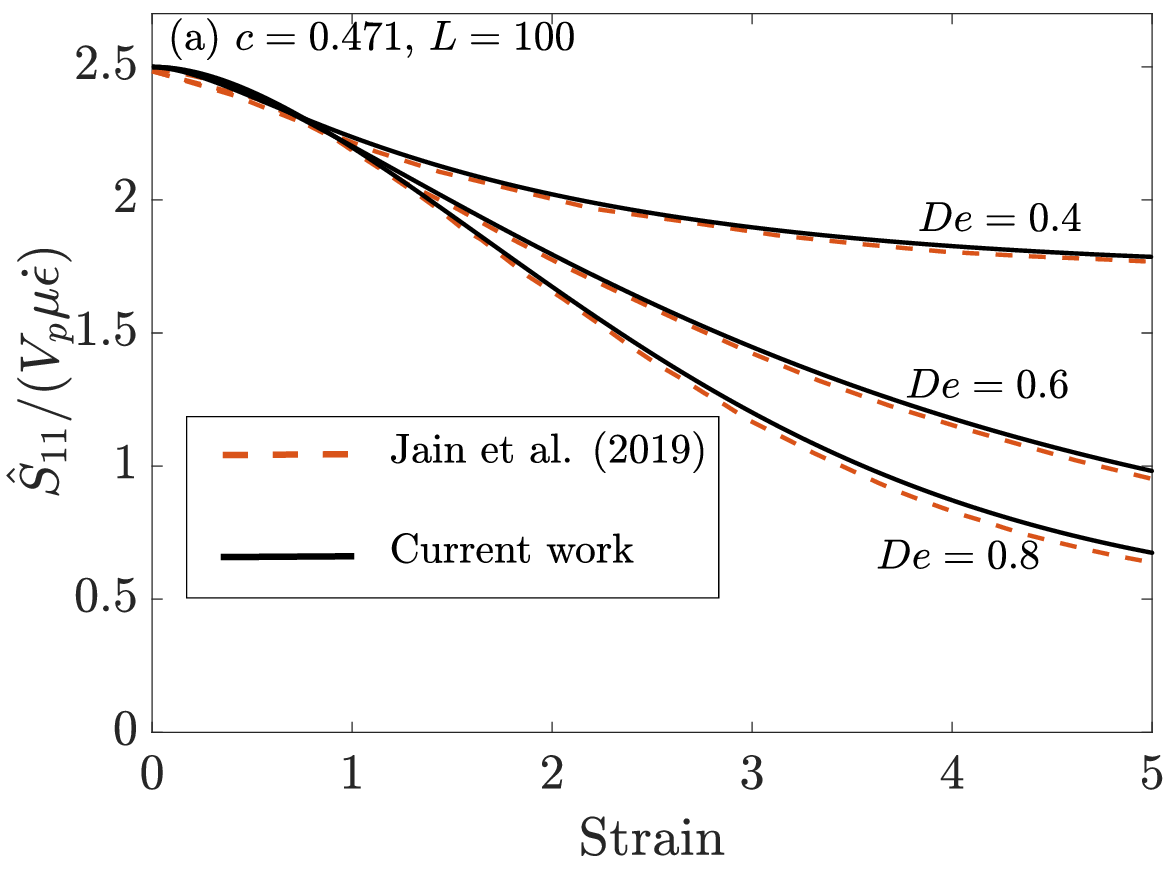}\label{fig:JainExtensional}}\\
\subfloat{\includegraphics[width=0.49\textwidth]{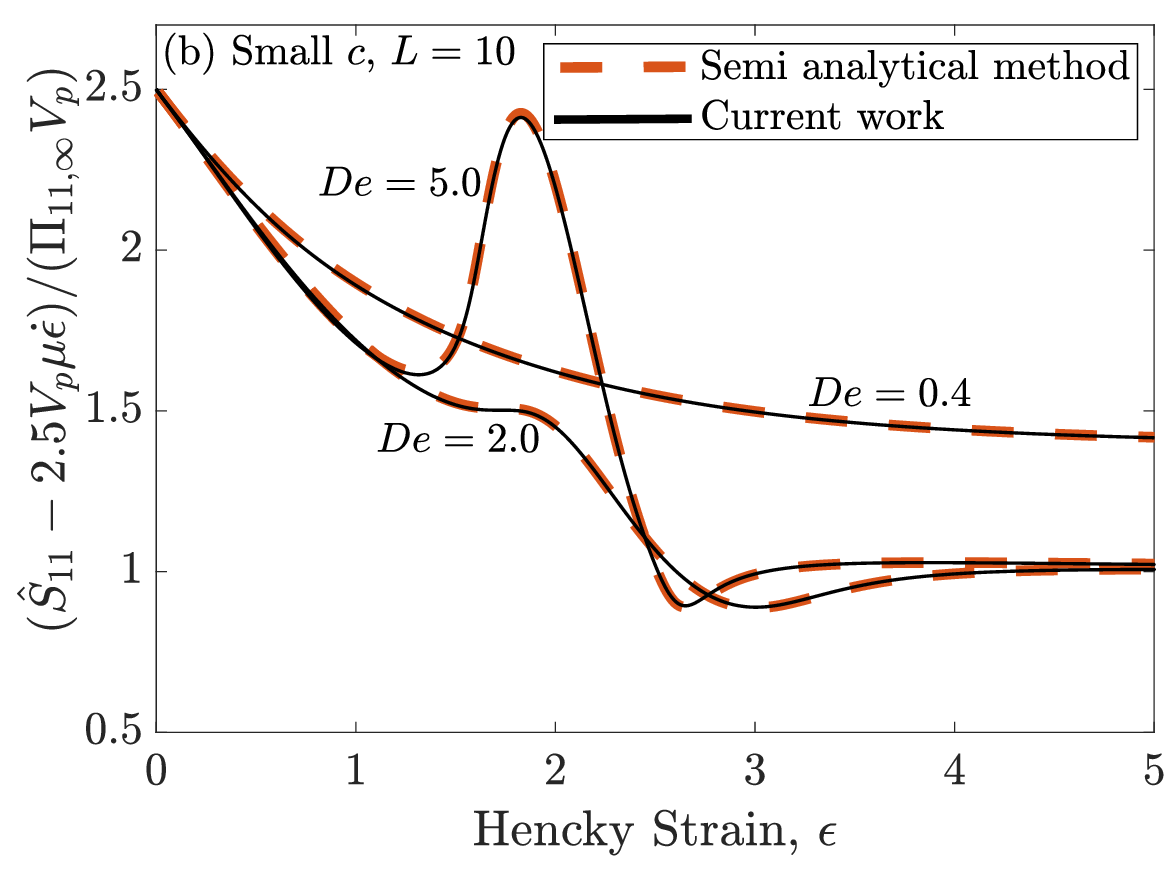}\label{fig:SemiAnalytL10}}\hfill
\subfloat{\includegraphics[width=0.49\textwidth]{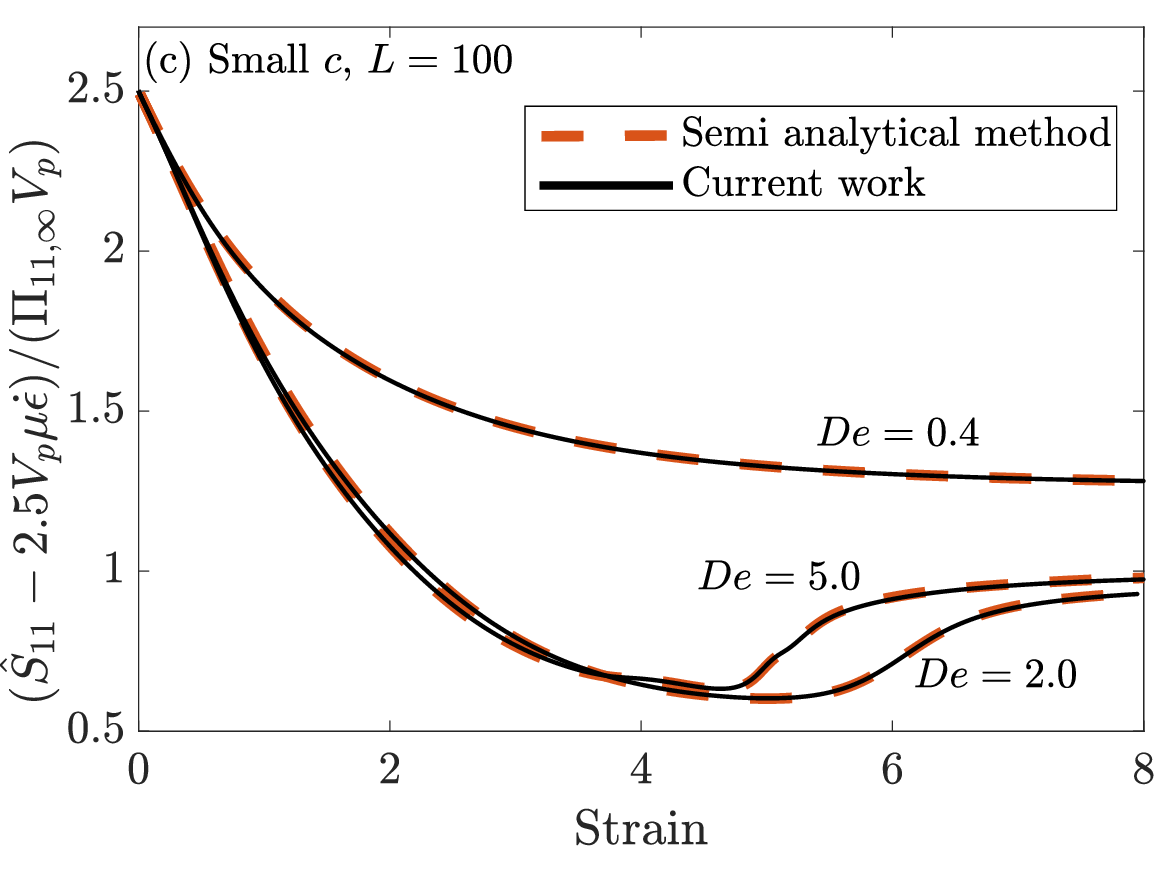}\label{fig:SemiAnalytL100}}
\caption {Stresslet {vs. strain} due to a sphere in uniaxial extensional flow of a FENE-P viscoelastic fluid. Comparison with the results of Jain et al. (2019)\cite{jain2019extensional} is shown in (a) with the viscoelastic fluid parameters: $L=100$, $c=0.471$ and various $De$. We compare our numerical results at $c=10^{-5}$ with those of our low $c$ semi-analytical method described in our forthcoming publication \cite{TransientPaper} at various $De$ and $L$ in (b) and (c). \label{fig:Extensional Stresslet}}
\end{figure}

Jain et al. (2019) \cite{jain2019extensional} considered the extensional rheology of a dilute suspension of spheres in a viscoelastic fluid, where the stresslet can be obtained by studying uni-axial extensional flow around a sphere. {As shown in figure \ref{fig:JainExtensional} the extensional component of the deviatoric stresslet, $\hat{\text{{S}}}_{11}$, (normalized with the product of particle volume, $V_p$, solvent's dynamic viscosity, $\mu$, and imposed extension rate, $\dot{\epsilon}$) for a FENE-P (equations \eqref{eq:ConstitutiveForce} and \eqref{eq:ConstitutiveSpecific}) viscoelastic fluid with $L=100$ and $c=0.471$ from our simulations qualitatively agrees with that of Jain et al. (2019) \cite{jain2019extensional} at all strain values for $De=0.4, 0.6,$ and 0.8. In this case, $De$ is the product of polymer relaxation time and imposed extension rate.} Due to the same reasons discussed above for shear rheology, the normalized extensional stresslet in these cases starts at 2.5, but then it reduces in magnitude as strain increases. In \cite{TransientPaper} we have used a semi-analytical method for extensional rheology of spheres in FENE-P fluid that allows us to perform a wider parameter study at much less computational cost. It is valid at a small polymer concentration, $c$, and we validate our numerical method described here using this semi-analytical method in figures \ref{fig:SemiAnalytL10} and \ref{fig:SemiAnalytL100}. We consider $c=10^{-5}$ and $De=0.4$, 2.0 and 5.0 at two different $L=10$ and 100 and show the variation of the non-Newtonian component of the deviatoric stresslet, $\hat{\text{{S}}}_{zz}-2.5V_p\mu\dot{\epsilon}$ normalized with the particle volume times extensional component of deviatoric undisturbed polymer stress, $\hat{\Pi}_{zz,\infty}V_p$ with Hencky strain. At zero Hencky strain, both the numerator and the denominator in this normalized stresslet are zero, but it has a finite limit of 2.5. Our numerical results capture all the features of the stresslet from this semi-analytical method, and the two are almost identical at all $De$, $L$, and Hencky strain shown. For all the {preceding} simulations concerning uni-axial extensional flow we have used $N_1=251$, $N_y=351$ and $N_z=25$. The flow is axi-symmetric; hence, we do not need many points in the azimuthal ($\xi_3$) direction. Therefore another benefit of using prolate spheroidal coordinates is that in simulating a strong, but axisymmetric flow around an aligned axisymmetric particle, we can maintain a small CPU time for the simulation by having fewer points in $\xi_3$ and increasing the resolution in the radial ($\xi_1$) and polar ($\xi_2$) direction. In this case, good resolution in the polar ($\xi_2$) direction is essential to capture the large polymer stress gradients around the extensional axis.

\vspace{0.2in}
Simulating viscoelastic fluids around particles is a challenging numerical problem, particularly at large polymer relaxation times or Deborah numbers, $De$. This is because of the large polymer stretch possible in this scenario creating large gradients in polymer stress that require good resolution around both the particle surface and in specific regions where large polymer stretch is observed. {By comparing and showing an agreement of results from our numerical method with the state-of-the-art simulation results available for relatively low aspect ratio spheroids ($\kappa=4$) and spheres, we have demonstrated that our numerical method is suitable for studying such flows.} Furthermore, due to the chosen coordinate system, we are well poised to investigate higher particle aspect ratios where the required numerical simulations will be more challenging. However, the physical effects due to viscoelasticity are expected to be more interesting.

\section{Conclusions}\label{sec:Conclusions}
{We have presented a finite difference numerical method to solve the flow of viscoelastic liquids around a prolate spheroid in a body conforming coordinate system. We can simulate much larger particle aspect ratios than previous computational studies as the particle surface is exactly modeled as one of the coordinate surfaces in the prolate spheroidal coordinates used. This is the inner boundary of the computational domain where the required no-slip/ no-penetration condition on the particle is imposed. The outer boundary of the computational domain is nearly spherical. It represents the far-field where appropriate boundary conditions can be imposed for any constant or time-varying combination of linear flows. This allows us to study a wide range of highly resolved particle shapes ranging from a spherical particle to a large aspect ratio prolate spheroidal fiber. Our method is valid for zero to moderate fluid inertia.} Various components within the numerical methods are inspired by the existing numerical techniques originally developed for varying applications. Schur complement method developed \cite{furuichi2011development} to solve zero-inertia, large viscosity gradient flows in the Earth's mantle is used to solve the mass and momentum equations. To remove the coordinate system generated singularity on the polar axes of the prolate spheroidal coordinate system, we used L'Hopital's rule as first demonstrated for finite difference methods developed in cylindrical coordinate system \cite{verzicco1996finite,morinishi2004fully,desjardins2008high}. Stability of the convective derivatives is obtained by using the higher-order upwinding central schemes \cite{nourgaliev2007high} that were originally used for interface tracking in multiphase flows and lead to low numerical diffusion. To overcome the violation of free-stream preservation in a curvilinear coordinate system, we simulate the deviation of the relevant flow variables from the known far-field flow that is undisturbed by the particle's presence. This also simplifies the governing equations. {For the case of zero particle and fluid inertia in viscoelastic liquids, using a novel resistivity formulation, we develop a computational technique to satisfy the torque- and force-free constraints on the particle in a non-iterative manner, thus saving computational resources.} We demonstrate our method on a variety of flows of Newtonian (with and without inertia) and viscoelastic fluids around spheres and prolate spheroids and find good agreement with existing numerical and theoretical results. The capability to handle a variety of fluids, boundary conditions, and particle shapes opens numerous avenues that can be explored with our numerical methodology. Flexibility in choosing different imposed flows on the outer boundary within a simulation allows our numerical method to study the flow around a particle or dilute suspensions of particles in industrial processes where they experience different linear flows in time. Asymptotic theories for flows around particles are generally developed for the parameter range that is difficult to test numerically due to requirements such as very large domain size or large particle aspect ratio. Such challenges are overcome by our numerical method.

\section{Acknowledgment}
Authors would like to thank Gaojin Li,  Olivier Desjardins and Mahdi Esmaily for fruitful discussions.\\
Funding: This work was supported by the National Science Foundation [grant number 2206851] and the National Aeronautics and Space Administration [grant number 80NSSC23K0348].
\bibliographystyle{model1-num-names}
\bibliography{NumericsProlateSpheroidal}

\begin{thebibliography}{80}
\expandafter\ifx\csname natexlab\endcsname\relax\def\natexlab#1{#1}\fi
\providecommand{\url}[1]{\texttt{#1}}
\providecommand{\href}[2]{#2}
\providecommand{\path}[1]{#1}
\providecommand{\DOIprefix}{doi:}
\providecommand{\ArXivprefix}{arXiv:}
\providecommand{\URLprefix}{URL: }
\providecommand{\Pubmedprefix}{pmid:}
\providecommand{\doi}[1]{\href{http://dx.doi.org/#1}{\path{#1}}}
\providecommand{\Pubmed}[1]{\href{pmid:#1}{\path{#1}}}
\providecommand{\bibinfo}[2]{#2}
\ifx\xfnm\relax \def\xfnm[#1]{\unskip,\space#1}\fi
\bibitem[{Mutiso et~al.(2013)Mutiso, Sherrott, Rathmell, Wiley, and
  Winey}]{mutiso2013integrating}
\bibinfo{author}{R.~M. Mutiso}, \bibinfo{author}{M.~C. Sherrott},
  \bibinfo{author}{A.~R. Rathmell}, \bibinfo{author}{B.~J. Wiley},
  \bibinfo{author}{K.~I. Winey},
\newblock \bibinfo{title}{Integrating simulations and experiments to predict
  sheet resistance and optical transmittance in nanowire films for transparent
  conductors},
\newblock \bibinfo{journal}{ACS nano} \bibinfo{volume}{7}
  (\bibinfo{year}{2013}) \bibinfo{pages}{7654--7663}.
\bibitem[{Yin et~al.(2010)Yin, Huang, Bu, Wang, and Xiong}]{yin2010inkjet}
\bibinfo{author}{Z.~Yin}, \bibinfo{author}{Y.~Huang}, \bibinfo{author}{N.~Bu},
  \bibinfo{author}{X.~Wang}, \bibinfo{author}{Y.~Xiong},
\newblock \bibinfo{title}{Inkjet printing for flexible electronics: Materials,
  processes and equipments},
\newblock \bibinfo{journal}{Chinese Science Bulletin} \bibinfo{volume}{55}
  (\bibinfo{year}{2010}) \bibinfo{pages}{3383--3407}.
\bibitem[{Barbati et~al.(2016)Barbati, Desroches, Robisson, and
  McKinley}]{barbati2016complex}
\bibinfo{author}{A.~C. Barbati}, \bibinfo{author}{J.~Desroches},
  \bibinfo{author}{A.~Robisson}, \bibinfo{author}{G.~H. McKinley},
\newblock \bibinfo{title}{Complex fluids and hydraulic fracturing},
\newblock \bibinfo{journal}{Annual review of chemical and biomolecular
  engineering} \bibinfo{volume}{7} (\bibinfo{year}{2016})
  \bibinfo{pages}{415--453}.
\bibitem[{Breitenbach(2002)}]{breitenbach2002melt}
\bibinfo{author}{J.~Breitenbach},
\newblock \bibinfo{title}{Melt extrusion: from process to drug delivery
  technology},
\newblock \bibinfo{journal}{European journal of pharmaceutics and
  biopharmaceutics} \bibinfo{volume}{54} (\bibinfo{year}{2002})
  \bibinfo{pages}{107--117}.
\bibitem[{Huang et~al.(2003)Huang, Zhang, Kotaki, and
  Ramakrishna}]{huang2003review}
\bibinfo{author}{Z.-M. Huang}, \bibinfo{author}{Y.-Z. Zhang},
  \bibinfo{author}{M.~Kotaki}, \bibinfo{author}{S.~Ramakrishna},
\newblock \bibinfo{title}{A review on polymer nanofibers by electrospinning and
  their applications in nanocomposites},
\newblock \bibinfo{journal}{Composites science and technology}
  \bibinfo{volume}{63} (\bibinfo{year}{2003}) \bibinfo{pages}{2223--2253}.
\bibitem[{Nakajima et~al.(1994)Nakajima, Kajiwara, and
  McIntyre}]{nakajima1994advanced}
\bibinfo{author}{T.~Nakajima}, \bibinfo{author}{K.~Kajiwara},
  \bibinfo{author}{J.~E. McIntyre}, \bibinfo{title}{Advanced fiber spinning
  technology}, \bibinfo{publisher}{Woodhead Publishing}, \bibinfo{year}{1994}.
\bibitem[{Chae and Kumar(2008)}]{chae2008making}
\bibinfo{author}{H.~G. Chae}, \bibinfo{author}{S.~Kumar},
\newblock \bibinfo{title}{Making strong fibers},
\newblock \bibinfo{journal}{Science} \bibinfo{volume}{319}
  (\bibinfo{year}{2008}) \bibinfo{pages}{908--909}.
\bibitem[{Ho et~al.(1993)Ho, Keller, Odell, and Ottewill}]{ho1993preparation}
\bibinfo{author}{C.~Ho}, \bibinfo{author}{A.~Keller},
  \bibinfo{author}{J.~Odell}, \bibinfo{author}{R.~Ottewill},
\newblock \bibinfo{title}{Preparation of monodisperse ellipsoidal polystyrene
  particles},
\newblock \bibinfo{journal}{Colloid and Polymer Science} \bibinfo{volume}{271}
  (\bibinfo{year}{1993}) \bibinfo{pages}{469--479}.
\bibitem[{Subramony(2017)}]{subramony2017employing}
\bibinfo{author}{S.~Subramony},
\newblock \bibinfo{title}{Employing shear induced hydrodynamic lift to achieve
  sieve-free separation based on size in cross-flow filtration}
  (\bibinfo{year}{2017}).
\bibitem[{Bird and Giacomin(2016)}]{bird2016polymer}
\bibinfo{author}{R.~B. Bird}, \bibinfo{author}{A.~J. Giacomin},
\newblock \bibinfo{title}{Polymer fluid dynamics: Continuum and molecular
  approaches},
\newblock \bibinfo{journal}{Annual review of chemical and biomolecular
  engineering} \bibinfo{volume}{7} (\bibinfo{year}{2016})
  \bibinfo{pages}{479--507}.
\bibitem[{Saffman(1965)}]{saffman1965lift}
\bibinfo{author}{P.~G. Saffman},
\newblock \bibinfo{title}{The lift on a small sphere in a slow shear flow},
\newblock \bibinfo{journal}{Journal of fluid mechanics} \bibinfo{volume}{22}
  (\bibinfo{year}{1965}) \bibinfo{pages}{385--400}.
\bibitem[{Bagchi and Balachandar(2002)}]{bagchi2002effect}
\bibinfo{author}{P.~Bagchi}, \bibinfo{author}{S.~Balachandar},
\newblock \bibinfo{title}{Effect of free rotation on the motion of a solid
  sphere in linear shear flow at moderate re},
\newblock \bibinfo{journal}{Physics of Fluids} \bibinfo{volume}{14}
  (\bibinfo{year}{2002}) \bibinfo{pages}{2719--2737}.
\bibitem[{Shi and Rzehak(2019)}]{shi2019lift}
\bibinfo{author}{P.~Shi}, \bibinfo{author}{R.~Rzehak},
\newblock \bibinfo{title}{Lift forces on solid spherical particles in unbounded
  flows},
\newblock \bibinfo{journal}{Chemical Engineering Science} \bibinfo{volume}{208}
  (\bibinfo{year}{2019}) \bibinfo{pages}{115145}.
\bibitem[{Mittal and Iaccarino(2005)}]{mittal2005immersed}
\bibinfo{author}{R.~Mittal}, \bibinfo{author}{G.~Iaccarino},
\newblock \bibinfo{title}{Immersed boundary methods},
\newblock \bibinfo{journal}{Annu. Rev. Fluid Mech.} \bibinfo{volume}{37}
  (\bibinfo{year}{2005}) \bibinfo{pages}{239--261}.
\bibitem[{Griffith and Patankar(2020)}]{griffith2020immersed}
\bibinfo{author}{B.~E. Griffith}, \bibinfo{author}{N.~A. Patankar},
\newblock \bibinfo{title}{Immersed methods for fluid--structure interaction},
\newblock \bibinfo{journal}{Annual review of fluid mechanics}
  \bibinfo{volume}{52} (\bibinfo{year}{2020}) \bibinfo{pages}{421--448}.
\bibitem[{Jain and Shaqfeh(2021)}]{jain2021transient}
\bibinfo{author}{A.~Jain}, \bibinfo{author}{E.~S. Shaqfeh},
\newblock \bibinfo{title}{Transient and steady shear rheology of particle-laden
  viscoelastic suspensions},
\newblock \bibinfo{journal}{Journal of Rheology} \bibinfo{volume}{65}
  (\bibinfo{year}{2021}) \bibinfo{pages}{1269--1295}.
\bibitem[{Shaqfeh(2019)}]{shaqfeh2019rheology}
\bibinfo{author}{E.~S. Shaqfeh},
\newblock \bibinfo{title}{On the rheology of particle suspensions in
  viscoelastic fluids},
\newblock \bibinfo{journal}{AIChE Journal} \bibinfo{volume}{65}
  (\bibinfo{year}{2019}).
\bibitem[{Alves et~al.(2021)Alves, Oliveira, and Pinho}]{alves2021numerical}
\bibinfo{author}{M.~Alves}, \bibinfo{author}{P.~Oliveira},
  \bibinfo{author}{F.~Pinho},
\newblock \bibinfo{title}{Numerical methods for viscoelastic fluid flows},
\newblock \bibinfo{journal}{Annual Review of Fluid Mechanics}
  \bibinfo{volume}{53} (\bibinfo{year}{2021}) \bibinfo{pages}{509--541}.
\bibitem[{Fattal and Kupferman(2004)}]{fattal2004constitutive}
\bibinfo{author}{R.~Fattal}, \bibinfo{author}{R.~Kupferman},
\newblock \bibinfo{title}{Constitutive laws for the matrix-logarithm of the
  conformation tensor},
\newblock \bibinfo{journal}{Journal of Non-Newtonian Fluid Mechanics}
  \bibinfo{volume}{123} (\bibinfo{year}{2004}) \bibinfo{pages}{281--285}.
\bibitem[{Richter et~al.(2010)Richter, Iaccarino, and
  Shaqfeh}]{richter2010simulations}
\bibinfo{author}{D.~Richter}, \bibinfo{author}{G.~Iaccarino},
  \bibinfo{author}{E.~S. Shaqfeh},
\newblock \bibinfo{title}{Simulations of three-dimensional viscoelastic flows
  past a circular cylinder at moderate reynolds numbers},
\newblock \bibinfo{journal}{Journal of fluid mechanics} \bibinfo{volume}{651}
  (\bibinfo{year}{2010}) \bibinfo{pages}{415}.
\bibitem[{Santelli et~al.(2021)Santelli, Orlandi, and
  Verzicco}]{santelli2021finite}
\bibinfo{author}{L.~Santelli}, \bibinfo{author}{P.~Orlandi},
  \bibinfo{author}{R.~Verzicco},
\newblock \bibinfo{title}{A finite--difference scheme for three--dimensional
  incompressible flows in spherical coordinates},
\newblock \bibinfo{journal}{Journal of computational physics}
  \bibinfo{volume}{424} (\bibinfo{year}{2021}) \bibinfo{pages}{109848}.
\bibitem[{Verzicco and Orlandi(1996)}]{verzicco1996finite}
\bibinfo{author}{R.~Verzicco}, \bibinfo{author}{P.~Orlandi},
\newblock \bibinfo{title}{A finite-difference scheme for three-dimensional
  incompressible flows in cylindrical coordinates},
\newblock \bibinfo{journal}{Journal of Computational Physics}
  \bibinfo{volume}{123} (\bibinfo{year}{1996}) \bibinfo{pages}{402--414}.
\bibitem[{Morinishi et~al.(2004)Morinishi, Vasilyev, and
  Ogi}]{morinishi2004fully}
\bibinfo{author}{Y.~Morinishi}, \bibinfo{author}{O.~V. Vasilyev},
  \bibinfo{author}{T.~Ogi},
\newblock \bibinfo{title}{Fully conservative finite difference scheme in
  cylindrical coordinates for incompressible flow simulations},
\newblock \bibinfo{journal}{Journal of Computational Physics}
  \bibinfo{volume}{197} (\bibinfo{year}{2004}) \bibinfo{pages}{686--710}.
\bibitem[{Desjardins et~al.(2008)Desjardins, Blanquart, Balarac, and
  Pitsch}]{desjardins2008high}
\bibinfo{author}{O.~Desjardins}, \bibinfo{author}{G.~Blanquart},
  \bibinfo{author}{G.~Balarac}, \bibinfo{author}{H.~Pitsch},
\newblock \bibinfo{title}{High order conservative finite difference scheme for
  variable density low mach number turbulent flows},
\newblock \bibinfo{journal}{Journal of Computational Physics}
  \bibinfo{volume}{227} (\bibinfo{year}{2008}) \bibinfo{pages}{7125--7159}.
\bibitem[{Yang et~al.(2016)Yang, Krishnan, and Shaqfeh}]{yang2016numerical}
\bibinfo{author}{M.~Yang}, \bibinfo{author}{S.~Krishnan},
  \bibinfo{author}{E.~S. Shaqfeh},
\newblock \bibinfo{title}{Numerical simulations of the rheology of suspensions
  of rigid spheres at low volume fraction in a viscoelastic fluid under shear},
\newblock \bibinfo{journal}{Journal of Non-Newtonian Fluid Mechanics}
  \bibinfo{volume}{233} (\bibinfo{year}{2016}) \bibinfo{pages}{181--197}.
\bibitem[{d'Avino et~al.(2014)d'Avino, Hulsen, Greco, and
  Maffettone}]{d2014bistability}
\bibinfo{author}{G.~d'Avino}, \bibinfo{author}{M.~Hulsen},
  \bibinfo{author}{F.~Greco}, \bibinfo{author}{P.~Maffettone},
\newblock \bibinfo{title}{Bistability and metabistability scenario in the
  dynamics of an ellipsoidal particle in a sheared viscoelastic fluid},
\newblock \bibinfo{journal}{Physical Review E} \bibinfo{volume}{89}
  (\bibinfo{year}{2014}) \bibinfo{pages}{043006}.
\bibitem[{Binagia et~al.(2020)Binagia, Phoa, Housiadas, and
  Shaqfeh}]{binagia2020swimming}
\bibinfo{author}{J.~P. Binagia}, \bibinfo{author}{A.~Phoa},
  \bibinfo{author}{K.~D. Housiadas}, \bibinfo{author}{E.~S. Shaqfeh},
\newblock \bibinfo{title}{Swimming with swirl in a viscoelastic fluid},
\newblock \bibinfo{journal}{Journal of Fluid Mechanics} \bibinfo{volume}{900}
  (\bibinfo{year}{2020}).
\bibitem[{Koch and Subramanian(2006)}]{koch2006stress}
\bibinfo{author}{D.~L. Koch}, \bibinfo{author}{G.~Subramanian},
\newblock \bibinfo{title}{The stress in a dilute suspension of spheres
  suspended in a second-order fluid subject to a linear velocity field},
\newblock \bibinfo{journal}{Journal of non-newtonian fluid mechanics}
  \bibinfo{volume}{138} (\bibinfo{year}{2006}) \bibinfo{pages}{87--97}.
\bibitem[{Einarsson and Mehlig(2017)}]{einarsson2017spherical}
\bibinfo{author}{J.~Einarsson}, \bibinfo{author}{B.~Mehlig},
\newblock \bibinfo{title}{Spherical particle sedimenting in weakly viscoelastic
  shear flow},
\newblock \bibinfo{journal}{Physical Review Fluids} \bibinfo{volume}{2}
  (\bibinfo{year}{2017}) \bibinfo{pages}{063301}.
\bibitem[{Harlen and Koch(1993)}]{harlen1993simple}
\bibinfo{author}{O.~Harlen}, \bibinfo{author}{D.~L. Koch},
\newblock \bibinfo{title}{Simple shear flow of a suspension of fibres in a
  dilute polymer solution at high deborah number},
\newblock \bibinfo{journal}{Journal of Fluid Mechanics} \bibinfo{volume}{252}
  (\bibinfo{year}{1993}) \bibinfo{pages}{187--207}.
\bibitem[{Leal(1975)}]{leal1975slow}
\bibinfo{author}{L.~Leal},
\newblock \bibinfo{title}{The slow motion of slender rod-like particles in a
  second-order fluid},
\newblock \bibinfo{journal}{Journal of Fluid Mechanics} \bibinfo{volume}{69}
  (\bibinfo{year}{1975}) \bibinfo{pages}{305--337}.
\bibitem[{Cox(1970)}]{cox1970motion}
\bibinfo{author}{R.~Cox},
\newblock \bibinfo{title}{The motion of long slender bodies in a viscous fluid
  part 1. general theory},
\newblock \bibinfo{journal}{Journal of Fluid mechanics} \bibinfo{volume}{44}
  (\bibinfo{year}{1970}) \bibinfo{pages}{791--810}.
\bibitem[{Cox(1971)}]{cox1971motion}
\bibinfo{author}{R.~Cox},
\newblock \bibinfo{title}{The motion of long slender bodies in a viscous fluid.
  part 2. shear flow},
\newblock \bibinfo{journal}{Journal of Fluid Mechanics} \bibinfo{volume}{45}
  (\bibinfo{year}{1971}) \bibinfo{pages}{625--657}.
\bibitem[{Batchelor(1970)}]{batchelor1970slender}
\bibinfo{author}{G.~Batchelor},
\newblock \bibinfo{title}{Slender-body theory for particles of arbitrary
  cross-section in stokes flow},
\newblock \bibinfo{journal}{Journal of Fluid Mechanics} \bibinfo{volume}{44}
  (\bibinfo{year}{1970}) \bibinfo{pages}{419--440}.
\bibitem[{D’Avino et~al.(2010)D’Avino, Tuccillo, Maffettone, Greco, and
  Hulsen}]{d2010numerical}
\bibinfo{author}{G.~D’Avino}, \bibinfo{author}{T.~Tuccillo},
  \bibinfo{author}{P.~Maffettone}, \bibinfo{author}{F.~Greco},
  \bibinfo{author}{M.~Hulsen},
\newblock \bibinfo{title}{Numerical simulations of particle migration in a
  viscoelastic fluid subjected to shear flow},
\newblock \bibinfo{journal}{Computers \& fluids} \bibinfo{volume}{39}
  (\bibinfo{year}{2010}) \bibinfo{pages}{709--721}.
\bibitem[{Chwang and Wu(1975)}]{chwang1975hydromechanics}
\bibinfo{author}{A.~T. Chwang}, \bibinfo{author}{T.~Y.-T. Wu},
\newblock \bibinfo{title}{Hydromechanics of low-reynolds-number flow. part 2.
  singularity method for stokes flows},
\newblock \bibinfo{journal}{Journal of Fluid mechanics} \bibinfo{volume}{67}
  (\bibinfo{year}{1975}) \bibinfo{pages}{787--815}.
\bibitem[{Pimenta and Alves(2017)}]{pimenta2017stabilization}
\bibinfo{author}{F.~Pimenta}, \bibinfo{author}{M.~Alves},
\newblock \bibinfo{title}{Stabilization of an open-source finite-volume solver
  for viscoelastic fluid flows},
\newblock \bibinfo{journal}{Journal of Non-Newtonian Fluid Mechanics}
  \bibinfo{volume}{239} (\bibinfo{year}{2017}) \bibinfo{pages}{85--104}.
\bibitem[{Gresho and Sani(1987)}]{gresho1987pressure}
\bibinfo{author}{P.~M. Gresho}, \bibinfo{author}{R.~L. Sani},
\newblock \bibinfo{title}{On pressure boundary conditions for the
  incompressible navier-stokes equations},
\newblock \bibinfo{journal}{International Journal for Numerical Methods in
  Fluids} \bibinfo{volume}{7} (\bibinfo{year}{1987})
  \bibinfo{pages}{1111--1145}.
\bibitem[{Sani et~al.(2006)Sani, Shen, Pironneau, and
  Gresho}]{sani2006pressure}
\bibinfo{author}{R.~L. Sani}, \bibinfo{author}{J.~Shen},
  \bibinfo{author}{O.~Pironneau}, \bibinfo{author}{P.~Gresho},
\newblock \bibinfo{title}{Pressure boundary condition for the time-dependent
  incompressible navier--stokes equations},
\newblock \bibinfo{journal}{International Journal for Numerical Methods in
  Fluids} \bibinfo{volume}{50} (\bibinfo{year}{2006})
  \bibinfo{pages}{673--682}.
\bibitem[{Cai et~al.(2014)Cai, Nonaka, Bell, Griffith, and
  Donev}]{cai2014efficient}
\bibinfo{author}{M.~Cai}, \bibinfo{author}{A.~Nonaka}, \bibinfo{author}{J.~B.
  Bell}, \bibinfo{author}{B.~E. Griffith}, \bibinfo{author}{A.~Donev},
\newblock \bibinfo{title}{Efficient variable-coefficient finite-volume stokes
  solvers},
\newblock \bibinfo{journal}{Communications in Computational Physics}
  \bibinfo{volume}{16} (\bibinfo{year}{2014}) \bibinfo{pages}{1263--1297}.
\bibitem[{Furuichi et~al.(2011)Furuichi, May, and
  Tackley}]{furuichi2011development}
\bibinfo{author}{M.~Furuichi}, \bibinfo{author}{D.~A. May},
  \bibinfo{author}{P.~J. Tackley},
\newblock \bibinfo{title}{Development of a stokes flow solver robust to large
  viscosity jumps using a schur complement approach with mixed precision
  arithmetic},
\newblock \bibinfo{journal}{Journal of Computational Physics}
  \bibinfo{volume}{230} (\bibinfo{year}{2011}) \bibinfo{pages}{8835--8851}.
\bibitem[{Subramanian and Koch(2005)}]{subramanian2005inertial}
\bibinfo{author}{G.~Subramanian}, \bibinfo{author}{D.~L. Koch},
\newblock \bibinfo{title}{Inertial effects on fibre motion in simple shear
  flow},
\newblock \bibinfo{journal}{Journal of Fluid Mechanics} \bibinfo{volume}{535}
  (\bibinfo{year}{2005}) \bibinfo{pages}{383--414}.
\bibitem[{Jiang et~al.(2014)Jiang, Shu, and Zhang}]{jiang2014free}
\bibinfo{author}{Y.~Jiang}, \bibinfo{author}{C.-W. Shu},
  \bibinfo{author}{M.~Zhang},
\newblock \bibinfo{title}{Free-stream preserving finite difference schemes on
  curvilinear meshes},
\newblock \bibinfo{journal}{Methods and applications of analysis}
  \bibinfo{volume}{21} (\bibinfo{year}{2014}) \bibinfo{pages}{1--30}.
\bibitem[{Visbal and Gaitonde(2002)}]{visbal2002use}
\bibinfo{author}{M.~R. Visbal}, \bibinfo{author}{D.~V. Gaitonde},
\newblock \bibinfo{title}{On the use of higher-order finite-difference schemes
  on curvilinear and deforming meshes},
\newblock \bibinfo{journal}{Journal of Computational Physics}
  \bibinfo{volume}{181} (\bibinfo{year}{2002}) \bibinfo{pages}{155--185}.
\bibitem[{Graham(2018)}]{graham2018microhydrodynamics}
\bibinfo{author}{M.~D. Graham}, \bibinfo{title}{Microhydrodynamics, Brownian
  motion, and complex fluids}, volume~\bibinfo{volume}{58},
  \bibinfo{publisher}{Cambridge University Press}, \bibinfo{year}{2018}.
\bibitem[{Afonso et~al.(2009)Afonso, Oliveira, Pinho, and
  Alves}]{afonso2009log}
\bibinfo{author}{A.~Afonso}, \bibinfo{author}{P.~J. Oliveira},
  \bibinfo{author}{F.~Pinho}, \bibinfo{author}{M.~Alves},
\newblock \bibinfo{title}{The log-conformation tensor approach in the
  finite-volume method framework},
\newblock \bibinfo{journal}{Journal of Non-Newtonian Fluid Mechanics}
  \bibinfo{volume}{157} (\bibinfo{year}{2009}) \bibinfo{pages}{55--65}.
\bibitem[{d’Avino et~al.(2010)d’Avino, Maffettone, Greco, and
  Hulsen}]{d2010viscoelasticity}
\bibinfo{author}{G.~d’Avino}, \bibinfo{author}{P.~Maffettone},
  \bibinfo{author}{F.~Greco}, \bibinfo{author}{M.~Hulsen},
\newblock \bibinfo{title}{Viscoelasticity-induced migration of a rigid sphere
  in confined shear flow},
\newblock \bibinfo{journal}{Journal of Non-Newtonian Fluid Mechanics}
  \bibinfo{volume}{165} (\bibinfo{year}{2010}) \bibinfo{pages}{466--474}.
\bibitem[{Zhong et~al.(2022)Zhong, He, Yang, Bi, and Yang}]{zhong2022modeling}
\bibinfo{author}{H.~Zhong}, \bibinfo{author}{Y.~He}, \bibinfo{author}{E.~Yang},
  \bibinfo{author}{Y.~Bi}, \bibinfo{author}{T.~Yang},
\newblock \bibinfo{title}{Modeling of microflow during viscoelastic polymer
  flooding in heterogenous reservoirs of daqing oilfield},
\newblock \bibinfo{journal}{Journal of Petroleum Science and Engineering}
  \bibinfo{volume}{210} (\bibinfo{year}{2022}) \bibinfo{pages}{110091}.
\bibitem[{Hulsen et~al.(2005)Hulsen, Fattal, and Kupferman}]{hulsen2005flow}
\bibinfo{author}{M.~A. Hulsen}, \bibinfo{author}{R.~Fattal},
  \bibinfo{author}{R.~Kupferman},
\newblock \bibinfo{title}{Flow of viscoelastic fluids past a cylinder at high
  weissenberg number: stabilized simulations using matrix logarithms},
\newblock \bibinfo{journal}{Journal of Non-Newtonian Fluid Mechanics}
  \bibinfo{volume}{127} (\bibinfo{year}{2005}) \bibinfo{pages}{27--39}.
\bibitem[{Kopp(2008)}]{kopp2008efficient}
\bibinfo{author}{J.~Kopp},
\newblock \bibinfo{title}{Efficient numerical diagonalization of hermitian
  3$\times$ 3 matrices},
\newblock \bibinfo{journal}{International Journal of Modern Physics C}
  \bibinfo{volume}{19} (\bibinfo{year}{2008}) \bibinfo{pages}{523--548}.
\bibitem[{Saad and Schultz(1986)}]{saad1986gmres}
\bibinfo{author}{Y.~Saad}, \bibinfo{author}{M.~H. Schultz},
\newblock \bibinfo{title}{Gmres: A generalized minimal residual algorithm for
  solving nonsymmetric linear systems},
\newblock \bibinfo{journal}{SIAM Journal on scientific and statistical
  computing} \bibinfo{volume}{7} (\bibinfo{year}{1986})
  \bibinfo{pages}{856--869}.
\bibitem[{Notay(2010)}]{notay2010aggregation}
\bibinfo{author}{Y.~Notay},
\newblock \bibinfo{title}{An aggregation-based algebraic multigrid method},
\newblock \bibinfo{journal}{Electronic transactions on numerical analysis}
  \bibinfo{volume}{37} (\bibinfo{year}{2010}) \bibinfo{pages}{123--146}.
\bibitem[{Napov and Notay(2012)}]{napov2012algebraic}
\bibinfo{author}{A.~Napov}, \bibinfo{author}{Y.~Notay},
\newblock \bibinfo{title}{An algebraic multigrid method with guaranteed
  convergence rate},
\newblock \bibinfo{journal}{SIAM journal on scientific computing}
  \bibinfo{volume}{34} (\bibinfo{year}{2012}) \bibinfo{pages}{A1079--A1109}.
\bibitem[{Notay(2008)}]{agmg}
\bibinfo{author}{Y.~Notay}, \bibinfo{title}{Agmg software and documentation.},
  \bibinfo{year}{2008}. \URLprefix \url{http://agmg.eu}.
\bibitem[{Golub and Van~Loan(2013)}]{golub2013matrix}
\bibinfo{author}{G.~H. Golub}, \bibinfo{author}{C.~F. Van~Loan},
  \bibinfo{title}{Matrix computations}, \bibinfo{publisher}{JHU press},
  \bibinfo{year}{2013}.
\bibitem[{Rapaport and Rapaport(2004)}]{rapaport2004art}
\bibinfo{author}{D.~C. Rapaport}, \bibinfo{author}{D.~C.~R. Rapaport},
  \bibinfo{title}{The art of molecular dynamics simulation},
  \bibinfo{publisher}{Cambridge university press}, \bibinfo{year}{2004}.
\bibitem[{Yu and Shao(2007)}]{yu2007direct}
\bibinfo{author}{Z.~Yu}, \bibinfo{author}{X.~Shao},
\newblock \bibinfo{title}{A direct-forcing fictitious domain method for
  particulate flows},
\newblock \bibinfo{journal}{Journal of computational physics}
  \bibinfo{volume}{227} (\bibinfo{year}{2007}) \bibinfo{pages}{292--314}.
\bibitem[{Padhy et~al.(2013)Padhy, Shaqfeh, Iaccarino, Morris, and
  Tonmukayakul}]{padhy2013simulations}
\bibinfo{author}{S.~Padhy}, \bibinfo{author}{E.~Shaqfeh},
  \bibinfo{author}{G.~Iaccarino}, \bibinfo{author}{J.~Morris},
  \bibinfo{author}{N.~Tonmukayakul},
\newblock \bibinfo{title}{Simulations of a sphere sedimenting in a viscoelastic
  fluid with cross shear flow},
\newblock \bibinfo{journal}{Journal of Non-Newtonian Fluid Mechanics}
  \bibinfo{volume}{197} (\bibinfo{year}{2013}) \bibinfo{pages}{48--60}.
\bibitem[{Kim and Karrila(2013)}]{kim2013microhydrodynamics}
\bibinfo{author}{S.~Kim}, \bibinfo{author}{S.~J. Karrila},
  \bibinfo{title}{Microhydrodynamics: principles and selected applications},
  \bibinfo{publisher}{Courier Corporation}, \bibinfo{year}{2013}.
\bibitem[{Falgout and Yang(2002)}]{falgout2002hypre}
\bibinfo{author}{R.~D. Falgout}, \bibinfo{author}{U.~M. Yang},
\newblock \bibinfo{title}{hypre: A library of high performance
  preconditioners},
\newblock in: \bibinfo{booktitle}{International Conference on computational
  science}, \bibinfo{organization}{Springer}, \bibinfo{year}{2002}, pp.
  \bibinfo{pages}{632--641}.
\bibitem[{Chung et~al.(2002)}]{chung2002computational}
\bibinfo{author}{T.~Chung}, et~al., \bibinfo{title}{Computational fluid
  dynamics}, \bibinfo{publisher}{Cambridge university press},
  \bibinfo{year}{2002}.
\bibitem[{Fornberg(1988)}]{fornberg1988generation}
\bibinfo{author}{B.~Fornberg},
\newblock \bibinfo{title}{Generation of finite difference formulas on
  arbitrarily spaced grids},
\newblock \bibinfo{journal}{Mathematics of computation} \bibinfo{volume}{51}
  (\bibinfo{year}{1988}) \bibinfo{pages}{699--706}.
\bibitem[{Nourgaliev and Theofanous(2007)}]{nourgaliev2007high}
\bibinfo{author}{R.~R. Nourgaliev}, \bibinfo{author}{T.~G. Theofanous},
\newblock \bibinfo{title}{High-fidelity interface tracking in compressible
  flows: unlimited anchored adaptive level set},
\newblock \bibinfo{journal}{Journal of Computational Physics}
  \bibinfo{volume}{224} (\bibinfo{year}{2007}) \bibinfo{pages}{836--866}.
\bibitem[{Constantinescu and Lele(2002)}]{constantinescu2002highly}
\bibinfo{author}{G.~S. Constantinescu}, \bibinfo{author}{S.~Lele},
\newblock \bibinfo{title}{A highly accurate technique for the treatment of flow
  equations at the polar axis in cylindrical coordinates using series
  expansions},
\newblock \bibinfo{journal}{Journal of Computational Physics}
  \bibinfo{volume}{183} (\bibinfo{year}{2002}) \bibinfo{pages}{165--186}.
\bibitem[{Jeffery(1922)}]{jeffery1922motion}
\bibinfo{author}{G.~B. Jeffery},
\newblock \bibinfo{title}{The motion of ellipsoidal particles immersed in a
  viscous fluid},
\newblock \bibinfo{journal}{Proceedings of the Royal Society of London. Series
  A, Containing papers of a mathematical and physical character}
  \bibinfo{volume}{102} (\bibinfo{year}{1922}) \bibinfo{pages}{161--179}.
\bibitem[{Oberbeck(1876)}]{oberbeck1876ueber}
\bibinfo{author}{A.~Oberbeck},
\newblock \bibinfo{title}{Ueber station{\"a}re fl{\"u}ssigkeitsbewegungen mit
  ber{\"u}cksichtigung der inneren reibung. (on steady state fluid flow and the
  calculation of the drag.)}  (\bibinfo{year}{1876}).
\bibitem[{Happel and Brenner(2012)}]{happel2012low}
\bibinfo{author}{J.~Happel}, \bibinfo{author}{H.~Brenner}, \bibinfo{title}{Low
  Reynolds number hydrodynamics: with special applications to particulate
  media}, volume~\bibinfo{volume}{1}, \bibinfo{publisher}{Springer Science \&
  Business Media}, \bibinfo{year}{2012}.
\bibitem[{Siewert et~al.(2014)Siewert, Kunnen, Meinke, and
  Schr{\"o}der}]{siewert2014orientation}
\bibinfo{author}{C.~Siewert}, \bibinfo{author}{R.~Kunnen},
  \bibinfo{author}{M.~Meinke}, \bibinfo{author}{W.~Schr{\"o}der},
\newblock \bibinfo{title}{Orientation statistics and settling velocity of
  ellipsoids in decaying turbulence},
\newblock \bibinfo{journal}{Atmospheric research} \bibinfo{volume}{142}
  (\bibinfo{year}{2014}) \bibinfo{pages}{45--56}.
\bibitem[{Andersson and Jiang(2019)}]{andersson2019forces}
\bibinfo{author}{H.~I. Andersson}, \bibinfo{author}{F.~Jiang},
\newblock \bibinfo{title}{Forces and torques on a prolate spheroid:
  Low-reynolds-number and attack angle effects},
\newblock \bibinfo{journal}{Acta Mechanica} \bibinfo{volume}{230}
  (\bibinfo{year}{2019}) \bibinfo{pages}{431--447}.
\bibitem[{Dabade et~al.(2015)Dabade, Marath, and
  Subramanian}]{dabade2015effects}
\bibinfo{author}{V.~Dabade}, \bibinfo{author}{N.~K. Marath},
  \bibinfo{author}{G.~Subramanian},
\newblock \bibinfo{title}{Effects of inertia and viscoelasticity on sedimenting
  anisotropic particles},
\newblock \bibinfo{journal}{Journal of Fluid Mechanics} \bibinfo{volume}{778}
  (\bibinfo{year}{2015}) \bibinfo{pages}{133--188}.
\bibitem[{Jiang et~al.(2021)Jiang, Zhao, Andersson, Gustavsson, Pumir, and
  Mehlig}]{jiang2021inertial}
\bibinfo{author}{F.~Jiang}, \bibinfo{author}{L.~Zhao}, \bibinfo{author}{H.~I.
  Andersson}, \bibinfo{author}{K.~Gustavsson}, \bibinfo{author}{A.~Pumir},
  \bibinfo{author}{B.~Mehlig},
\newblock \bibinfo{title}{Inertial torque on a small spheroid in a stationary
  uniform flow},
\newblock \bibinfo{journal}{Physical Review Fluids} \bibinfo{volume}{6}
  (\bibinfo{year}{2021}) \bibinfo{pages}{024302}.
\bibitem[{Brenner(1999)}]{brenner1999screening}
\bibinfo{author}{M.~P. Brenner},
\newblock \bibinfo{title}{Screening mechanisms in sedimentation},
\newblock \bibinfo{journal}{Physics of fluids} \bibinfo{volume}{11}
  (\bibinfo{year}{1999}) \bibinfo{pages}{754--772}.
\bibitem[{Segre(2007)}]{segre2007inertial}
\bibinfo{author}{P.~Segre},
\newblock \bibinfo{title}{Inertial screening in sedimentation},
\newblock \bibinfo{journal}{arXiv preprint arXiv:0709.0995}
  (\bibinfo{year}{2007}).
\bibitem[{Bergougnoux and Guazzelli(2021)}]{bergougnoux2021dilute}
\bibinfo{author}{L.~Bergougnoux}, \bibinfo{author}{{\'E}.~Guazzelli},
\newblock \bibinfo{title}{Dilute sedimenting suspensions of spheres at small
  inertia},
\newblock \bibinfo{journal}{Journal of Fluid Mechanics} \bibinfo{volume}{914}
  (\bibinfo{year}{2021}).
\bibitem[{D’Avino et~al.(2008)D’Avino, Hulsen, Snijkers, Vermant, Greco,
  and Maffettone}]{d2008rotation}
\bibinfo{author}{G.~D’Avino}, \bibinfo{author}{M.~A. Hulsen},
  \bibinfo{author}{F.~Snijkers}, \bibinfo{author}{J.~Vermant},
  \bibinfo{author}{F.~Greco}, \bibinfo{author}{P.~L. Maffettone},
\newblock \bibinfo{title}{Rotation of a sphere in a viscoelastic liquid
  subjected to shear flow. part i: Simulation results},
\newblock \bibinfo{journal}{Journal of rheology} \bibinfo{volume}{52}
  (\bibinfo{year}{2008}) \bibinfo{pages}{1331--1346}.
\bibitem[{Batchelor(1970)}]{batchelor1970stress}
\bibinfo{author}{G.~Batchelor},
\newblock \bibinfo{title}{The stress system in a suspension of force-free
  particles},
\newblock \bibinfo{journal}{Journal of fluid mechanics} \bibinfo{volume}{41}
  (\bibinfo{year}{1970}) \bibinfo{pages}{545--570}.
\bibitem[{Koch et~al.(2016)Koch, Lee, and Mustafa}]{koch2016stress}
\bibinfo{author}{D.~L. Koch}, \bibinfo{author}{E.~F. Lee},
  \bibinfo{author}{I.~Mustafa},
\newblock \bibinfo{title}{Stress in a dilute suspension of spheres in a dilute
  polymer solution subject to simple shear flow at finite deborah numbers},
\newblock \bibinfo{journal}{Physical Review Fluids} \bibinfo{volume}{1}
  (\bibinfo{year}{2016}) \bibinfo{pages}{013301}.
\bibitem[{Einstein(1906)}]{einsteinoriginal}
\bibinfo{author}{A.~Einstein},
\newblock \bibinfo{title}{Eine neue bestimmung der moleküldimensionen},
\newblock \bibinfo{journal}{Annalen der Physik} \bibinfo{volume}{324}
  (\bibinfo{year}{1906}) \bibinfo{pages}{289--306}.
\bibitem[{Jain et~al.(2019)Jain, Einarsson, and Shaqfeh}]{jain2019extensional}
\bibinfo{author}{A.~Jain}, \bibinfo{author}{J.~Einarsson},
  \bibinfo{author}{E.~S. Shaqfeh},
\newblock \bibinfo{title}{Extensional rheology of a dilute particle-laden
  viscoelastic solution},
\newblock \bibinfo{journal}{Physical Review Fluids} \bibinfo{volume}{4}
  (\bibinfo{year}{2019}) \bibinfo{pages}{091301}.
\bibitem[{Sharma and Koch(2022)}]{TransientPaper}
\bibinfo{author}{A.~Sharma}, \bibinfo{author}{D.~L. Koch},
\newblock \bibinfo{title}{Extensional rheology of a dilute suspension of
  spheres in a viscoelastic liquid},
\newblock \bibinfo{journal}{In preparation}  (\bibinfo{year}{2022}).

\end{thebibliography}

\end{document}